\documentclass[11pt,a4paper]{iopart}

\usepackage[utf8]{inputenc}
\expandafter\let\csname equation*\endcsname\relax

\expandafter\let\csname endequation*\endcsname\relax 

\usepackage{amsfonts}
\usepackage{amssymb}
\usepackage{amsthm}
\usepackage{newlfont}
\usepackage{makeidx,bm}
\usepackage{graphicx}
\usepackage{iopams}
\usepackage{todonotes}
\usepackage{hyperref}
\usepackage{xcolor, colortbl}
\usepackage{multirow}
\usepackage{makecell}
\usepackage{rotating}
\usepackage{graphicx}
\usepackage{subcaption} 

\usepackage{amsmath}

\makeatletter
\newcommand{\reqnomode}{\tagsleft@true}
\newcommand{\leqnomode}{\tagsleft@false}
\makeatother

\newcommand{\Int}{\int\limits}

\def\bea{\begin{eqnarray}}
\def\eea{\end{eqnarray}}
\def\ba{\begin{array}}
\def\ea{\end{array}}

\def\la{\langle}
\def\ra{\rangle}
\def \dr{D_R}

\usepackage{etoolbox}

\makeatletter
\def\@mkboth#1#2{}
\newlength\appendixwidth
\preto\appendix{\addtocontents{toc}{\protect\patchl@section}}
\newcommand{\patchl@section}{%
  \settowidth{\appendixwidth}{\textbf{Appendix }}%
  \addtolength{\appendixwidth}{1.5em}%
  \patchcmd{\l@section}{1.5em}{\appendixwidth}{}{\ddt}%
}
\makeatother

\begin{document}

\title[]{Tagged particle behavior in a harmonic chain of direction reversing active Brownian particles}
\author{Shashank Prakash}
\address{Raman Research Institute, Bengaluru 560080, India\\
Email: shashankp@rrimail.rri.res.in}
\author{Urna Basu}
\address{S. N. Bose National Centre for Basic Sciences, Kolkata 700106, India\\
Email: urna@bose.res.in}
\author{Sanjib Sabhapandit}
\address{Raman Research Institute, Bengaluru 560080, India\\
Email: sanjib@rri.res.in}

\date{July 2022}

\begin{abstract}
   We study the tagged particle dynamics in a harmonic chain of direction reversing active Brownian particles, with spring constant $k$, rotation diffusion coefficient $\dr$, and directional reversal rate $\gamma$. We exactly compute the tagged particle position variance for quenched and annealed initial orientations of the particles. For well-separated time scales, $k^{-1}, \dr^{-1}$ and $\gamma^{-1}$, the strength of spring constant $k$ relative to $\dr$ and $\gamma$ gives rise to different coupling limits and for each coupling limit there are short, intermediate, and long time regimes. In the thermodynamic limit, we show that, to the leading order, the tagged particle variance exhibits an algebraic growth $t^{\nu}$, where the value of the exponent $\nu$ depends on the specific regime. For a quenched initial orientation, the exponent $\nu$ crosses over from $3$ to $1/2$, via intermediate values $5/2$ or $1$, depending on the specific coupling limits. On the other hand, for the annealed initial orientation, $\nu$ crosses over from $2$ to $1/2$ via an intermediate value $3/2$ or $1$ for strong coupling limit and weak coupling limit respectively.  An additional time scale $t_N=N^2/k$ emerges for a system with a finite number of oscillators $N$. We show that the behavior of the tagged particle variance across $t_N$  can be expressed in terms of a crossover scaling function, which we find exactly. Moreover, we study the velocity autocorrelation. Finally, we characterize the stationary state behavior of the separation between two consecutive particles by calculating the corresponding spatio-temporal correlation function. 
\end{abstract}

\noindent\rule{\hsize}{2pt}
\tableofcontents
\noindent\rule{\hsize}{2pt}

\maketitle

\section{Introduction}
 
A system of interacting particles undergoing Newtonian dynamics results in a diffusive motion for a tagged particle with the mean-squared displacement growing linearly with time \cite{sfms1, lebowitz1967kinetic_1-d-hs2, lebowitz1972velocity_1-d-hs3,roy_hs4,roy2_hs5}. The tagged particle dynamics is still diffusive for a collection of interacting diffusive particles, in dimensions $d\geq2$, albeit with a lower diffusivity. However, in one-dimensional interacting diffusive systems, the tagged particle dynamics show a subdiffusive behavior. Systems with symmetric dynamics generically show a $\sqrt{t}$ growth for mean-squared displacement, which, for example, has been observed for hard-particle diffusive gases~\cite{sfms2,van1983diffusion_hr1,percus1974_hr2,sfmpd1_hr3,levitt1973_hr-4}, symmetric simple exclusion process~\cite{Arratia_ip_sep0,sep-book,spitzer_sep-book2,ferrari1986_sep1,richards1977theory_sep2}, random average process~\cite{IPP-rap1,rajesh2001_ravp-3}, and systems with harmonic interactions~\cite{TPDin1dchain,buttiker1980_o-hi1}, to name a few. On the other hand, systems with asymmetric dynamics give rise to subdiffusive behavior with different exponents~\cite{asep-3,asep-0, asep-1, asep-4,asep-5,asep-2}. This classical problem of tagged particle motion in one-dimensional interacting systems is still of much interest and various aspects of it are still being investigated. Examples include the study of typical and large deviations of the displacement fluctuations~\cite{kollmann_n_sfd-0,krapivsky-tridib2014-prl,sadhu2015large,hegde_n_sfd-2,grabsch2023universal_new_sfd-2,IPP-sfd,Galanti_2014}, duality relations~\cite{rizkallah2023duality}, effects of pinning~\cite{shamik-rp-1}, disorder~\cite{ben2009_n-sfd-1}, and memory~\cite{IAP-msd-2,IAP-msd-1}, and entropy production~\cite{stc-ther_ipp}.

Though much has been known about the passive systems, the study of tagged particle dynamics in an interacting system of active particles is still nascent. Active particles self-propel by extracting energy from the surroundings at the individual level, generating a directed motion with a typical persistence time~\cite{ame1,ame2,ame3,romanczuk2012active_n-am, bechinger2016active_n-am2, fodor2018statistical_n-am3}. Theoretical attempts to characterize active motion often use minimal statistical models,  where the position of an active particle evolves via an overdamped Langevin equation 
$\dot{\boldsymbol{r}}(t)=\boldsymbol{v}(t)$. Different stochastic dynamics of the self-propulsion velocity $\boldsymbol{v}(t)$ leads to different models for active particles. Run-and-tumble particle (RTP)~\cite{mme-rt1,mme-rt2,mme-rt3}, active Ornstein-Uhlenbeck process (AOUP)~\cite{szamel2014}, and active Brownian particle (ABP)~\cite{mme-abp2,mme-abp3}  are some of the basic models for which the single particle dynamics have been extensively studied~\cite{mme-rt4,santra2023long_n-rtp,sevilla2020two_n-rtp,basu2020exact_n-rtp,mme-abp1,santra2022universal,martin2021aoup_n}. While at times much larger than the persistence time these models generically show a diffusive behavior, the signatures of activity are manifest at short-times, where the dynamics is strongly non-diffusive. This non-diffusive behavior is expected to affect the dynamics of a tagged particle in interacting active systems. Indeed, the variance of the position of a tagged particle, in a chain of such active particles, shows a crossover~\cite{prashant,marconi2024active,chaki2019enhanced_hi} from a superdiffusive behavior, $t^{\mu}$ with $\mu>1/2$, to the subdiffusive behavior $\sqrt{t}$, typical of passive systems. Furthermore, the interplay between self-propulsion and interaction gives rise to novel behaviors such as  formation of active crystals~\cite{caprini2020time}, actuation in active solids~\cite{baconnier2022selective}, and generalized active Einstien relation~\cite{caprini2024emergent}.

Recently, it has been found that an active particle dynamics, namely, direction reversing active Brownian particle (DRABP), where the self-propulsion velocity $\boldsymbol{v}(t)$ has multiple time-scales, display several novel features, in the intermediate time regimes, due to the interplay between these time-scales~\cite{Ion1,ion2-drabp2}. It is natural to ask, how the presence of the additional time-scales changes the tagged particle behavior in an interacting system of such particles. In this paper, we comprehensively characterize the tagged particle dynamics in a harmonic chain of DRABPs with periodic boundary conditions, by studying the position-variance for quenched and annealed initial orientations. We find that the presence of an additional time-scale leads to the emergence of new dynamical regimes with distinct super-diffusive growth of the variance  [see Tables.~\ref{fig:table_FIC}-\ref{fig:table_UIC}]. We also study the effect of the finite size of the chain on the variance and compute the scaling function that describes the crossover across the finite-size time scale. While the fluctuations of the tagged particle grow with time, the separations between the adjacent tagged particles reach a stationary state, which we characterize analytically by computing the spatio-temporal correlation among the separation variables.

The paper is organized as follows: In section~\ref{mar}, we define the model and present a summary of the main results. The basic setup of the calculation is presented in Sec.~\ref{pv}. Sections~\ref{fio_pv}, and \ref{uio_pv} are devoted to the detailed derivation of the tagged particle variance for quenched and annealed initial orientations, respectively.  We explore the finite-size effects in Sec.~\ref{fse}. In Sec.~\ref{vc-sec} we investigate the velocity autocorrelation for the annealed initial condition. We discuss the statistics of the separations between two consecutive particles in Sec.~\ref{sec:separations}. Finally, we conclude with some open questions in Sec.~\ref{sec-conclusion}.

\section{Model and Results} \label{mar}

The position vector $\vec r(t)$  of a single direction reversing active Brownian particle (DRABP) evolves via the Langevin equation, 
\begin{equation}
 \dot{\boldsymbol{r}}(t)= v_{0} \, \sigma(t) \, \hat n(t), \label{eq:single_drabp}
\end{equation}
where $v_{0}$ denotes the self-propulsion speed of the particle and $\sigma(t)$ is a dichotomous noise that alternates between $\pm 1$  with a constant rate $\gamma$. The orientation vector $\hat n(t)$ undergoes a  Brownian motion on the unit sphere. In two dimensions, $\hat n(t)= (\cos \theta(t), \, \sin \theta(t))$, where the orientation angle $\theta(t)$ undergoes a Brownian motion with diffusion constant $D_R$. Clearly, the dynamics of a single DRABP has two inherent time-scales $\gamma^{-1}$ and $\dr^{-1}$. The reversal time-scale $\gamma^{-1}$ indicates the typical time between two direction reversal events. On the other hand, the rotational diffusion time-scale $\dr^{-1}$ indicates the characteristic time beyond which the orientation vector $\hat{n}(t)$ becomes uncorrelated, i.e., $\langle \hat{n}(t)\cdot \hat{n}(t') \rangle  \sim e^{-\dr |t-t'|}$. 

It has been shown that the position fluctuations of a DRABP show distinctly different behavior in different dynamical regimes characterized by the two active time-scales $D_R^{-1}$ and $\gamma^{-1}$ \cite{Ion1}. From a strongly non-diffusive and anisotropic behavior at early-times $t \ll (\gamma^{-1}, D_R^{-1})$, the DRABP crosses over to an effective diffusive behavior at late-times $t \gg (\gamma^{-1}, D_R^{-1})$. In fact, in the limit $v_0 \to \infty$ and $(D_R + 2 \gamma) \to \infty$, while keeping 
\begin{align}
D_\text{eff}= \frac{v_0^2}{2(D_R + 2 \gamma)}, \label{eq:deff_def}
\end{align}

fixed (which corresponds to the late-time regime), the DRABP typically behaves like a passive Brownian particle, with a Gaussian position distribution,
\begin{equation}
   \mathcal{P}(\vec r,t) = \frac 1{4 \pi D_\text{eff}t} \exp{\left( -\frac{|\vec r|^2}{4 D_\text{eff} t} \right)}.
\end{equation}
Most interestingly, for $\gamma \gg D_R$, a non-trivial scaling distribution emerges in the intermediate regime $\gamma^{-1} \ll t \ll  D_R^{-1}$.

In this work, we consider a periodic chain of $N$ harmonically coupled direction reversing active Brownian particles, where each particle is uniquely identified by an index $\alpha=0,1,2, \dotsc, N-1$. In particular,  we focus on the fluctuations of $x$-components of positions $\{ x_\alpha(t) \}$. The potential energy is given by, 
\begin{align}
 U(x_0,x_1, \dotsc x_{N-1}) = \frac k 2 \sum_{\alpha=0}^{N-1} (x_{\alpha} - x_{\alpha+1})^2, 
\end{align}
with the periodic boundary condition $x_{N}=x_0$. The presence of the harmonic coupling  introduces another time-scale $k^{-1}$, in addition to the two active time-scales $D_R^{-1}$ and $\gamma^{-1}$. This coupling time-scale $k^{-1}$ signifies the typical time beyond which the particles feel the interaction with other particles. In other words, for $t \ll k^{-1}$, the system behaves as a collection of independent DRABPs.

The position  of the $\alpha$-th particle evolves via the Langevin equation,
\begin{align}
\dot{x}_{\alpha}(t)&=k[x_{\alpha +1}(t)+x_{\alpha -1}(t) -2x_{\alpha}(t)] + \xi_{\alpha}(t), \label{eomx} 
\end{align}
where,
\begin{align}
\xi_{\alpha}(t)&= v_{0}\, \sigma_{\alpha}(t) \cos\theta_{\alpha}(t),\label{eq:xi}
\end{align}
denotes the active noise. The dichotomous noises $\{\sigma_{\alpha}(t)\}$ independently alternate between $\pm 1$  with a constant rate $\gamma$, and
the orientations $\{ \theta_\alpha(t)\}$ undergo independent Brownian motions,
\begin{equation}
 \dot{\theta}_{\alpha}(t)=\sqrt{2 D_R}\,\,{\eta}_\alpha(t),
 \label{od}
\end{equation}
where $\{\eta_\alpha(t) \}$ denote independent Gaussian white noises with  $\langle{\eta}_\alpha(t) \rangle =0$, and $\langle \eta_\alpha(t) \eta_\beta(t') \rangle = \delta_{\alpha,\beta} \delta(t-t')$. We consider the initial condition $x_{\alpha}(0)=0$ and $\sigma_{\alpha}(0)=1$ for all $\alpha$ and initial orientations $\{ \theta_{\alpha}(0)\}$ to be either quenched or annealed. For the quenched case, we set the initial orientations of all the particles to a specific value $\theta_0$, i.e., $\{ \theta_{\alpha}(0)=\theta_0\}$. On the other hand, the initial orientation of each particle is drawn independently from a uniform distribution in $[0,2\pi]$ for the annealed case.  

The objective of this work is to characterize the fluctuations of the position $x_\alpha(t)$ of a tagged particle through its mean and variance. The presence of the three distinct time scales leads to a set of different dynamical regimes, each characterized by a distinct behavior of the position variance, ranging from subdiffusive to superdiffusive. These leading order dynamical behaviors, in each regime, also depend on the initial orientations of the particles. Tables \ref{fig:table_FIC} and \ref{fig:table_UIC} present a summary of the distinct dynamical behaviors, in thermodynamic limit, i.e., $N \to \infty$, for the quenched and annealed initial orientations, respectively. In particular, we find that, at short-time regime, i.e., at times much smaller than all the time scales of the system, $\langle x_\alpha^2(t) \rangle_c \sim t^3$, reminiscent of the short time regime of the independent DRABP. On the other hand, at late times, i.e., at a time much larger than all the time scales, $\langle x_\alpha^2(t) \rangle_c \sim \sqrt{t}$, similar to the behavior of a tagged particle in a harmonic chain of Brownian particles~\cite{TPDin1dchain}. Furthermore, we study the effect of finite-size on the position variance. An additional time scale, $t_N=N^2/k$, appears due to the finiteness of the system, and we observe a crossover in variance from subdiffusive behavior for $t \ll t_N$ to diffusive behavior for $t \gg t_N$. This crossover is captured by a scaling behavior of the variance $ \langle x_\alpha^2(t) \rangle_c =  D_{\text{eff}}\sqrt{t/k}\,f(t/t_N)$, where the crossover function is given by \eref{scaling_fn1} exactly. 

We also study the velocity autocorrelation $\la v_{\alpha}(t_1) v_{\alpha}(t_2) \ra$, where we focus on the stationary state behavior. We find that at late times, i.e., $|t_1- t_2| \gg \{ k^{-1}, (\dr+2\gamma)^{-1} \} $ the correlation function decays as a power-law given by $|t_1-t_2|^{-3/2}$.

Finally, we investigate the statistics of the separations between the adjacent particles $y_\alpha(t)= x_{\alpha+1}(t) - x_\alpha(t)$, which eventually reaches a stationary state. In the passive limit, i.e., when $D_R \to \infty$, and  $v_0 \to \infty$, keeping $D_\text{eff}$ fixed,  the stationary state of $\{ y_\alpha\}$ is given by the Boltzmann distribution,
\begin{align}   
P(\{ y_\alpha\}) \propto \exp{\left[-\frac k{2D_\text{eff}}\sum_{\alpha=0}^{N-1} y_\alpha^2\right ]}, \label{eq:ydist_passive}
\end{align}
 for a thermodynamically large system. For finite $D_R$, i.e., in the active regime, the system reaches a nonequilibrium steady state which is no longer given by the above Boltzmann distribution. We characterize the signatures of activity in this stationary state by computing the spatio-temporal two-point correlation,
\begin{math}
C(\beta, \tau) = \lim_{t \to \infty}\la y_0(t)y_\beta(t + \tau) \ra. 
\end{math} 
We find that, in the thermodynamic limit, $N \to \infty$
\begin{equation}
    C(\beta,\tau)= \sum_{n=0}^{\infty} \tilde{C}_n(\beta) \, \frac{\tau^n}{n!} ,
\end{equation}
where the coefficients $\tilde C_n(\beta)$ can be computed explicitly [see Sec.~\ref{sec:separations}].
Moreover, in the large activity limit, the equal-time spatial correlation decays exponentially, 
\begin{align}
     C(\beta,0) = \frac {D_\text{eff}}{k} \frac{e^{- |\beta|/\sqrt{\mu}}}{2 \sqrt{\mu}},~~ \text{with} \quad \mu = \frac {k}{\dr+ 2 \gamma}. \label{eq:C_b0_active}
\end{align}
%
%
On the other hand, for large $\tau \gg (\dr+2 \gamma)^{-1}$, we have,
\begin{align}
   C(\beta,\tau) &=  \frac{D_{\text{eff}} }{2k \sqrt{\pi k }} \frac{e^{-\beta^2/4k\tau}}{\sqrt{\tau}}.  
   \label{2tcrA}
\end{align}

In the following sections, we provide detailed derivations of the results mentioned above.

\begin{center}
 \begin{table}[t]
   \resizebox{17cm}{!}{
    \renewcommand{\arraystretch}{1.3}
   \hskip 0cm
    \begin{tabular}{|p{0.2cm}|p{1.7cm}|c|c|c|c|}
    
    \hline
   & & $t \ll \tau_{1} \ll \tau_2 \ll \tau_{3} $ & $\tau_{1} \ll t \ll \tau_2 \ll \tau_{3} $ & $\tau_{1} \ll \tau_2 \ll t \ll \tau_{3} $  & $\tau_{1} \ll \tau_2  \ll \tau_{3} \ll t $   \\
   & & \scriptsize \textbf{Short-time regime} & \scriptsize \textbf{Early-intermediate regime} & \scriptsize \textbf{Late-intermediate regime} & \scriptsize \textbf{Long-time regime} 
     \\ \hline

       &$\tau_{1}=k^{-1}$ \newline $\tau_{2}=D^{-1}_R$ \newline $\tau_{3}= \gamma^{-1}$& \cellcolor{gray!30}   & \cellcolor{pink!80}& \multicolumn{-2}{c}{\cellcolor{blue!30}} & \cellcolor{blue!30}\\
         \cline{2-2} \cline{5-5}
         
         \multirow{-2.4}{*} {\begin{turn}{-270} \tiny \textbf{Strong Coupling}\end{turn} } &$\tau_{1}=k^{-1}$ \newline
          $\tau_{2}=\gamma^{-1}$ \newline $\tau_{3}=D^{-1}_R$ & \cellcolor{gray!30} & \multirow{-2.5}{*} { \cellcolor{pink!80}  \makecell{R-II \\ \\ $t^{5/2}$  \\ \\ \text{[see \eref{eir-2}]} } } & \multirow{1}{*} { \cellcolor{blue!50}  \makecell{ R-III \\$\sqrt{t}$ \\ \text{[see \eref{lir_lot}]}  } } &\cellcolor{blue!30}\\
          \cline{1-1} \cline{2-2} \cline{4-5} 
        
         & $\tau_{1}=D^{-1}_R$ \newline $\tau_{2}=\gamma^{-1}$ \newline $\tau_{3}=k^{-1}$& \cellcolor{gray!30} & \multicolumn{-2}{c}{\cellcolor{yellow!40} } & \cellcolor{yellow!40} & \cellcolor{blue!30}\\
         \cline{2-2} \cline{4-4}

      \multirow{-2.2}{*} {\begin{turn}{-270} \tiny \textbf{Weak Coupling}\end{turn} } & $\tau_{1}=\gamma^{-1}$ \newline $\tau_{2}=D^{-1}_R$ \newline $\tau_{3}=k^{-1}$ & \cellcolor{gray!30}  & \multirow{-0.9}{*} { \cellcolor{yellow!80}  \makecell{ R-VI \\ $t$ \\ \text{[see \eref{eir_ydk}]} } } &  \multirow{-3}{*} { \cellcolor{yellow!40}  \makecell{ R-V \\ \\$t$  \\ \\ \text{[see \eref{dyk_ir_2}]} } } &\cellcolor{blue!30}\\
       \cline{1-1} \cline{2-2} \cline{4-5}

      & $\tau_{1}=D^{-1}_R$ \newline
          $\tau_{2}=k^{-1}$ \newline
          $\tau_{3}=\gamma^{-1}$& \cellcolor{gray!30}  & \multirow{-0.9}{*} { \cellcolor{yellow!40}  \makecell{ R-VII \\ $t$ \\  \text{[see \eref{dyk_ir_2}}]} }& \multicolumn{-2}{c}{\cellcolor{blue!30}}&\cellcolor{blue!30}\\
         \cline{2-2} \cline{4-4}
        
       \multirow{-2.5}{*} {\begin{turn}{-270} \tiny \textbf{Moderate Coupling}\end{turn} }  & $\tau_{1}=\gamma^{-1}$ \newline $\tau_{2}=k^{-1}$ \newline $\tau_{3}=D^{-1}_R$ &  \multirow{-12}{*} {\cellcolor{gray!30} \makecell{
        R-I \\ \\ $t^3$  \\ \\  \text{[see \eref{stb}]}  } } & \multirow{-0.9}{*} { \cellcolor{yellow!80}  \makecell{R-VIII \\$t$ \\  \text { [see \eref{eir_ydk}] } } } & \multirow{1}{*} { \cellcolor{blue!50}  \makecell{ R-IX \\$\sqrt{t}$ \\  \text{[see \eref{lir_lot}]}  } } & \multirow{-12}{*} {\cellcolor{blue!30}  \makecell{ R-IV \\ \\ $\sqrt{t}$ \\ \\  \text{ [see \eref{eq:longt_final}]}  } }   \\
          \hline 
    \end{tabular} } 
     
    \caption{Tabular representation of the leading order behavior of the variance in the different dynamical regimes starting with quenched initial orientation ($\theta_0 \ne \pi/2$). The nine distinct regimes are indicated by R-I, R-II, \dots, R-IX. The first column indicates the relative time-scales. The second column defines the three different time scales and the topmost row indicates the specific regime considered. The equations describing the theoretical predictions of the variance in the corresponding regimes are also mentioned. }
    \label{fig:table_FIC}
       
\end{table}
\end{center}

\begin{center}
 \begin{table}[t]
    \centering
    \resizebox{13cm}{!} {
    \renewcommand{\arraystretch}{1.3}
   \hskip-0cm
    \begin{tabular}{|c|p{3.2cm}|c|c|c|}
    \hline
   & $\tau_{1} \ll \tau_2 $ & $t \ll \tau_{1} \ll \tau_2 $ & $\tau_{1} \ll t \ll \tau_2  $ & $\tau_{1} \ll \tau_2 \ll t  $    \\
  & &  \scriptsize \textbf{Short-time regime} & \scriptsize \textbf{ Intermediate regime} &  \scriptsize \textbf{Long-time regime} \\ 
  
  \hline
   
&    & \cellcolor{gray!30} &  \cellcolor{pink!80} &\cellcolor{blue!30} \\ 

& $\tau_{1}=k^{-1}$ \newline $\tau_{2}= (D_{R}+2\gamma)^{-1}$   & \cellcolor{gray!30} & \cellcolor{pink!80}  & \cellcolor{blue!30} \\
 
\multirow{-3.6}{*} {\begin{turn}{-270} \tiny \textbf{Strong Coupling}\end{turn} }       &  & \cellcolor{gray!30} & \multirow{-3.6}{*} { \cellcolor{pink!80}  \makecell{ R-II \\ $t^{3/2}$ \\ \text{[see \eref{kdy_ir_ric}]}  } }  & \cellcolor{blue!30} \\
    \cline{0-1}  \cline{3-3} 

&  & \cellcolor{gray!30} & \cellcolor{yellow!40}  & \cellcolor{blue!30} \\   

 & $\tau_{1}=(D_{R}+2\gamma)^{-1}$ \newline $\tau_{2}=k^{-1}$ & \cellcolor{gray!30} & \cellcolor{yellow!40}  & \cellcolor{blue!30} \\ 
      
 \multirow{-3.2}{*} {\begin{turn}{-270} \tiny \textbf{Weak Coupling}\end{turn} }   &    &  \multirow{-7.5}{*} {\cellcolor{gray!30} \makecell{ R-I \\ \\ $t^2$  \\ \\ \text{[see \eref{smalltime_ric}]} } } & \multirow{-3.6}{*} { \cellcolor{yellow!40}  \makecell{ R-III \\ $t$ \\ \text{[see \eref{dyk_ir_ric}]}  } }  & \multirow{-7.5}{*} {\cellcolor{blue!30}  \makecell{R-IV \\ \\ $\sqrt{t}$ \\ \\ \text{[see \eref{eq:longt_final_ric}]} } }  \\
 
\hline 
\end{tabular} }
    \caption{Tabular representation of the leading order behavior of the variance in the different dynamical regimes starting with annealed initial orientation. The four distinct regimes are indicated by R-I, R-II, R-III, and R-IV. The first column indicates the relative time-scales. The second column defines the two time-scales and the topmost row indicates the specific regime considered. The equations describing the theoretical predictions of the variance in the corresponding regimes are also mentioned.}
           \label{fig:table_UIC}       
\end{table}
\end{center}

\section{Variance of the position of a tagged particle}  \label{pv}

In this section, we set up the formalism to compute the variance of the position $x_{\alpha}$ of a given tagged particle with a fixed $\alpha$. To compute the variance,
\begin{equation}
    \big \langle  x_{\alpha}^2(t) \big \rangle_{c}= \big \langle  x_{\alpha}^2(t) \big \rangle - \big \langle  x_{\alpha}(t) \big \rangle^2, \label{eq:vardef}
\end{equation}
it is convenient to go to the normal modes of the harmonic chain. The decoupled  normal modes satisfy a set of $N$ first-order differential equations, 
\begin{equation}
 \dot {\Tilde{x}}_{s}(t)= -a_{s} \Tilde{x}_{s}(t) + \Tilde{\xi}_{s}(t),~~~ \text{with} \quad  a_{s}=4k\sin^2\Big({\frac{\pi s}{N}}\Big),
 \label{eomF}
\end{equation}
where $\{  \tilde x_s(t); s= 0,1, \dotsc N-1 \}$ denote the discrete Fourier transformation (DFT) of $\{  x_\alpha(t)\}$ and $\{\tilde \xi_s(t)\}$ denotes the DFT of the active noise \eref{eq:xi}. We note that any arbitrary $ \{ f_{\alpha}(t) \}$ and its DFT $\{\Tilde{f}_{s}(t)\}$ with respect to $\alpha$, are related by,
\begin{align}
 \Tilde{f}_{s}(t)= \frac{1}{N} \sum_{\alpha=0}^{N-1} \exp{\left (-\frac{i2\pi s \alpha}{N}\right)} f_{\alpha}(t) \quad\text{and}\quad 
 f_{\alpha}(t) =\sum_{s=0}^{N-1} \exp{\left(\frac{i2\pi s \alpha}{N} \right)} \Tilde{f}_{s}(t). \label{ft}   
\end{align}
The set of equations \eqref{eomF} can be formally solved to obtain,
\begin{equation}
  \Tilde{x}_{s}(t)= \Tilde{x}_{s}(0) e^{-a_{s}t}+ e^{-a_{s}t}  \int_{0}^{t} e^{a_{s} t_1} \Tilde{\xi}_{s}(t_1) \, dt_1, \label{sF}
\end{equation}
where $\Tilde{x}_{s}(0)$ denotes the DFT of the initial position profile. For annealed initial orientation, $\cos{\theta_{\alpha}(0)}$ has a symmetric distribution, resulting in $\big \langle x_{\alpha}(t) \big \rangle=0$, for initial position $\{x_{\alpha}(0)=0\}$. On the other hand, for the quenched initial orientation, $\{\theta_{\alpha}(0)=\theta_0\}$ with $\{x_{\alpha}(0)=0\}$, taking the average over \eref{sF} and inverting it back via \eref{ft}, we get,
\begin{equation}
    \big \langle x_{\alpha}(t) \big \rangle=v_0\, t\cos{\theta_{0}}\,e^{-(\dr+2\gamma)t}. \label{eq:mean}
\end{equation}



Using \eqref{ft}, the variance of the tagged particle position \eqref{eq:vardef} can be expressed as, 
\begin{equation}
  \big \langle   x^2_{\alpha}(t) \big \rangle_{c}= \sum_{s=0}^{N-1} \sum_{s'=0}^{N-1}  \exp{\Bigg[\frac{i2\pi \alpha (s-s')}{N}\Bigg]}\big \langle   \Tilde{x}_{s}(t) \Tilde{x}_{s'}^*(t) \big \rangle_{c},
  \label{cifs}
  \end{equation}
where
\begin{equation}
     \big \langle   \Tilde{x}_{s}(t) \Tilde{x}_{s'}^*(t) \big \rangle_{c}=  \big \langle   \Tilde{x}_{s}(t) \Tilde{x}_{s'}^*(t) \big \rangle - \big \langle   \Tilde{x}_{s}(t) \big \rangle \big \langle  \Tilde{x}_{s'}^*(t) \big \rangle ,  
\end{equation}
is the correlation of the Fourier modes. Here, $\Tilde{x}_{s'}^*(t) $ denotes the complex conjugate of $\Tilde{x}_{s'}(t)$ and $\big \langle ...  \big \rangle$ denotes the statistical average over the active noise.  

The correlation $\big \langle   \Tilde{x}_{s}(t) \Tilde{x}_{s'}^*(t) \big \rangle_{c}$ can be expressed in terms of the correlations of the Fourier transforms of the active noise, using Eq.~\eref{sF}, 
\begin{equation}
  \big \langle   \Tilde{x}_{s}(t) \Tilde{x}^*_{s'}(t) \big \rangle_{c}=  e^{-(a_{s}+a_{s'})t}  \int_{0}^{t} dt_{1} \int_{0}^{t} dt_{2}\, e^{a_{s}t_{1}+a_{s'}t_{2}} \, \big \langle   \Tilde{\xi}_{s}(t_{1}) \Tilde{\xi}^*_{s'}(t_{2}) \big \rangle _{c}. 
\label{cfs}
\end{equation}
The noise correlation in Fourier space, in turn, could be written in terms of the correlation of the noise in real space,
\begin{equation}
 \big \langle \Tilde{\xi}_{s}(t_{1}) \Tilde{\xi}_{s'}^*(t_{2})\big \rangle _{c}= \frac{1}{N^2} \sum_{\alpha=0}^{N-1} 
\exp{\Bigg[-\frac{i2\pi \alpha (s-s')}{N}\Bigg]}
 \big \langle   \xi_{\alpha}(t_{1}) \xi_{\alpha}(t_{2}) \big \rangle_{c}.
 \label{noise_IFT}
  \end{equation}

We note that the active noise $\xi_\alpha(t)$ for different particles are independent and the  two-time correlation,
$G(t_{1},t_{2})=\big \langle   \xi_{\alpha}(t_{1}) \xi_{\alpha}(t_{2}) \big \rangle_{c}$ is independent of $\alpha$ for both the quenched and annealed initial conditions we have considered here. In particular, for the quenched initial orientation [see ~\ref{t-t-corr} for the detailed calculation], we have,
\begin{align}
G(t_{1},t_{2}) =\frac{v_{0}^2}{2}\Big[ & e^{-(D_{R}+2\gamma) \mid t_{1}-t_{2} \mid} + \cos{ 2\theta_{0}}e^{-\big(D_{R}(  t_{1}+ t_{2}+ 2 \min [ t_{1},t_{2} ]) +2\gamma \mid t_{1}-t_{2} \mid \big)}
\cr &- 2\cos^2\theta_{0}e^{-(D_{R}+2\gamma)(t_{1}+ t_{2})} \Big].
\label{ficc}
\end{align}
On the other hand, for the annealed initial orientation [see ~\ref{t-t-corr} for a detailed calculation],
\begin{align}
G(t_{1},t_{2}) =\frac{v_{0}^2}{2} e^{-(D_{R}+2\gamma) \mid t_{1}-t_{2} \mid}.
\label{g_ric}
\end{align}
Note that, in the passive limit, $v_0^2 \to \infty$, $\dr \to \infty$, and $\gamma \to \infty$, \eref{ficc} and \eref{g_ric} reduce to 
\begin{equation}
    G(t_1,t_2)= 2 D_{\text{eff}} \, \delta (t_1-t_2),
    \label{g_pc}
\end{equation}
with $v_0^2/(\dr+2\gamma)= 2 D_{\text{eff}} $ finite.

We can express the noise auto-correlation in the Fourier space in \eref{noise_IFT}, in terms of $G(t_1,t_2)$, yielding,
\begin{equation}
  \big \langle \Tilde{\xi}_{s}(t_{1}) \Tilde{\xi}_{s'}^*(t_{2})\big \rangle _{c}= \frac{ G(t_{1},t_{2})}{N} \delta_{s,s'}.
 \label{noise_IFT2}
  \end{equation}
Using \eref{noise_IFT2} in \eref{cfs} we get,
\begin{equation}
  \big \langle   \Tilde{x}_{s}(t) \Tilde{x}_{s'}^*(t) \big \rangle_{c}=  \frac{\delta_{s,s'}}{N} e^{-2a_{s}t} \Int_{0}^{t} \,dt_{1} \Int_{0}^{t} dt_{2} \, e^{a_{s}(t_{1}+t_{2})} G(t_{1},t_{2}). \label{xs_xsp}
\end{equation}
Using the above equation in \eref{cifs}, we get a simplified expression for the 
variance of the position of the tagged particle,
\begin{equation}
  \big \langle   x^2_{\alpha}(t) \big \rangle_{c}= \sum_{s=0}^{N-1} \big \langle \big |\,\tilde x_{s}(t)\big |^2 \big \rangle_{c}.  
  \label{cifa_1}
  \end{equation}
  
In the next two sections, we explicitly compute this variance for the quenched and annealed initial orientations.

\section{Quenched initial orientation}   \label{fio_pv}

\begin{figure} [t]
            \centering
            \includegraphics[scale=0.7]{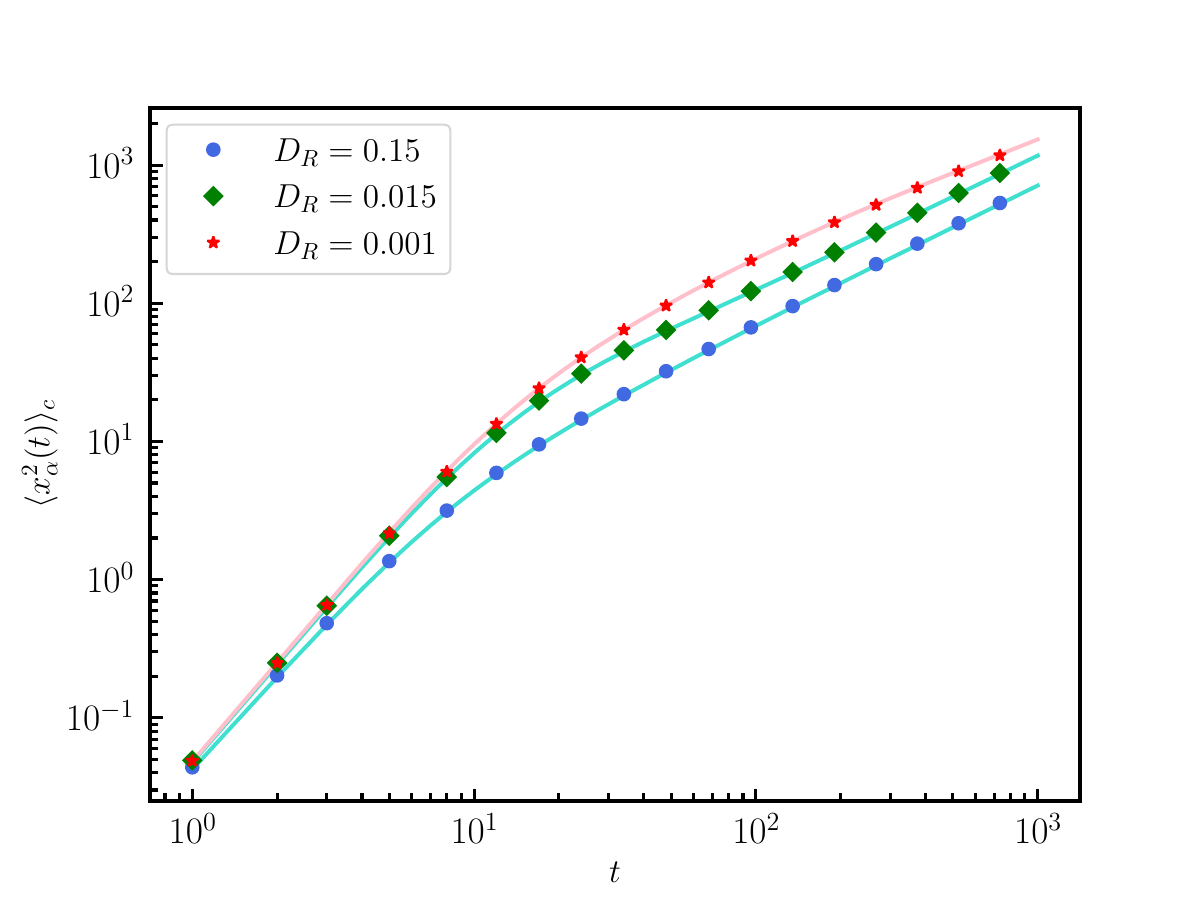}
            \caption{Comparison between the theoretical expression for variance in \eref{sum} (shown by the solid lines) with numerical simulation for different values of $D_{R}$ keeping $k=1$ and $\gamma=0.1$ fixed. The symbols indicate the data obtained from numerical simulations with $N=4$ and $v_0=1$.}
           \label{fig:sum-1}
\end{figure}

For the quenched initial orientation $\{ \theta_\alpha(0) = \theta_0\}$, using \eref{ficc} and \eref{xs_xsp} in \eref{cifa_1}, we get,
\begin{align}\label{sum}
 & \big \langle x^2_{\alpha}(t) \big \rangle_{c} = \frac{v_{0}^2}{N}\sum_{s=0}^{N-1}  \bigg [   \frac{1}{2(\dr+2\gamma)}\Bigg( \frac{1-e^{-2 a_{s}t}}{a_{s}}  +  \frac{e^{-(D_{R}+2\gamma)t} e^{-a_{s}t}-1}{\dr+2\gamma+ a_{s}} 
+  \frac{e^{-a_{s}t}(e^{-(\dr+2\gamma)t} -e^{-a_{s}t})}{\dr+2\gamma- a_{s}} \Bigg)
\cr & + \frac{\cos{2\theta_{0}}}{2(\dr-2\gamma)} \Bigg( \frac{e^{-4\dr t}-e^{-2a_{s}t}}{2\dr- a_{s}}  +  \frac{e^{-(\dr+2\gamma)t} e^{-a_{s}t}-e^{-4\dr t}}{3\dr-2\gamma-a_{s} } 
 + \frac{e^{-a_{s}t}(e^{-a_{s}t}-e^{-(\dr+2\gamma)t} )}{ \dr+2\gamma-a_{s} } \Bigg) 
 \cr & - \frac{\cos^2{\theta_{0}}(e^{-(\dr+2\gamma)t}-e^{-a_{s} t})^{2}}{(\dr+2\gamma- a_{s} )^2}  \bigg],
\end{align}
where $a_s$ is defined in \eref{eomF}. The above equation is exact, and for any finite $N$, the position variance of the tagged particle can be computed by numerically evaluating the sum. In \fref{fig:sum-1} we compare this exact result with numerical simulations for a set of different parameters and find excellent agreement, as expected. Note that, in the passive limit the above equation reduces to the tagged particle variance in a harmonic chain of Brownian particles,
\begin{equation}
    \big \langle x^2_{\alpha}(t) \big \rangle_{c} = \frac{D_{\text{eff}}}{N}\sum_{s=0}^{N-1}  \frac{1-e^{-2 a_{s}t}}{a_{s}}.
    \label{hc_pp}
\end{equation}

It is particularly interesting to consider the limit of thermodynamically large system size, i.e., $N\rightarrow \infty$. In this limit, setting $2 \pi s/N =q$,
the sum in \eref{sum} can be converted into an integral, which is most conveniently expressed as,
\begin{align}
\label{bqeqn2_1}
& \big \langle  x_{\alpha}^2(t) \big \rangle_{c} = v_0^2\int_{-\pi}^{\pi}  \,\frac{dq}{2 \pi}    \bigg [   \frac{1}{2(\dr+2\gamma)}\Bigg( \frac{1-e^{-2 b_{q}t}}{b_{q}}  +  \frac{e^{-(D_{R}+2\gamma)t} e^{-b_{q}t}-1}{\dr+2\gamma+ b_{q}} 
+  \frac{e^{-b_{q}t}(e^{-(\dr+2\gamma)t} -e^{- b_{q}t})}{\dr+2\gamma- b_{q}} \Bigg) \cr & + \frac{\cos{2\theta_{0}}}{2(\dr-2\gamma)} \Bigg( \frac{e^{-4\dr t}-e^{-2b_{q}t}}{2\dr- b_{q}}  +  \frac{e^{-(\dr+2\gamma)t} e^{-b_{q}t}-e^{-4\dr t}}{3\dr-2\gamma-b_{q}} 
 + \frac{e^{-b_{q}t}(e^{-b_{q}t}-e^{-(\dr+2\gamma)t})}{ \dr+2\gamma-b_{q}} \Bigg) \cr & - \frac{\cos^2{\theta_{0}}(e^{-(\dr+2\gamma)t}-e^{-b_{q}t})^{2}}{(\dr+2\gamma-b_{q})^2}  \bigg], 
\end{align}  
where, we have defined,
\begin{equation}
    b_q= 4k\sin^2 \left( {\frac{q}{2}} \right). 
    \label{eq:bq}
\end{equation}
The first term and part of the second term in \eref{bqeqn2_1} can be integrated exactly, giving,
\begin{align}
B_1 (t) \equiv \frac{v_0^2}{2(\dr+2\gamma)} \int_{-\pi}^{\pi}  \,\frac{dq}{2 \pi}       \frac{1-e^{-2 b_{q}t}}{b_{q} } = 2 D_\text{eff}\,   \Big [I_0(4 k t)+I_1(4 k t) \Big ] t \,   e^{-4 k t}, \label{eq:term1}
\end{align}
and,
\begin{align}
B_2 \equiv \frac{v_0^2}{2(\dr+2\gamma)} \int_{-\pi}^{\pi} \frac{dq}{2\pi} \frac{1}{\big[(D+2\gamma)+ b_{q} \big]} =  \frac {D_\text{eff}}{\sqrt{(\dr + 2\gamma)(\dr + 2\gamma+ 4 k)}}, \label{eq:term2}
\end{align}
where $I_n(z)$ denotes the modified Bessel function of first kind and $D_\text{eff}$ is defined in \eref{eq:deff_def}. This simplifies \eref{bqeqn2_1} to,
\begin{align}
& \big \langle x^2_{\alpha}(t) \big \rangle_{c} = B_1(t) - B_2 +v_0^2\int_{-\pi}^{\pi}  \,\frac{dq}{2 \pi}    \bigg [  \frac{e^{-b_{q}t}}{2(\dr+2\gamma)} \Bigg \{   \frac{e^{-(D_{R}+2\gamma)t} }{\dr+2\gamma+ b_{q}} 
\cr 
&  +  \frac{e^{-(\dr+2\gamma)t} -e^{- b_{q}t}}{\dr+2\gamma- b_{q}} \Bigg \} + \frac{\cos{2\theta_{0}}}{2(\dr-2\gamma)} \Bigg \{ \frac{e^{-4\dr t}-e^{-2b_{q}t}}{2\dr- b_{q} }  +  \frac{e^{-(\dr+2\gamma)t} e^{-b_{q}t}-e^{-4\dr t}}{ 3\dr-2\gamma-b_{q}} 
\cr 
& - \frac{e^{-b_{q}t}(e^{-(\dr+2\gamma)t}-e^{-b_{q}t})}{\dr+2\gamma-b_{q} } \Bigg \} - \frac{\cos^2{\theta_{0}}(e^{-(\dr+2\gamma)t}-e^{-b_{q}t})^{2}}{\big[\dr+2\gamma- b_{q} \big]^2}  \bigg]. 
\label{bqeqn2}
\end{align} 
Note that, in the passive limit we have $\big \langle x^2_{\alpha}(t) \big \rangle_{c}= B_1(t)$.


 We now analyze the above equation in various dynamical regimes emerging from the interplay of the time-scales  $\dr^{-1}$, $\gamma^{-1}$ and $k^{-1}$. For any given distinct values of $\{\dr^{-1}, \gamma^{-1}, k^{-1}\}$, there are three different time-scales, the smallest $\tau_1 = \min (\dr^{-1}, \gamma^{-1}, k^{-1})$, the largest $\tau_3 = \max (\dr^{-1}, \gamma^{-1}, k^{-1})$, and $\tau_2$, given by the third one. We consider the most interesting scenario where the time-scales are well separated, 
such that $\tau_1 \ll  \tau_2 \ll  \tau_3$. Correspondingly, for each ordering of $\dr$, $\gamma$ and $k$, there are four dynamical regimes, 
\begin{enumerate}
\item Short-time regime:  $t \ll \tau_1 $
\item Early-intermediate regime : $\tau_1 \ll t \ll \tau_2$
\item Late-intermediate regime : $\tau_2 \ll t \ll \tau_3$
\item Long-time regime:  $t \gg \tau_3$.
\end{enumerate}

To analyse the behaviour of the tagged particle variance in the various regimes, it is convenient to recast \eref{bqeqn2} using a change of variable $z^2 = b_q t$, as,
\begin{align}
\big \langle  x^2_{\alpha}(t)  \big \rangle_{c} &= B_1(t) - B_2 + \frac{v_0^2}{2 \pi  \sqrt{kt}} \int_{-\sqrt{4kt}}^{\sqrt{4kt}}  \,\ dz  \bigg [      \frac{e^{-(D_{R}+2\gamma)t} e^{-z^{2}}}{2(D_{R}+2\gamma)\big[(D_{R}+2\gamma)+ (z^{2}/t) \big]} \cr 
&+ \frac{e^{-(D_{R}+2\gamma)t} e^{-z^{2}}-e^{-2 z^{2}}}{2(D_{R}+2\gamma)\big[(D_{R}+2\gamma)- (z^{2}/t) \big]} 
+\frac{\cos{2\theta_{0}}(e^{-4D_{R}t}-e^{-2z^{2}})}{(D_{R}-2\gamma)(4D_{R}-2 (z^{2}/t))}  \cr 
&+  \frac{\cos{2\theta_{0}}(e^{-(D_{R}+2\gamma)t} e^{-z^{2}}-e^{-4D_{R}t})}{2(D_{R}-2\gamma) \big[(3D_{R}-2\gamma)-(z^{2}/t)\big] } + \frac{\cos{2\theta_{0}}(e^{-2 z^{2}}-e^{-(D_{R}+2\gamma)t} e^{-z^{2}})}{2(D_{R}-2\gamma) \big[(D_{R}+2\gamma)-(z^{2}/t)\big] } \cr 
&- \frac{ \cos^2{\theta_{0}}(e^{-(D_{R}+2\gamma)t}-e^{-z^{2}})^{2}}{\big[(D_{R}+2\gamma)- (z^{2}/t) \big]^2}   \bigg] \frac{1 }{ \sqrt[]{1-z^{2}/(4kt)} }.
\label{zeqn}
\end{align}

Before the intermediate regimes, we first discuss the short-time and long-time regimes, where the dynamical behaviors are, in fact, independent of the ordering of the time-scales. 

\subsection{Short-time regime $(t \ll \tau_1)$ (R-I)} 
\begin{figure} [t]
            \centering
            \includegraphics[scale=0.7]{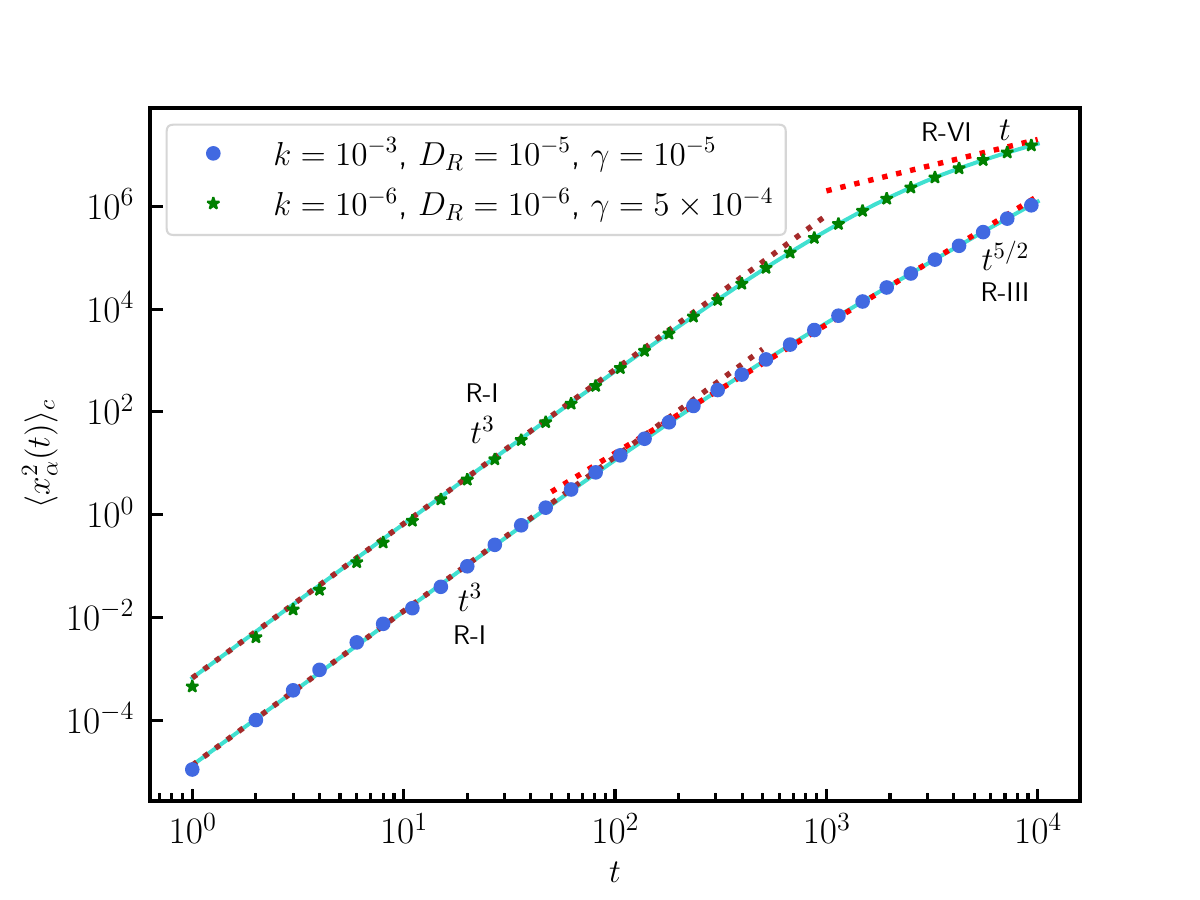}
            \caption{The crossover behavior of the variance from the short-time regime to the early-intermediate regime for quenched initial orientation. The solid lines are obtained by numerically integrating \eref{bqeqn2}. The dotted lines correspond to the asymptotic behavior given by \eref{stb}, \eref{eq:st_coup_EI_1}, and \eref{eir_ydk}. The symbols indicate the data obtained from numerical simulations with $N=500$ and $v_0=1$.
            }
           \label{st_fg}
\end{figure}

In this regime the time  $t$ is much smaller than all the time-scales present in the system. Hence, expanding \eref{bqeqn2} in Taylor series of $t$, we get the leading order behavior of the tagged particle variance,
\begin{align}
 \big \langle  x^2_{\alpha}(t)  \big \rangle_{c} = \frac{2v_{0}^2t^{3}}{3} \big[ D_{R} \sin^2{\theta_{0}}+2\gamma\cos^2{\theta_{0}}\big] + O(t^4).
 \label{stb}
 \end{align}
Note that, this leading order behavior is independent of $k$ and is the same as that of a single DRABP in the short-time regime ($t \ll (k^{-1},\dr^{-1}, \gamma^{-1}$) \cite{Ion1}, since the particles do not feel the effect of the harmonic coupling. However, the effect of coupling shows up in the next order correction as,
\begin{align}
\frac{v^2_0}{12} \bigg[(D_{R}-2\gamma)[7D_{R}+6(k+\gamma)]\cos{2\theta_{0}} -3(D_{R}+2\gamma)[D_{R}+2(k+\gamma)]\bigg]t^{4}.
\end{align}
 In \fref{st_fg} we compare the numerical simulation for the short-time behavior for different orders of time scale with the asymptotic behavior \eref{stb} and get excellent agreement.

\subsection{Long-time regime $(t \gg \tau_3)$ (R-IV)} In the long-time regime, $t$ is much larger than all the time-scales of the system, and consequently, we have, $\dr t \gg 1, \gamma t \gg 1$ as well as $k t \gg 1$. Using these limits in \eref{zeqn}, we get,
\begin{align}
\big \langle   x^2_{\alpha}(t) \big \rangle_{c} &= B_1(t) - B_2 + v_0^2\int_{-\sqrt{4 kt}}^{\sqrt{4 kt}}  \,\frac{dz\, e^{-2z^2}}{2 \pi \sqrt{k t}}     \bigg [ \frac{\cos{2\theta_{0}}}{2(\dr-2\gamma) \big[(\dr +2\gamma)-(z^2/t)\big] }    \cr &   -  \frac{\cos{2\theta_{0}}}{2(\dr-2\gamma)(2\dr- (z^2/t)) }   -  \frac{1}{2(\dr+2\gamma)\big[(\dr+2\gamma)- (z^2/t) \big]} \cr 
&- \frac{\cos^2{\theta_{0}}}{\big[(\dr+2\gamma)-  (z^2/t) \big]^2}  \bigg]\frac 1{\sqrt{1-z^2/(4 kt)}}.
\label{eq:longtime}
\end{align}
Moreover, the leading order behaviour of $B_1(t)$ in \eref{eq:term1} is given by,
\begin{align}
B_1(t) = \frac{D_\text{eff}}{\sqrt{\pi k}}  \Big(\sqrt{2t}  - \frac 1{16k \sqrt{2t}}  \Big )+O(t^{-3/2}).
\label{c1t_eq}
\end{align}

\begin{figure} [t]
            \centering
            \includegraphics[scale=0.7]{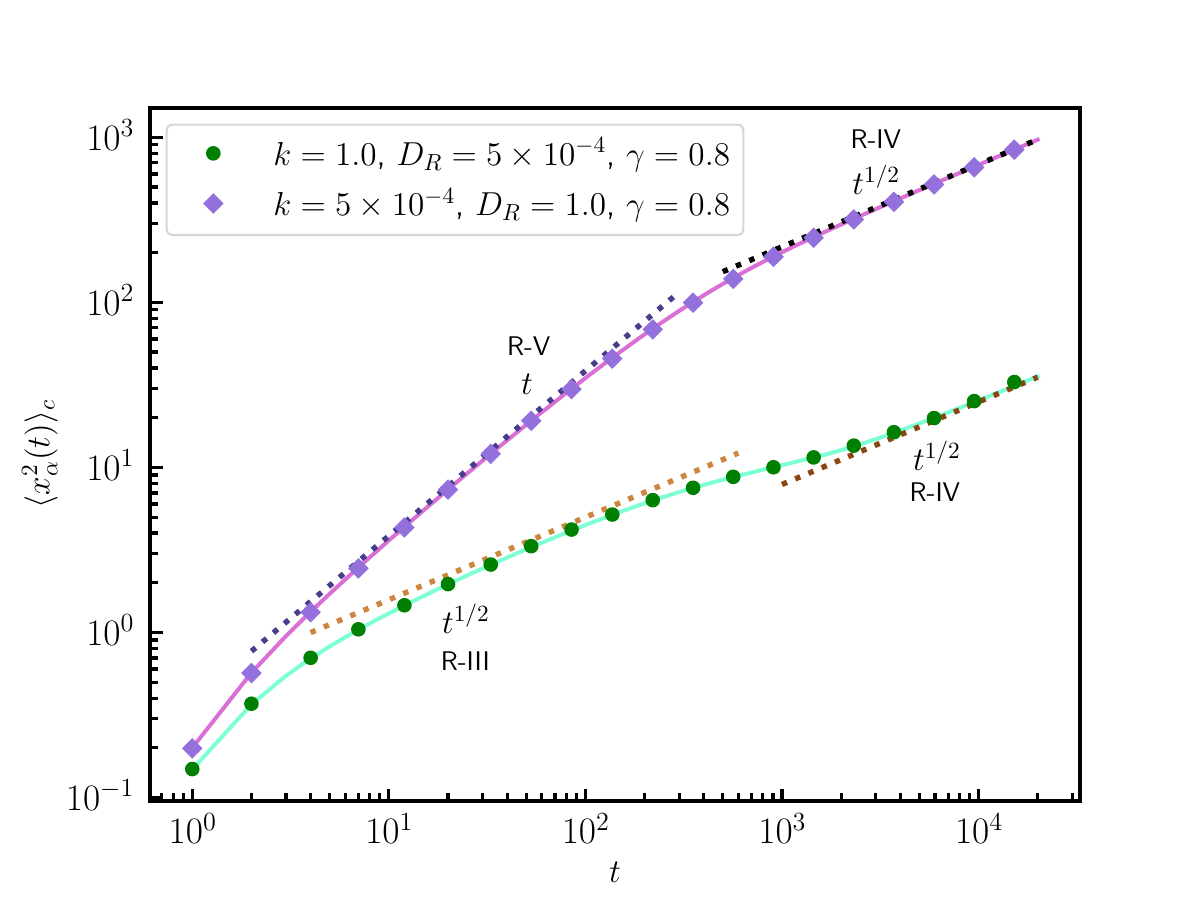}
            \caption{The crossover behavior of the variance from the late-intermediate regime to long-time regime for quenched initial orientation for $k \ll \dr$ and $k \gg \dr$ with fixed $\gamma$. The solid lines are obtained by numerically integrating \eref{bqeqn2}. The dotted lines indicate the asymptotic behavior given by \eref{eq:longt_final}, \eref{lir_lot}, and \eref{Lir_lot}. The symbols indicate the data obtained from numerical simulations with $N=500$ and $v_0=1$.}
           \label{fig:li-lt_ric}
\end{figure}

Clearly, the integral in \eref{eq:longtime} is dominated by contributions from near $z=0$. In fact, to get the leading order contribution, it suffices to set $z^2/t \to 0$, and extend the limits of the integral to $\pm \infty$. Carrying out the resulting Gaussian integral and using the large-time behaviour of $B_1(t)$, we finally get the long-time behaviour of the variance of the tagged particle position, 
\begin{align}
\big \langle  x^2_{\alpha}(t)  \big \rangle_{c} &= D_\text{eff}\sqrt{\frac{2t}{\pi k}}- \frac {D_\text{eff}}{\sqrt{(\dr + 2\gamma)(\dr + 2\gamma+ 4 k)}}\cr
& - \frac{D_\text{eff}}{\sqrt{2\pi k t}} \Bigg [ \frac 1{16k } +\frac{4D_{R}+(D_{R}-2\gamma)\cos{2\theta_{0}}}{4 \dr(\dr+2 \gamma)} \Bigg ]   +O(t^{-3/2}). \label{eq:longt_final}
\end{align}
Note that, similar to the short-time regime, the leading order behavior in the long-time regime is also independent of the ordering of the time scales.  Since in this regime, the time is much larger than both the active time-scales, the behavior of the tagged particle is governed by \eref{hc_pp} and is same as the tagged particle behavior in a harmonic chain of passive particles~\cite{TPDin1dchain}.  The correction terms, however, carry signatures of activity. In \fref{fig:li-lt_ric} we illustrate the $\sqrt{t}$ growth in the long-time regime obtained from the numerical simulation along with the analytical prediction \eref{eq:longt_final}.

In the following, we explore the behavior of the tagged particle in the remaining twelve dynamical regimes. It is convenient to group them according to the relative strength of the coupling $k$ with respect to the activity parameters. We consider the three scenarios (i) Strong-coupling limit $[k \gg (\dr, \gamma)]$, (ii) Weak-coupling limit $[k \ll (\dr, \gamma)]$ and (iii) Moderate-coupling limit $[\text{min}(\dr, \gamma) \ll k \ll \text{max}(\dr, \gamma)]$ below.

\subsection[Strong-coupling limit]{Strong-coupling limit $[k \gg (\dr, \gamma)]$} \label{scl-sc} The strong-coupling limit refers to the scenarios when the coupling strength $k$ is larger than both the activity parameters $\dr$ and $\gamma$. Correspondingly, we have,
\begin{equation}
   \tau_1 = k^{-1} \quad \ll \quad \tau_2=\text{min}(\dr^{-1},\gamma^{-1}) \quad \ll \quad \tau_3=\text{max}(\dr^{-1},\gamma^{-1}).
   \label{tau_defn_sc}
\end{equation}
 In the intermediate regimes, $t \gg \tau_1$ and the leading order behavior of the tagged particle variance can be obtained by taking the limit $kt \to \infty$ in \eref{zeqn}. In this limit, the integrals can be performed explicitly [see \eref{AI1}-\eref{AI5} in \ref{aap:int}], yielding, 
\begin{align}
 &\la x_\alpha(t)^2 \ra = B_{1}(t)-B_{2}+ \frac{D_{\text{eff}}}{2\sqrt{(\dr +2\gamma)kt}} \Bigg \{ \sqrt{t}\,\text{erfc} \left(\sqrt{(\dr+2\gamma)t} \right) \cr 
 &+  \frac{(\dr +2 \gamma)^{3/2} \cos{2\theta_{0}}  }{(\dr-2\gamma)} \left(  \frac{\text{erf}\left(\sqrt{(2\gamma-3\dr) t} \right)}{\sqrt{(2\gamma-3\dr)}} - \frac{\text{erfi}\left(\sqrt{4 \dr t} \right)}{\sqrt{2 \dr}} \right) \sqrt{t} \, e^{-4\dr t}  
 \cr 
 &+ \Bigg [ \left (1- \frac{\cos{2 \theta_{0}}(\dr+2\gamma)}{(\dr-2\gamma)} \right) \left[ \text{erfi}\left(\sqrt{(\dr+2\gamma)t} \right) -  \text{erfi}\left(\sqrt{2(\dr+2\gamma)t} \right) \right]  \cr
 & + \cos^2{\theta_{o}}  \bigg( \left[(\dr+2\gamma)t+\frac 12 \right]\text{erfi} \left(\sqrt{(\dr+2\gamma)t} \right) \cr 
 &- \left[(\dr+2\gamma)t+\frac 14 \right] \text{erfi}\left(\sqrt{2(\dr+2\gamma)t} \right) \bigg) \Bigg] \sqrt{t} \, e^{-2(\dr+2\gamma)t}  \cr 
 & + \frac {2t}{\sqrt{\pi}}\sqrt{2(\dr+2\gamma)} \cos^2{\theta_{o}} \left(1-\sqrt{2}e^{-(\dr+2\gamma)t} \right) + O\left(\frac 1 {kt}\right) \Bigg\}.  \label{scEq}
\end{align}
The leading behaviors in the early and late-intermediate regimes can be obtained by taking the appropriate limits in the above equations. \\

\noindent{{\bf{Early intermediate regime}} ($\tau_{1} \ll t \ll \tau_{2} \ll  \tau_{3}$) (R-II).--} We obtain the leading order behavior in the early-intermediate regime by considering $\dr t \ll 1$ and $\gamma t \ll 1$ in \eref{scEq} as, 
\begin{align}
\la x_\alpha(t)^2 \ra \simeq \frac{16}{15}  (\sqrt{2}-1)[D_{R} \sin^2{\theta_{0}}+2\gamma\cos^2{\theta_{0}}] \frac{v_{0}^2t^{5/2}}{ \sqrt{\pi k}}.\label{eq:st_coup_EI_1}
\end{align}
Clearly, for arbitrary initial orientation and any finite $D_{R}$ and $\gamma$, the leading order behavior of the tagged particle variance $\la x_\alpha(t)^2 \ra \sim t^{5/2}$. Note that, $\gamma = 0$ corresponds to the ABP case, where  $t^{5/2}$ behavior has been observed for $\theta_0= \pi/2$~\cite{prashant}.

In the special case $\gamma=0$ and $\theta_{0}=0$, the leading order term vanishes and we need to consider the subleading corrections to \eref{eq:st_coup_EI_1}, which come from two sources: (i) the next order terms in $\dr t$ and $\gamma t$ from \eref{scEq} and (ii) subleading corrections in $1/(kt)$ in the expansion of $[1-z^2/(4kt)]^{-1/2}$ in \eref{zeqn} in the limit $\dr t \ll 1$ and $\gamma t \ll 1$. This results in,
\begin{align}
\la x_\alpha(t)^2 \ra = \frac{v_0^2 \sqrt{\tau_1}}{\sqrt{\pi} \, \tau_2}\, t^{5/2} \left[ L_1 + L_2 \frac{t}{\tau_2}  + \frac{15 \sqrt 2}{96} L_1 \frac{\tau_1}{t} + \cdots  \right], \label{eir-2}
\end{align}
where $\tau_1$, $\tau_2$, and $\tau_3$ are defined in \eref{tau_defn_sc} and the leading order coefficient $L_{1}$ has been already obtained in \eref{eq:st_coup_EI_1}. It is more conveniently expressed as, 
 \begin{align}
    L_1 &= \frac{16(\sqrt 2 -1)}{15} \times 
    \begin{cases}
    \displaystyle
       \sin^2\theta_0 + 2  \frac{\tau_2}{\tau_3} \cos^2 \theta_0  & \text{for}~~ \dr > \gamma \\[3mm]
      \displaystyle \frac{\tau_2}{\tau_3} \sin^2\theta_0 + 2 \cos^2 \theta_0 & \text{for} ~~ \dr< \gamma.
    \end{cases}
    \end{align}
 The coefficient $L_{2}$, appearing in the subleading term, is given by,
    \begin{align}
    L_2 & = \frac{64(\sqrt 2 -1)}{105}  \times 
    \begin{cases}
    \displaystyle
       \left (1 - \frac{2\tau_2}{\tau_3}\right) \left[1 + \frac{7-\sqrt{2}}{2}  \frac{\tau_2}{\tau_3} \cos^2 \theta_0 \right] & \text{for}~~ \dr > \gamma \\[3mm]
      \displaystyle \left(\frac{\tau_2}{\tau_3}-2\right) \left[\frac{\tau_2}{\tau_3} + \frac{7-\sqrt{2}}{2}  \cos^2 \theta_0 \right] & \text{for} ~~ \dr< \gamma.
    \end{cases}
\end{align}

Note that, for any given values of $\dr$ and $\gamma$, depending on the relative strength of $t/\tau_{2}$ and $\tau_{1}/t$, the two subleading terms could be of the same order. Interestingly, for the special case $\gamma=0$ and $\theta_{0}=0$, the leading order behavior is given by 
\begin{equation}
   \la x_\alpha(t)^2 \ra= \frac{64(\sqrt{2}-1)}{105}\frac{v_{0}^2 \dr^2}{\sqrt{\pi k}}\, t^{7/2},
\end{equation}
which is consistent with the result obtained earlier for the ABP case with $\theta_0=0$~\cite{prashant}.

\begin{figure} [t]
            \centering
            \includegraphics[scale=0.7]{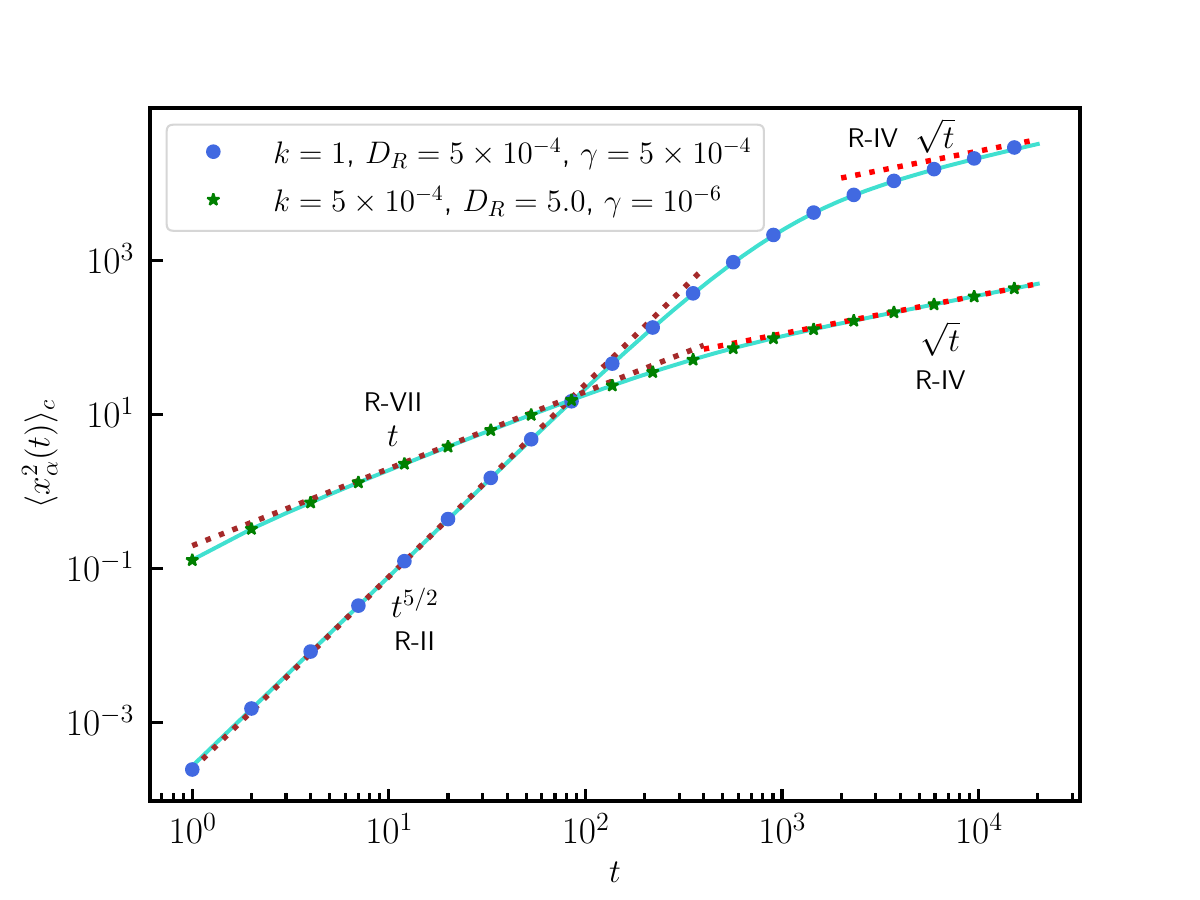}
            \caption{The crossover behavior of the variance from the early-intermediate regime to the late-intermediate regime for quenched initial orientation. The solid lines are obtained by numerically integrating \eref{bqeqn2}. The dotted lines correspond to the asymptotic behavior given by \eref{eq:longt_final}, \eref{eq:st_coup_EI_1}, and \eref{dyk_ir_2}. The symbols indicate the data obtained from numerical simulations with $N=500$ and $v_0=1$.}
           \label{fig:eir_lir_q}
\end{figure}

The $t^{5/2}$ growth of variance in the early intermediate regime for the strong coupling limit is shown in \fref{st_fg}  and \fref{fig:eir_lir_q} .\\

\noindent{{\bf Late intermediate regime} ($\tau_{1} \ll \tau_{2} \ll t  \ll  \tau_{3}$) (R-III).--}
Next, we consider the late-intermediate regimes $\tau_1 \ll \tau_2 \ll t \ll \tau_3$. In this regime, the time is larger than one of the activity time-scales and the system behaves like a passive chain, for which the tagged particle behavior is given by \eref{hc_pp}. We expect a similar subdiffusive behavior. For $\tau_2 = \dr^{-1}$ and $\tau_{3}=\gamma^{-1}$, i.e., $\dr \gg \gamma$, the leading order behavior can be obtained by taking the limit $D_{R}t \gg 1$, $\gamma t \ll 1$ in \eref{scEq}, as,
\begin{align}
\big \langle  x^2_{\alpha}(t)  \big \rangle_{c} &\simeq  \frac{v_{0}^2 \sqrt{t}}{ \dr \sqrt{2\pi k}}.
\label{ltdyk}
\end{align}
The subleading corrections to \eref{ltdyk} can be obtained by considering the next order terms in \eref{scEq}. This results in,
\begin{align}
\big \langle  x^2_{\alpha}(t)  \big \rangle_{c} &=  \frac{v_{0}^2 \sqrt{t}}{ \dr \sqrt{2\pi k}} \Bigg[1-\sqrt{\frac{\pi \tau_{2}}{2 t}} + O(\tau_{2}/t) \Bigg].
\label{ltdyk_c}
\end{align}
Note that, in this case, the tagged particle variance becomes independent of the initial orientation $\theta_{0}$ due to the large rotational diffusion constant. Incidentally, the above equation can also be obtained by considering the limit $\gamma \ll \dr \ll k$ in \eref{eq:longt_final}.

On the other hand, for $\tau_2= \gamma^{-1}$ and $\tau_{3}=\dr^{-1}$, i.e., $\gamma \gg \dr$, we have $D_{R}t \ll 1$, $\gamma t \gg 1$. Considering these limits in \eref{scEq}, the leading order behavior is given by, 
\begin{align}
\big \langle  x^2_{\alpha}(t)  \big \rangle_{c} \simeq \frac{v_{0}^2 \cos^2{\theta_{0}}}{\gamma \sqrt{2\pi k}}\sqrt{t}.
\label{lir_lot}
\end{align}
Again the subleading corrections to \eref{lir_lot} can be obtained by considering the next order terms in \eref{scEq} which results in,
\begin{align}
\big \langle  x^2_{\alpha}(t)  \big \rangle_{c} &= \frac{v_{0}^2 \cos^2{\theta_{0}}}{\gamma \sqrt{2\pi k}}\sqrt{t}\Bigg[ 1- \frac {1}{2} \sqrt{ \frac{\pi \tau_2}{t}} - \frac{4}{3} (1-\tan^2{\theta_{0}}) \frac{t}{\tau_3} + O(\tau_{2}/t) \Bigg]. 
\label{eq:late-intermediate_eqn_kyd_t}
\end{align}
Clearly, for  arbitrary initial orientation, irrespective of the relative strength of $D_{R}$ and $\gamma$, the leading order behavior of the tagged particle variance $\la x_\alpha(t)^2 \ra \sim t^{1/2}$, except for the special case $\gamma=0$ and $\theta_{0}=\pi/2$, where this leading order term vanishes. In that case, we need to consider the subleading corrections, $\la x_\alpha(t)^2 \ra \sim t^{3/2}$.

Note that, in strong coupling limit, the subleading contributions coming from the expansion of $[1-z^2/(4kt)]^{-1/2}$ in \eref{zeqn} is $O(\tau_{1}/t) \ll O(\tau_{2}/t)$. The crossover behavior of the variance from the early-intermediate regime to the late-intermediate regime and the late-intermediate regime to long-time behavior in the strong-coupling limit is shown in \fref{fig:eir_lir_q} and \fref{fig:li-lt_ric} respectively.

\subsection[Weak coupling limit]{Weak coupling limit $[k \ll (\dr, \gamma)]$}  \label{wcl-sc} The weak-coupling limit refers to the scenarios when the coupling strength $k$ is smaller than both the activity parameters $\dr$ and $\gamma$ and correspondingly the largest time-scale $\tau_3 = k^{-1}$. The other two time-scales are $\tau_1=\min(\dr^{-1},\gamma^{-1})$, $\tau_2=\max(\dr^{-1},\gamma^{-1})$. In this case, we can expand \eref{bqeqn2} in powers of $t/\tau_{3}$, $\tau_{2}/\tau_{3}$ and $\tau_{1}/\tau_{2}$. This expansion can be most conveniently expressed in terms of the parameters $k$, $\dr$,$\gamma$ as, 

\begin{align}
&\big \langle  |x_{\alpha}(t)|^2 \big \rangle_{c} = B_1(t) - B_2 + D_{\text{eff}} \sum_{n=0}^{\infty} \sum_{m=0}^{\infty} A(m+n) \frac{(-4 k t)^n}{n!} \Bigg \{ \frac{e^{-(\dr+2\gamma)t}}{\dr+2\gamma} \left( \frac{-4k}{\dr+2\gamma} \right)^m \cr 
& + \frac{2(\dr \sin^2{\theta_{0}} -2\gamma \cos^2{\theta_{0}})}{\dr^2-4\gamma^2} (e^{-(\dr+2\gamma)t} - 2^n )\left( \frac{4k}{\dr+2\gamma} \right)^m -\frac{(\dr+2\gamma)\cos{2\theta_{0}}}{\dr-2\gamma} \bigg[ \frac{2^{n-1}}{\dr}\left(\frac{2k}{\dr} \right)^m \cr
& - \frac{e^{-(\dr+2\gamma)t}}{3\dr-2\gamma} \left( \frac{4k}{3\dr-2\gamma} \right)^m \bigg] + \frac{2 (m+1) \cos^2{\theta_{0}}}{\dr+2\gamma} \left( 2e^{-(\dr+2\gamma)t} -2^n \right) \left( \frac{4k}{\dr+2\gamma} \right)^m   \Bigg \} \cr
& + D_{\text{eff}} \sum_{m=0}^{\infty} A(m)  \Bigg \{  \frac{\dr+2\gamma}{\dr-2\gamma}  e^{-4\dr t} \cos{2\theta_{0}} \left[ \frac{1}{2\dr} \left( \frac{2k}{\dr} \right)^m -\frac{1}{3\dr-2\gamma} \left( \frac{4k}{\dr+2\gamma} \right)^m  \right] \cr
& - \frac{2(m+1)\cos^2{\theta_{0}} e^{-2(\dr+2\gamma)t}}{\dr+2\gamma} \left( \frac{4k}{\dr+2\gamma} \right)^m \Bigg \},\label{bqeqnee}
\end{align}
where $A(m)$ is a constant, given by
\begin{align}
    A(m)&=\frac{1}{2 \pi} \int_{-\pi}^{\pi}  \,dq  \sin^{2m}{(q/2)} = \frac{\Gamma \left(m+\frac{1}{2}\right)}{ \sqrt{\pi } \, m!}. 
\end{align}
 Although the sum over $m$ in \eref{bqeqnee} can be performed explicitly [see \eref{exp_sum1}-\eref{expansion_sum} in~\ref{aap:int}], the sum over $n$ is hard to evaluate in a closed form. Nevertheless, \eref{bqeqnee} is exact for all parameters. In fact, the expansion as a power series in  $kt$, as given in \eref{bqeqnee}, is particularly useful in the intermediate regime of the weak-coupling limit, where $kt\ll 1$. In this limit, taking $\exp{[-(\dr+2\gamma)t]} \to 0$, \eref{bqeqnee} reduces to a much simpler form 
\begin{align}
&\big \langle  | x_{\alpha}(t) |^2 \big \rangle_{c} = B_1(t) - B_2 - \frac{2 D_{\text{eff}}}{\dr-2\gamma}\sum_{n=0}^{\infty} \sum_{m=0}^{\infty} A(m+n) \frac{(-8 k t)^n}{n!} \cr 
& \qquad \times  \Bigg \{ \left[  \frac{\dr}{\dr+2\gamma} + m \, \cos^2{\theta_{0}} \right]  \left( \frac{4k}{\dr+2\gamma} \right)^m  +  \frac{ (\dr+2\gamma) \cos{2\theta_{0}}}{4\dr} \left(\frac{2k}{\dr} \right)^m \Bigg \} \cr
& +  e^{-4\dr t}  \frac{D_{\text{eff}}}{2\dr} \frac{\dr+2\gamma}{\dr-2\gamma} \cos{2\theta_{0}} \sum_{m=0}^{\infty} A(m)    \left[ \left( \frac{2k}{\dr} \right)^m -\frac{2 \dr}{3\dr-2\gamma} \left( \frac{4k}{\dr+2\gamma} \right)^m  \right].\quad 
\label{bqeqn2_ea}
\end{align}

Unlike the strong-coupling limit, the behavior of the tagged particle variance in the intermediate regime in the weak-coupling limit depends on the relative strength of $\dr$ and $\gamma$. Hence we choose a particular order of $\dr,\gamma$ and then examine \eref{bqeqn2_ea} to get the leading order behavior of the position variance. In the following, we analyze the early intermediate and the late intermediate regimes separately.  \\

\noindent{{\bf Early intermediate regime}} ($\tau_{1} \ll t \ll \tau_{2} \ll  \tau_{3}$).--  For $\dr \gg \gamma$, we have $\tau_{1}=\dr^{-1}$ and $\tau_{2}=\gamma^{-1}$ (R-V). Considering the limit $D_{R}t \gg 1$,$\gamma t \ll 1$ and $kt \ll 1$ in \eref{bqeqn2_ea} we get the leading order behavior for this case as,
\begin{align}
    \big \langle x_{\alpha}(t)^2\big \rangle_{c} &= \frac{v_{0}^2 \, t}{\dr} \bigg [  1 -  \frac{(6+\cos{2\theta_{0}})}{4 D_{R} t} - 2 k t + \dotsb \bigg].
 \label{dyk_ir_2}
\end{align}

On the other hand, for $\gamma \gg \dr$, we have $\tau_{1}=\gamma^{-1}$ and $\tau_{2}=\dr^{-1}$ (R-VI). Considering the limit $D_{R}t \ll 1$,$\gamma t \gg 1$ and $kt \ll 1$ in \eref{bqeqn2_ea} we get the leading order behavior for this case as, 
\begin{align}
    \big \langle   x_{\alpha}(t) ^2 \big \rangle_{c}= \frac{v_{0}^2 \, t }{\gamma} \cos^2 \theta_0 \bigg [   1
 -  \frac{3}{4\gamma t} - 2 k t  -  \dr t (1-\tan^2{\theta_{0}}) + \dotsb \bigg].
 \label{eir_ydk}
\end{align}
Interestingly, for $\theta_{0}=\pi/2$, the tagged particle behaves ballistically at the leading order, i.e., $ \big \langle   x_{\alpha}(t) ^2 \big \rangle_{c} \simeq v_{0}^2 (\dr/\gamma )  t^2 $. \\

\begin{figure} [t]
            \centering
            \includegraphics[scale=0.7]{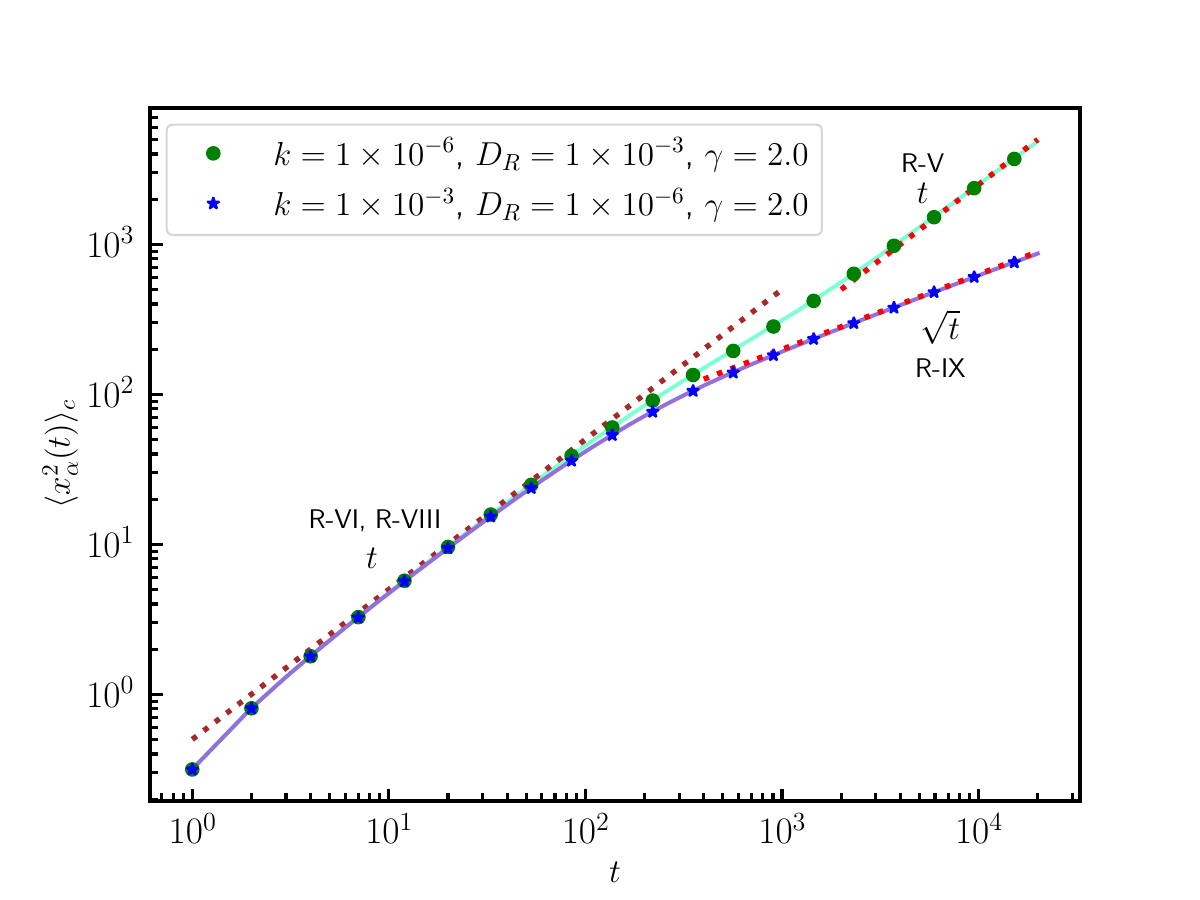}
            \caption{The crossover behavior of the variance from the early-intermediate regime to the late-intermediate regime for quenched initial orientation. The solid lines are obtained by numerically integrating \eref{bqeqn2}. The dotted lines correspond to the asymptotic behavior given by \eref{lir_lot}, \eref{dyk_ir_2}, and  \eref{eir_ydk}. The symbols indicate the data obtained from numerical simulations with $N=100$ and $v_0=1$.}
           \label{fig:eir_lir_q-ii}
\end{figure}

\noindent{\bf Late intermediate regime} ({$ \tau_{1} \ll \tau_{2} \ll t \ll  \tau_{3}$}) (R-V).-- In this regime, both $D_{R}t \gg 1$ and $\gamma t \gg 1$, while $kt \ll 1$. Considering these limits in \eref{bqeqn2_ea}, the leading order behavior  of the variance can be written as,
\begin{align}
    \big \langle   x_{\alpha}(t) ^2 \big \rangle_{c} \simeq 2 D_{\text{eff}} t,
 \label{Lir_lot}
 \end{align}
independent of the relative strength of $\dr$ and $\gamma$. 
The sub-leading corrections, however, depend on the relative strength of $\dr$ and $\gamma$. In particular, for $\dr \gg \gamma$ the tagged particle behavior is the same as that in the early intermediate regime, given by \eref{dyk_ir_2}. On the other hand, for $\gamma \gg \dr$, we have,
\begin{align}
    \big \langle   x_{\alpha}(t) ^2 \big \rangle_{c}= \frac{v_{0}^2 t}{2\gamma} \bigg [   1
 -  \frac{3\cos^2{\theta_{0}}}{2\gamma t} - 2 k t + \dotsb \bigg]. 
 \label{lir_ydk}
\end{align} 
Note that, in this regime, $D_{\text{eff}}$, given by \eref{eq:deff_def}, becomes
\begin{align}
    D_{\text{eff}} \to \begin{cases}
       \displaystyle \frac{v_0^2}{2 \dr} &\text{for} ~~ \gamma \ll \dr \\[4mm]
       \displaystyle \frac{v_0^2}{4 \gamma } &\text{for} ~~ \gamma \gg \dr.
    \end{cases}
\end{align}
\Fref{fig:eir_lir_q-ii} shows the crossover of the 
 variance from the early intermediate regime to the late intermediate regime when $\gamma \gg \dr$, in the weak-coupling limit. In the intermediate regimes of weak coupling limit, $t$ crosses either one (early intermediate regime) or both (late intermediate regime) the activity time-scales, but stays below the coupling time scale ($k^{-1}$). Hence the tagged particle is in passive limit and its behavior is governed by \eref{hc_pp}. Since $t \ll k^{-1}$, the tagged particle does not experience harmonic coupling, and the leading order behavior is the same as an independent passive particle. For $\gamma \gg \dr$, though the behavior is diffusive in both early and late intermediate regimes, the diffusivity in the early intermediate regime is greater than that in the late intermediate regime. While the diffusivity remains the same in both the intermediate regimes for $\dr \gg \gamma$. 

\subsection[Moderate-coupling limit]{Moderate-coupling limit $[\min(\dr, \gamma) \ll k \ll \max(\dr,\gamma)]$} The moderate-coupling refers to the scenarios when $k$ lies between the two activity parameters $\dr$ and $\gamma$. Consequently the coupling strength $k$ determines the intermediate time scale, $\tau_{2}=k^{-1}$, while the other two time-scales are $\tau_1=\min(\dr^{-1},\gamma^{-1})$ and $\tau_3=\max(\dr^{-1},\gamma^{-1})$.\\

\noindent{\bf Early intermediate regime} ($\tau _{1} \ll t \ll  \tau_{2} \ll \tau_{3}$).-- The behavior of the tagged particle in the early intermediate regime can be most conveniently obtained from \eref{bqeqn2_ea}, which was obtained in the limit $kt \ll 1$ and $(\dr+2\gamma) t \gg 1$. Therefore, the scenarios corresponding to both the relative orders $\dr\gg \gamma$ (R-VII) and $\gamma \gg \dr$ (R-VIII) in the moderate coupling limit are the same as the corresponding scenarios in the weak coupling limit. Namely, $ \big \langle   x_{\alpha}(t) ^2 \big \rangle_{c}$ is given by \eref{dyk_ir_2} and \eref{eir_ydk}, for $\dr \gg \gamma$ and $\gamma \gg \dr$, respectively. \Fref{fig:eir_lir_q} and \ref{fig:eir_lir_q-ii} show this early intermediate regime behavior, for $\dr \gg \gamma$ and $\gamma \gg \dr$, respectively. \\

\noindent{\bf Late intermediate regime} ($\tau _{1}  \ll  \tau_{2} \ll t  \ll \tau_{3}$).-- In this regime, $kt \gg 1$, and hence, the leading order behavior of the variance can be extracted from \eref{scEq}. Therefore, the scenarios corresponding to both the relative orders $\dr\gg \gamma$ and $\gamma \gg \dr$ in the moderate coupling limit are the same as the corresponding scenarios in the strong coupling limit. Namely, the leading order behavior of $ \big \langle   x_{\alpha}(t) ^2 \big \rangle_{c}$ is given by \eref{lir_lot} and \eref{ltdyk}, for $\gamma \gg \dr$ and $\dr \gg \gamma$, respectively. However, the subleading corrections differ from those obtained in the strong coupling limit  and can be extracted directly from \eref{zeqn} by considering the appropriate limits. For $\dr \gg \gamma$ (R-IV), we get
\begin{align}
\big \langle  x^2_{\alpha}(t)  \big \rangle_{c} &=\frac{v_0^2\sqrt{t}}{\dr \sqrt{2 \pi k}}  \left[ 1-\frac{1}{32 k t} - \sqrt{\frac{2\pi k}{\dr}}\frac{1}{\sqrt{\dr t}} + \dotsb \right],
\label{lid_dky_t}
\end{align} 
while for $\gamma \gg \dr$ (R-IX), we get,
\begin{align}
\big \langle  x^2_{\alpha}(t)  \big \rangle_{c} &= \frac{v^{2}_{0}\cos^2{\theta_{0}}}{2 \gamma}\sqrt{\frac{2t}{\pi k}} \Bigg [ 1- \sqrt{\frac{\pi k}{8 \gamma^2 t }} -\frac{1}{32 k t} + \frac{4 (1-2\cos^2{\theta_{0}})}{3 \cos^2{\theta_{0}}} \dr t + \dotsb \Bigg].
\label{eq:lir_ykd}
\end{align}
Interestingly, for $\theta_{0}=\pi/2$, leading order behaviour changes from $\sqrt{t}$ to $t^{3/2}$. Figures~\ref{fig:eir_lir_q} and \ref{fig:eir_lir_q-ii} show the crossover behavior from the early-intermediate regime to the late-intermediate regime in the moderate-coupling limit, for $\dr \gg \gamma$ and $\gamma \gg \dr$, respectively.

\section{Annealed initial orientation}  \label{uio_pv}
In this section, we consider the scenario where the initial orientation $\{\theta_{\alpha}(0)\}$ are independently drawn from a uniform distribution of $[0, 2 \pi ]$. As a result, $G(t_1,t_2)$ is given by \eref{g_ric}. Putting \eref{g_ric} in \eref{xs_xsp} and using \eref{cifa_1} we get,
\begin{equation}
 \big \langle  x_{\alpha}^2(t) \big \rangle_{c} = \frac{D_\text{eff}}{N} \sum_{s=0}^{N-1}  \bigg [   \frac{1-e^{-2 a_{s}t}}{a_{s} }  +  \frac{e^{-(D_{R}+2\gamma)t} e^{-a_{s}t}-1}{\big(\dr+2\gamma+ a_{s} \big)}   +  \frac{e^{-a_{s}t}(e^{-(\dr+2\gamma)t} -e^{- a_{s}t})}{\big(\dr+2\gamma-a_{s} \big)}  \bigg] , 
 \label{ric_msd}
\end{equation}
where $a_{s}$ is defined in \eref{eomF}. \Eref{ric_msd} can be converted into an integral in the large $N$ limit as,
\begin{align}
\big \langle  x_{\alpha}^2(t) \big \rangle_{c} &=  D_\text{eff} \int_{0}^{2 \pi} \frac{dq}{2 \pi } \bigg [   \frac{1-e^{-2 b_{q}t}}{b_{q} }  +  \frac{e^{-(D_{R}+2\gamma)t} e^{-b_{q}t}-1}{\big(\dr+2\gamma+ b_{q} \big)}   +  \frac{e^{-b_{q}t}(e^{-(\dr+2\gamma)t} -e^{- b_{q}t})}{\big(\dr+2\gamma- b_{q} \big)}  \bigg],
\label{ric_bq}
\end{align}
where $b_{q}$ is given by \eref{eq:bq}. As before, the first and part of the second terms can be integrated exactly, given by \eref{eq:term1} and \eref{eq:term2} respectively. This simplifies \eref{ric_bq} to,

\bea 
\fl \big \langle  x_{\alpha}^2(t) \big \rangle_{c} = B_{1}(t)-B_{2}+D_\text{eff} \int_{0}^{2 \pi} \frac{dq}{2 \pi}   \bigg [   \frac{e^{-(D_{R}+2\gamma)t} e^{-b_{q}t}}{\big(\dr+2\gamma+ b_{q} \big)} +  \frac{e^{-b_{q}t}(e^{-(\dr+2\gamma)t} -e^{- b_{q}t})}{\big(\dr+2\gamma- b_{q} \big)}  \bigg]
\label{ric_bq2}
\eea

Note that, here the parameters $\dr$ and $\gamma$ appear together as $(\dr+2\gamma)$. Consequently, we have two distinct time scales $(\dr+2\gamma)^{-1}$ and $k^{-1}$, in contrast to the quenched initial condition, where there are three distinct time scales set by $\dr^{-1}$, $\gamma^{-1}$ and $k^{-1}$. In the following, we consider the three dynamical regimes emerging when the two scales, $\tau_1=\min((\dr+2\gamma)^{-1},k^{-1})$ and $\tau_{2}=\max((\dr+2\gamma)^{-1},k^{-1})$, are well separated:
\begin{enumerate}
\item [1.] Short-time regime:  $t \ll \tau_1 $
\item [2.] Long-time regime:  $t \gg \tau_2 $
\item [3.] Intermediate regime: $\tau_1 \ll t \ll \tau_2$ 
\end{enumerate}

 \begin{figure} [t]
            \centering
            \includegraphics[scale=0.7]{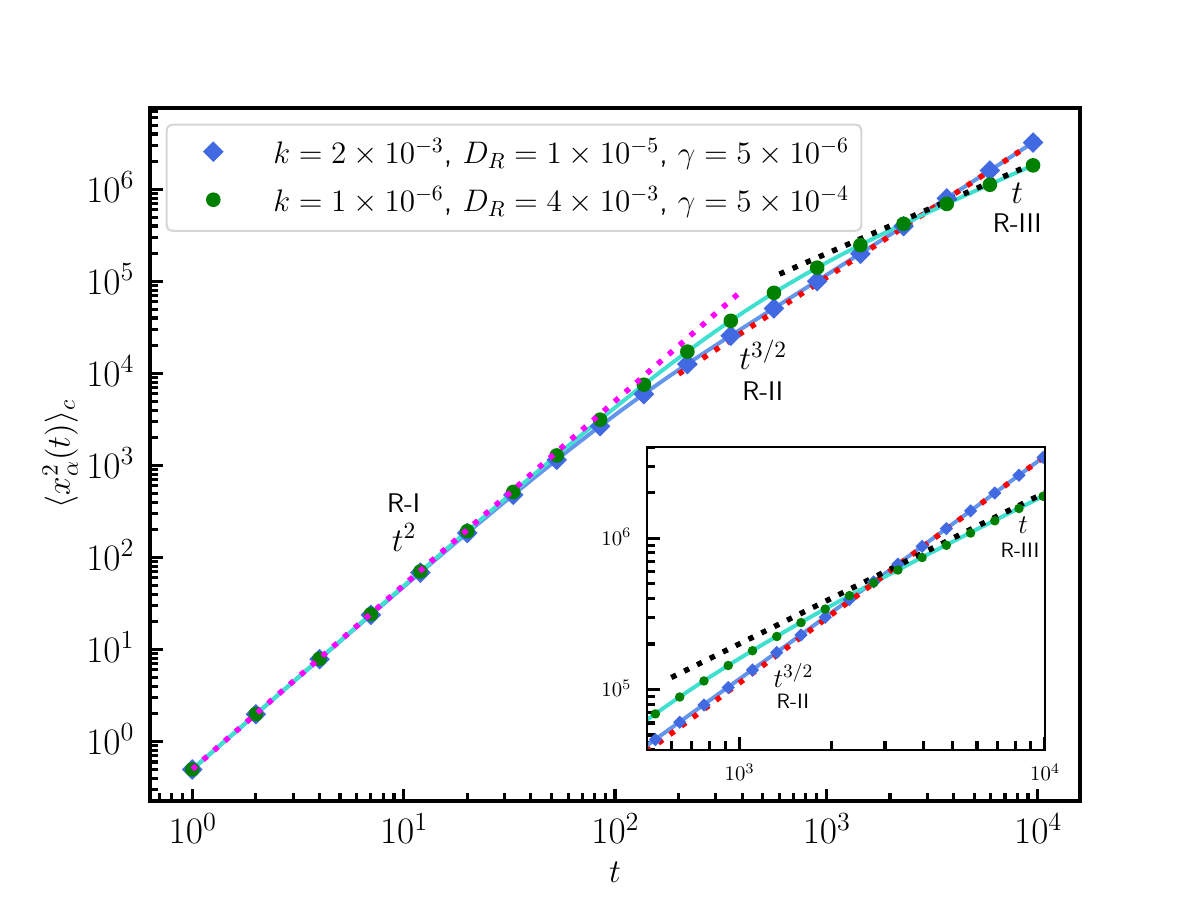}
            \caption{The crossover behavior of the variance from short-time regime to intermediate regime for annealed initial orientation. The solid lines are obtained by numerically integrating \eref{bqeqn2}. The dotted lines correspond to the asymptotic behavior given by \eref{smalltime_ric}, \eref{kdy_ir_ric}, and  \eref{dyk_ir_ric}. The symbols indicate the data obtained from numerical simulations with $N=500$ and $v_0=1$.}
           \label{fig:ric_st_ei}
\end{figure}

\subsection{Short-time regime $(t \ll \tau_1)$ (R-I)} At very short times, the particles do not feel the effect of the interaction. Consequently, each particle independently moves ballistically with the speed $v_{0}$ along its initial orientation $\theta_{\alpha}(0)$, which is drawn randomly from $(0, 2 \pi) $. Hence, the average displacement is zero, and the variance $\big \langle x_{\alpha}(t) ^2 \big \rangle_{c}  \approx v^2_{0} t^2 \big \langle \cos^2{\theta_{\alpha}(0)} \big \rangle=v^2_{0} t^2/2$. However, the effect of the interaction is expected to show up in the subleading corrections.  In fact, the leading order behavior and the subleading corrections can be systematically extracted by expanding \eref{ric_bq2} in power series of $t$,

\begin{equation}
 \big \langle x_{\alpha}(t) ^2 \big \rangle_{c} =  \frac{ v^2_0 t^{2} }{2} \left[1 - \frac{1}{3}(D_{R}+2\gamma+ 6k)t  + O(t^2)\right].
 \label{smalltime_ric}
 \end{equation}
Interestingly, in the short-time regime, the position fluctuation in the quenched case is smaller than the annealed case by a factor proportional to $t/\tau_2$ [see \eref{stb}]. \Fref{fig:ric_st_ei} shows the $t^2$ growth of the variance at short-times.

To analyze the behavior of the variance in the remaining two regimes it is convenient to recast \eref{ric_bq2} using a change of variable $z^2=b_{q}t$, 
\begin{align}
\big \langle   x_{\alpha}(t)^2 \big \rangle_{c} &= B_{1}(t)-B_{2}+ \frac{D_{\text{eff}}}{2 \pi  \sqrt{kt}}\int_{-\sqrt{4kt}}^{\sqrt{4kt}}  \,\ dz\,  e^{-z^{2}} \bigg [  \frac{e^{-(\dr+2\gamma)t}}{(\dr+2\gamma)+ z^{2}/t } \cr & \qquad \qquad +  \frac{e^{-(\dr+2\gamma)t} -e^{- z^{2}}}{(\dr+2\gamma)- z^{2}/t }\bigg] \frac{1}{ \sqrt{1-z^{2}/4kt} }.  
\label{zeqn_ric}
\end{align}

\subsection{Long-time regime $(t \gg \tau_2)$ (R-IV)} 
\begin{figure} [t]
            \centering
            \includegraphics[scale=0.7]{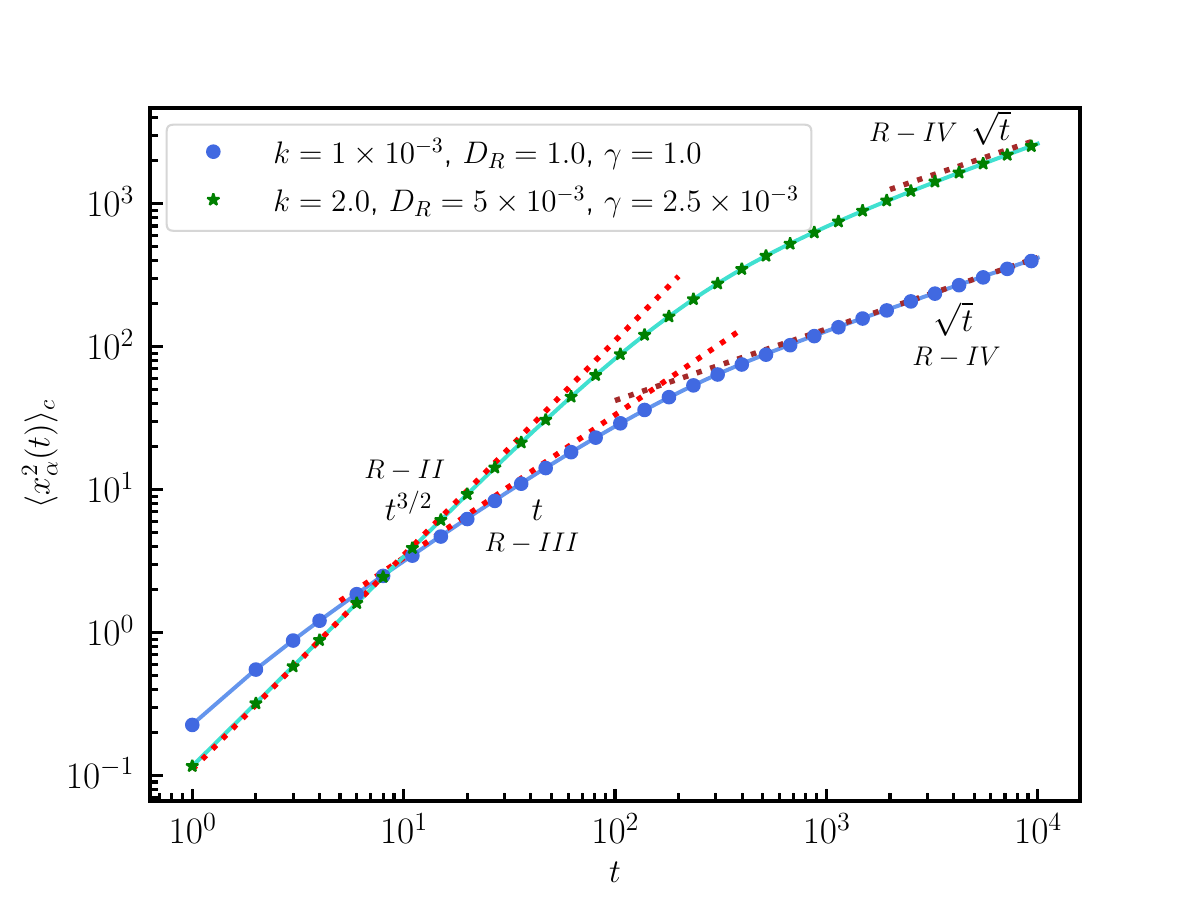}
            \caption{The crossover behavior of the variance from short-time regime to intermediate regime for annealed initial orientation. The solid lines are obtained by numerically integrating \eref{bqeqn2}. The dotted lines correspond to the asymptotic behavior given by \eref{eq:longt_final_ric}, \eref{kdy_ir_ric}, and  \eref{dyk_ir_ric}. The symbols indicate the data obtained from numerical simulations with $v_0=1$ and $N=500$ for weak-coupling limit, while N=800 for strong-coupling limit.}
           \label{fig:ric_ir_lt}
\end{figure}

In the long-time regime, $t$ is much larger than both the time-scales of the system. Therefore, setting $e^{-(\dr + 2\gamma)t} \to 0$ in \eref{zeqn_ric}, we get,
\begin{align}
\big \langle x_{\alpha}(t)^2 \big \rangle_{c} &= B_{1}(t)-B_{2}- \frac{D_\text{eff}}{2 \pi  \sqrt{kt}}\int_{-\sqrt{4kt}}^{\sqrt{4kt}}  \,\ dz   \frac{e^{-2 z^{2}}}{\big[(\dr+2\gamma)- z^{2}/t \big]} \frac{1}{ \sqrt{1-z^{2}/(4kt)} }. 
\label{zeqn_ric_lt}
\end{align}
The integral in \eref{zeqn_ric_lt} is dominated by the contribution from near $z=0$. Consequently, one can expand the integrand in powers of $z^2/[(\dr+2\gamma)t]$ and $z^2/(4kt)$ and set the domain of integral from $-\infty$ to $\infty$. Finally, we get the variance of the tagged particle in the long-time regime, 
\begin{align}
\big \langle  x^2_{\alpha}(t)  \big \rangle_{c} &= D_\text{eff}\sqrt{\frac{2t}{\pi k}}- \frac {D_\text{eff}}{\sqrt{(\dr + 2\gamma)(\dr + 2\gamma+ 4 k)}}\cr
& - \frac{D_\text{eff}}{\sqrt{2\pi k t}} \Bigg [ \frac 1{16k } +\frac{1}{2(\dr+2 \gamma)} \Bigg ]   +O(t^{-3/2}). \label{eq:longt_final_ric}
\end{align}
\Fref{fig:ric_ir_lt} shows the $\sqrt{t}$ behavior in the long-time regime.

The system eventually forgets the initial orientation, and the leading order behavior in the long time regime $\big \langle  x^2_{\alpha}(t)  \big \rangle_{c} \approx D_\text{eff}\sqrt{\frac{2t}{\pi k}}$ for both annealed and quenched initial orientations. The difference due to the initial orientations shows up in the subleading corrections [cf. \eref{eq:longt_final} and \eref{eq:longt_final_ric}].

\subsection{Intermediate regime} 
The behavior of the tagged particle in the intermediate regime ($\tau _{1}  \ll t \ll \tau_{2}$) depends on whether the coupling strength $k$ is larger or smaller than the activity parameter $(\dr+2\gamma)$. In the strong coupling limit $[k \gg (\dr+2\gamma)]$, the two time-scales are  $\tau_1=k^{-1}$ and $\tau_2=(\dr+2\gamma)^{-1}$, whereas in the weak coupling limit $[k \ll (\dr+2\gamma)]$, $\tau_1=(\dr+2\gamma)^{-1}$ and $\tau_2=k^{-1}$. In the following, we discuss the two limits separately.

\subsubsection{Strong-coupling limit $[k \gg (\dr, 2\gamma)]$ (R-II)}- In this limit, the intermediate regime is characterized by $kt \gg 1$ and $(\dr+2\gamma) t \ll 1$. The leading order behavior can be obtained from \eref{zeqn_ric} as,
\begin{align}
    \big \langle x^2_{\alpha}(t)\big \rangle_{c} &= \frac{2(\sqrt{2}-1)v^2_0 t^{3/2}}{3\sqrt{\pi k }} \Bigg[ 1 -\left(1-\frac{\sqrt{2}}{3} \right)(\dr+2\gamma)t  - \frac{3\sqrt{2}}{32kt} + \dotsb \Bigg].
 \label{kdy_ir_ric}
\end{align}
Figures~\ref{fig:ric_st_ei} and \ref{fig:ric_ir_lt} show this $t^{3/2}$ growth in the variance.

\subsubsection{Weak-coupling limit $[k \ll (\dr, 2\gamma )]$ (R-III)}- On the other hand in the weak-coupling limit, the intermediate regime is characterized by $kt \ll 1$ and $(\dr+2\gamma) t \gg 1$. The leading order behavior can be obtained from \eref{ric_bq2} as,
\begin{align}
    \big \langle x^2_{\alpha}(t)\big \rangle_{c} &= 2D_{\text{eff}} \, t \bigg [ 1  - \frac{1}{(\dr+2\gamma)t} - 2k t + \dotsb \bigg]
 \label{dyk_ir_ric}
\end{align}
Figures \ref{fig:ric_st_ei} and \ref{fig:ric_ir_lt} show this linear growth in variance.

Apart from the long-time regime, where the behavior is anyway expected to be independent of the initial condition, the tagged particle variance in all the other dynamical regimes shows rather different behavior for the quenched and annealed initial conditions. In particular, the effect of the initial condition is drastically felt in the intermediate dynamical regime of the strong coupling limit, where randomizing the initial orientation changes the behavior to  $t^{3/2}$ from the $t^{5/2}$ observed for a quenched initial condition. Moreover, at short-times also, the position variance of the tagged particle with the annealed initial orientation shows a $t^2$ growth, as opposed to the $t^3$ growth for a quenched initial orientation.

\section{Finite-size effects}  \label{fse}

There is an additional time scale $t_N$ associated with any finite system of size $N$. This finite-size time-scale $t_N$ diverges in the thermodynamic limit $N \to \infty$. Therefore, for a harmonic chain of $N$ DRABPs, the behaviors we studied in earlier sections are expected to hold for $t \ll t_N$, for a large but finite $N$. On the other hand,  for $t \gg t_N$ one expects a different behavior due to finite size effects. In this section, we study the finite-size effect and crossover behavior across $t_N$,  of the tagged particle variance.
 
We start with the exact expression for the position variance given by \eref{sum}, which holds for an arbitrary value of $N$. We assume that the system size is large enough so that $t_N$ is the largest time scale, i.e., $\{ \dr^{-1},\gamma^{-1},k^{-1} \} \ll t_N$. Hence to study the finite size effect and the cross-over behavior across $t_N$, we study the behavior in the regime $\{ \dr^{-1},\gamma^{-1},k^{-1} \} \ll t$. For $ t \gg \{ \dr^{-1},\gamma^{-1} \}$ and large $N$, \eref{sum} simplifies to, 
\begin{align}
 \big \langle  x^2_{\alpha}(t)  \big \rangle_{c} &= \frac{D_{\text{eff}}}{N} \sum_{s=-N/2}^{N/2}  \bigg [   \frac{1-e^{-2a_{s}t}}{a_{s}}- \frac{1}{\dr+2\gamma+ a_{s} }-\frac{e^{-2a_s t}}{\dr+2\gamma-a_s}  \cr
 &+ \frac{(\dr+2\gamma )e^{-2 a_s t}}{\dr+2\gamma-a_s} \left(\frac{\cos{2\theta_{0}}}{2D_{R}-a_s}-\frac{2\cos^2{\theta_{0}}}{D_{R}+2\gamma-a_s} \right)\bigg],
 \label{sum FN}
\end{align}
where $a_s$ is defined in \eref{eomF}. Note that, we have shifted the limits of summation using the fact that the sum in \eref{sum} is symmetric about $N/2$. Now, since $t \gg k^{-1}$, the summation in \eref{sum FN} is dominated by contributions from terms with $s \ll N$. Hence, we can use the approximation  $e^{-2a_s t} \simeq e^{-8\pi^2 s^2 t/t_N}$, where the finite-size time-scale $t_N=N^2/k$. Clearly, for $t \gg t_N$, $e^{-8\pi^2 s^2 t/t_N} \to 0$ for $s \ne 0$. Then, from \eref{sum FN}, we get, for $t \gg t_N$, 
\begin{align}
 \big \langle  x^2_{\alpha}(t)  \big \rangle_{c} &= \frac{D_{\text{eff}}}{N} \bigg[2t +\frac{\cos{2\theta_{0}}}{2D_{R}}-\frac{1+2\cos^2{\theta_{0}}}{(D_{R}+2\gamma)}\bigg] + \frac{v_{0}^2}{N}\sum_{s=1}^{N/2}  \bigg [   \frac{1}{a_{s}(D_{R}+2\gamma+a_{s})}\bigg].
 \label{sum FNs}
\end{align} 
Hence, for $t \gg t_N$, the leading order behavior of the variance grows linearly with time, as $\big \langle  x^2_{\alpha}(t)  \big \rangle_{c} \simeq 2 t D_{\text{eff}}/ N$, which is consistent with earlier studies done on the harmonic chain of ABP, RTP, and AOUP \cite{prashant}.

On the other hand, for $t \ll t_N$, we cannot ignore  $e^{-8 \pi^2 s^2 t/t_N}$. However, since $kt \gg 1$, the summation for terms containing $e^{-4 \pi^2 s^2 t/t_N}$ is still dominated by $s \ll N$. Therefore Eq. \eqref{sum} becomes,
\begin{align}
 \big \langle  x^2_{\alpha}(t)  \big \rangle_{c} = \frac{D_{\text{eff}}}{N}\sum_{s=-N/2}^{N/2}  \bigg [& \frac{1-e^{-8 \pi^2 s^2 t/t_N}}{4 \pi^2 s^2 t/t_N} - \frac{1}{D+2\gamma+ a_{s}} 
 \cr &+ \frac{\cos{2\theta_{0}}e^{-8 \pi^2 s^2 t/t_N}}{2D_{R}} -\frac{(2\cos^2{\theta_{0}}+1)e^{-8 \pi^2 s^2 t/t_N}}{2(D_{R}+2\gamma)}  \bigg].
 \label{sum FN-A2}
\end{align}
In the thermodynamic limit $N \to \infty$, \eref{eq:longt_final} is recovered by converting the sum into an integral by taking $q=2 \pi s /N$. 


\begin{figure} [t]
            \centering
            \includegraphics[scale=0.7]{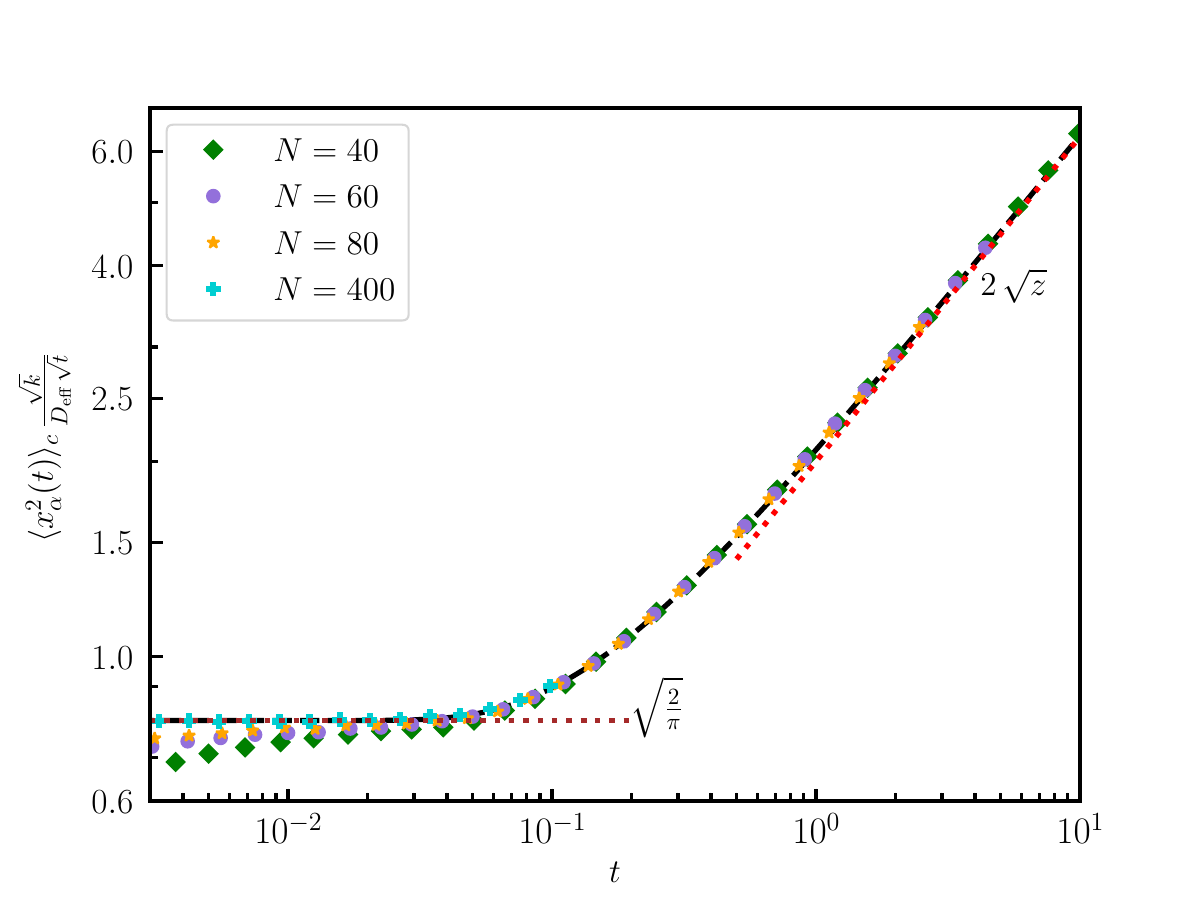}
            \caption{Comparison between the theoretical expression for scaling function \eref{scaling_fn1} (solid black line) with numerical simulation for different values of $N$, keeping $k=1$, $\dr=1$, $\gamma=1$ and $v_0=1$ fixed.}
           \label{fig:scaling_fn}
\end{figure}

To summarize, from \eref{eq:longt_final} and \eref{sum FNs}, we have the leading order behavior of the variance as
\begin{align}
\big \langle  x^2_{\alpha}(t)  \big \rangle_{c} \simeq D_\text{eff} \times \begin{cases}
    \sqrt{\frac{2t}{\pi k}} & \text{for} ~ t \ll t_N \\
    \frac{2 t}{N} & \text{for} ~ t \gg t_N. \\
\end{cases} 
\label{sfcn}
\end{align}
Thus, in the scaling limit, $t \to \infty$, $N \to \infty$, keeping $t/t_N$ fixed, suggests a scaling form, 
\begin{equation}
    \big \langle  x^2_{\alpha}(t)  \big \rangle_{c} = D_\text{eff} \sqrt{\frac{t}{k}}\, f \left(\frac{t}{t_N} \right),
\end{equation}
where the crossover function must have the limiting behavior, 
\begin{equation}
    f(z) \to \begin{cases}
           \sqrt{2/\pi} & \text{as} ~ z \to 0  \\
           2 \sqrt{z} & \text{as} ~ z \to \infty
          .     \end{cases} 
          \label{sFb}
\end{equation}
In fact, by taking the scaling limit in Eq.~\eqref{sum FN-A2}, we obtain the full scaling function as, 
\begin{equation}
f(z) = \frac{1}{4 \pi^2 \sqrt{z}} \sum_{s=-\infty}^{\infty} \frac{1-e^{-8 \pi^2 s^2 z }}{s^2 }. \label{scaling_fn1}
\end{equation}
In \fref{fig:scaling_fn} we compare this exact scaling function with the same obtained from numerical simulations for different values of $N$ and find an excellent agreement.

The limiting behavior of $f(z)$, mentioned in \eref{sFb}, can be obtained from the above equation. To see the large $z$ behavior, it is useful to separate the $s=0$ term and recast the above equation as,
\begin{equation}
  f(z) = 2 \sqrt{z} + \frac{1}{12 \sqrt{z}} -\frac{1}{2 \pi^2 \sqrt{z}} \sum_{s=1}^{\infty} \frac{e^{-8 \pi^2 s^2 z }}{s^2},
    \label{scaling_fn2}
\end{equation}
where we have used $\sum_{s=1}^{\infty} s^{-2}= \pi^2/6$ and the symmetry of the summand for $s \to -s$. Clearly, the first term gives the limiting behavior mentioned in \eref{sFb}, and the second term is the leading order correction. The higher-order corrections can be systematically obtained from the series.

On the other hand, in the limit $z \ll 1$, with a change of variable $2 \pi s \sqrt{z}=u$, the summation in \eref{scaling_fn1} can be converted to an integral as,
\begin{equation}
    f(z) \xrightarrow{z \to 0} \frac{1}{2 \pi} \int_{-\infty}^{\infty} \frac{1-e^{-2u^2}}{u^2} \, du = \sqrt{\frac{2}{\pi}},
    \label{szb}
\end{equation}
which is consistent with \eref{sFb}. However, it is not straightforward to obtain the corrections to the above limiting behavior systematically from \eref{scaling_fn1}. For this, one can use the Poisson summation formula \cite{zygmund}, to rewrite the sum in \eref{scaling_fn1} as,
\begin{equation}
  f(z)= \sqrt{\frac{2}{\pi}} + \sum_{m=1}^{\infty} \left[ 
2 \sqrt{\frac{2}{\pi }} \, \exp{\left(-\frac{m^2}{8 z}\right)} - \frac{m }{ \sqrt{z}} \, \mathrm{erfc}\left(\frac{m}{ 2 \sqrt{2z}}\right) \right], \label{scaling_fn3}
\end{equation}
which approaches the constant $\sqrt{2/\pi}$, as $z \to 0$. Furthermore, using the asymptotic expansion of $\mathrm{erfc}(x)$, we have a systematic series expansion about $z=0$,
\begin{equation}
    f(z)= \sqrt{\frac{2}{\pi}} \Bigg[ 1 + 8 z \, \sum_{m=1}^{\infty} \frac{1}{m^2} \exp{\left(-\frac{m^2}{8 z}\right)} \,  
    \sum_{n=0}^{\infty} \frac{(-4z)^{n}}{m^{2 n}} \, (2n+1)!!  \Bigg].
    \label{szab}
    \end{equation}

To summarize, we find that for $t \gg N^2/k$ the tagged particle performs a center of mass diffusion with an effective diffusion constant $D_{\text{eff}}/N$. The exact scaling function $f(z)$ ---  governing the crossover of the variance from sub-diffusive $ D_{\text{eff}}\,\sqrt{\frac{2t}{\pi k}}$ to diffusive $2\, (D_{\text{eff}}/N)\,t $ behavior across the finite-size time-scale --- is equivalently given by  \eref{scaling_fn2}, \eref{scaling_fn3}, and \eref{szab},  each suitable for evaluating the function in different limiting scenarios.

\section{Velocity autocorrelation} \label{vc-sec}

The velocity of the tagged particle can be defined as $v_\alpha(t) \equiv \dot x_\alpha(t)$ where $\dot x_\alpha(t)$ is given by equation \eref{eomx}. In this section, we investigate the autocorrelation function $ \big \langle   v_{\alpha}(t_1)v_{\alpha}(t_2)  \big \rangle$ for the annealed initial orientation, following \eref{ft}, can be expressed as,
\begin{equation}
  \big \langle   v_{\alpha}(t_1)v_{\alpha}(t_2) \big \rangle= \sum_{s=0}^{N-1} \sum_{s'=0}^{N-1}  \exp{\Bigg[\frac{i2\pi \alpha(s-s')}{N}\Bigg]}\big \langle   \Tilde{v}_{s}(t_1) \Tilde{v}_{s'}^*(t_2) \big \rangle_{c},
  \label{vifs}
  \end{equation}
where $ \Tilde{v}_{s}^*(t)$ is the Fourier transform of $v_\alpha(t)$. From \eref{eomF}, we can write,
\begin{align}
  \big \langle  \Tilde{ v}_{s}(t_1) \Tilde{v}_{s'}^{*}(t_2) \big \rangle = & \big \langle  \Tilde{\xi}_{s}(t_1) \Tilde{\xi}_{s'}^{*}(t_2) \big \rangle -  a_s \left[ \big \langle  \Tilde{x}_{s}(t_1) \Tilde{\xi}_{s'}^{*}(t_2) \big \rangle +  \big \langle  \Tilde{\xi}_{s}(t_1) \Tilde{x}_{s'}^{*}(t_2) \big \rangle \right] + a_s^2 \big \langle   \Tilde{x}_{s}(t_1) \Tilde{x}_{s'}^*(t_2) \big \rangle,  
  \label{vifs2}
  \end{align}
where $\big \langle  \Tilde{\xi}_{s}(t_1) \Tilde{\xi}_{s'}^{*}(t_2) \big \rangle$ is given by \eref{noise_IFT} and $a_s$ is defined in xxx. The other correlations appearing on the right-hand side of \eref{vifs2} can be computed using \eref{sF}. For $t_1 \ge t_2$, we have, 
\begin{align}
    \big \langle  \Tilde{x}_{s}(t_1) \Tilde{\xi}_{s'}^{*}(t_2) \big \rangle = & \frac{\delta_{s,s'}}{N}  \Int_{0}^{t_1} \,dt' \,  e^{-a_{s}(t_1-t')} G(t',t_2)  \cr
   =& \frac{v_0^2 }{2N} \delta_{s,s'} \Bigg[\frac{2(\dr+2\gamma)e^{-a_s(t_1-t_2)}-(a_s+\dr+2\gamma)e^{-(\dr+2\gamma)(t_1-t_2)}}{(\dr+2\gamma)^2-a_s^2}  \cr 
   & \qquad \qquad - \frac{e^{-a_st_1-(\dr+2\gamma)t_2}}{(a_s+\dr+2\gamma)} \Bigg],
   \label{vxi1}
   \end{align}
   where we have used the explicit form for $G(t_1,t_2)$ from \eref{g_ric}. Similarly, we have,
   \begin{align}
    \big \langle  \Tilde{\xi}_{s}(t_1) \Tilde{x}_{s'}^{*}(t_2)    \big \rangle &= \frac{\delta_{s,s'}}{N}  \Int_{0}^{t_2} \,dt' \,  e^{-a_{s}(t_2-t')} G(t_1,t') \cr
    &=   \frac{v_0^2}{2N} \delta_{s,s'} \left[ \frac{e^{-(\dr+2\gamma)(t_1-t_2)}-e^{-a_s t_2 -(\dr+2\gamma)t_1}}{(a_s+\dr+2\gamma)} \right],
    \label{vxi2}
\end{align}
and,
\begin{align}
   \big \langle   \Tilde{x}_{s}(t_1) \Tilde{x}_{s'}^*(t_2) \big \rangle  =& \frac{\delta_{s,s'}}{N}  \Int_{0}^{t_1} \,dt' \Int_{0}^{t_2} dt'' \, e^{-a_{s}(t_1+t_2-t'-t'')} G(t',t''), \cr 
   = &   \frac{v_0^2 }{2N} \delta_{s,s'} \Bigg[ \frac{(\dr+2\gamma)e^{-a_s(t_1-t_2)}- a_se^{-(\dr+2\gamma)(t_1-t_2)}}{ a_s \left[(\dr+2\gamma)^2-a_s^2 \right]} \cr 
   - & \frac{e^{-a_s(t_1+t_2)}}{a_s(\dr+2\gamma-a_s)} - \frac{e^{-(\dr+2\gamma)t_2-a_s t_1}+ e^{-(\dr+2\gamma)t_1-a_s t_2}}{ (\dr+2\gamma)^2-a_s^2} \Bigg ]. ~~~
    \label{v2s_xsp}
\end{align}
\begin{figure}[t]
\includegraphics[width=0.55\textwidth]{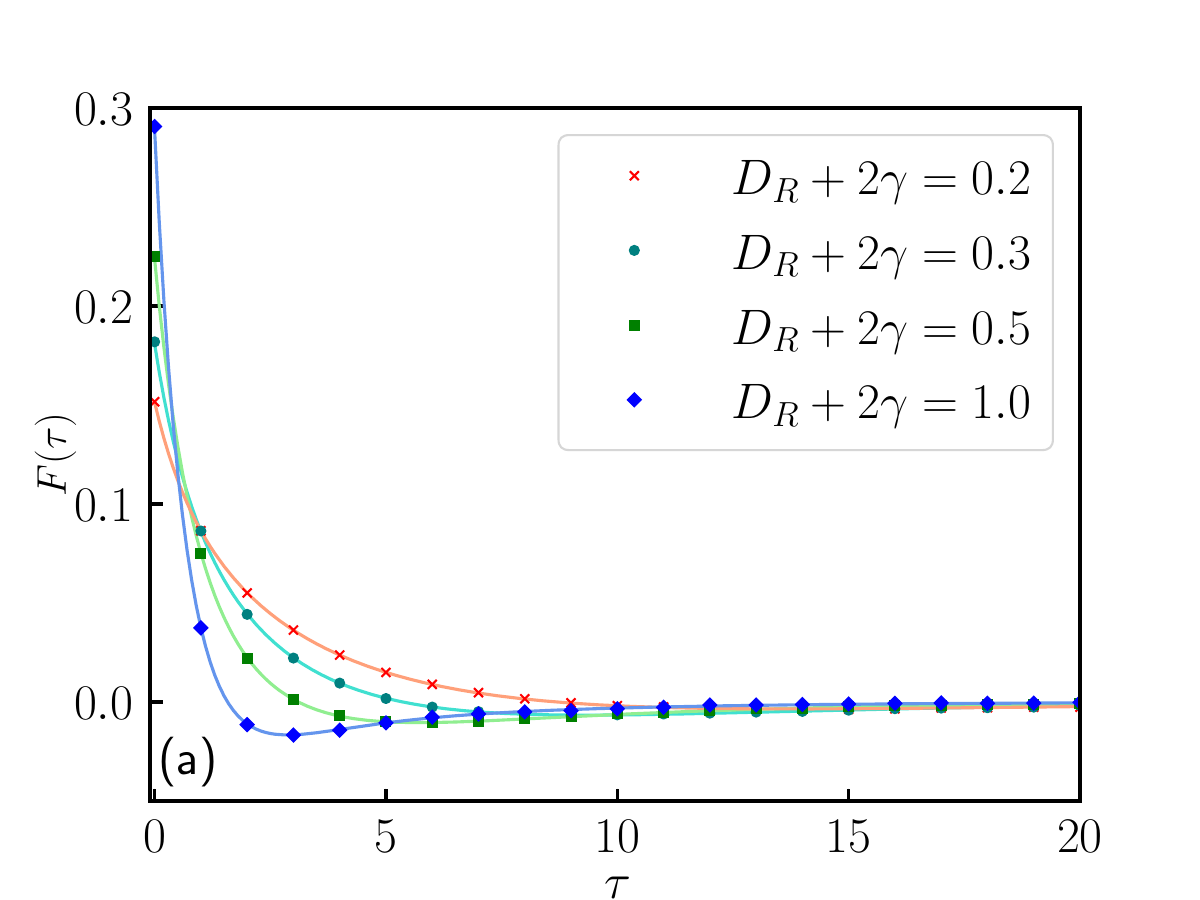} 
\hspace*{-0.8cm}
\includegraphics[width=0.55\textwidth]{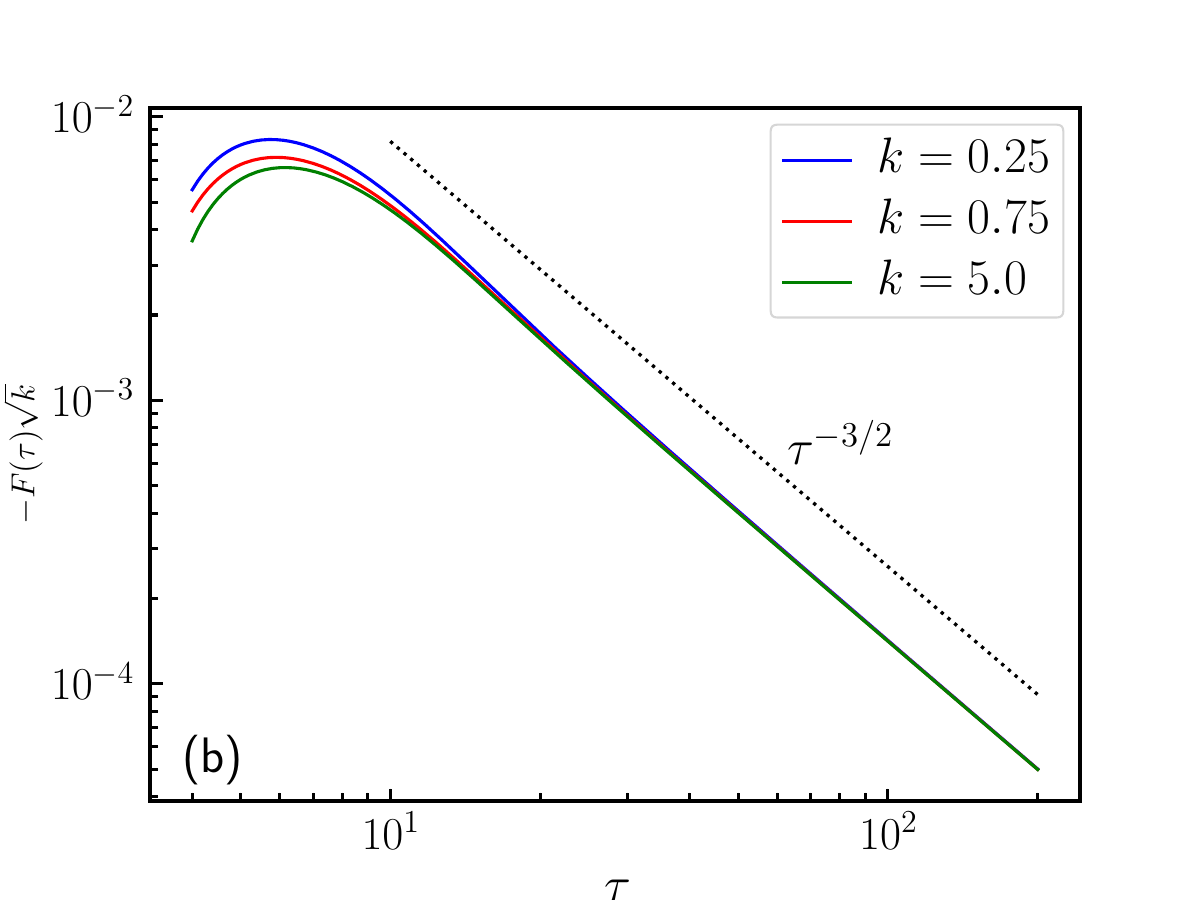}
 \caption{(a) Comparison between the theoretical expression for $F(\tau)$ (solid lines) \eref{ltvalpha-a} with the numerical simulation (symbols) for different values of $\dr+2\gamma$, keeping $k=0.5$, and $N=10$ fixed. (b) Behavior of  $-F(\tau) \sqrt{k}$ (symbol), where $F(\tau)$ given by \eref{ltvalpha}, for different values of $k$, keeping $\dr+2\gamma=0.5$ fixed . The dotted line indicates the $\tau^{-3/2}$ power law tail. For both plots (a) and (b), $v_0$ is set to 1. } \label{fig:vc_t}
\end{figure}

%


Similar expressions for the case $t_2 > t_1$ can be obtained by interchanging $t_1$ and $t_2$ in the above expressions. Using \eref{vxi1}-\eref{v2s_xsp} and \eref{vifs2}  in \eref{vifs} we have,
\begin{align}
    &  \big \langle   v_{\alpha}(t_1)v_{\alpha}(t_2) \big \rangle=   \frac{v_0^2}{2N}  \sum_{s=0}^{N-1} a_s\Bigg [ \frac{a_se^{-(\dr+2\gamma)|t_1-t_2|} -(\dr+2\gamma)e^{-a_s|t_1-t_2|}}{(\dr+2\gamma)^2 -a_s^2}  \cr 
    & +  \frac{ (\dr+2\gamma) \left(e^{-(\dr+2\gamma)t_2-a_s t_1}+ e^{-(\dr+2\gamma)t_1-a_s t_2} \right)}{ (\dr+2\gamma)^2 -a_s^2 }  - \frac{e^{-a_s(t_1+t_2)}}{(\dr+2\gamma-a_s)} \Bigg] \cr & +\frac{v_0^2}{2} e^{-(\dr+2\gamma)|t_1-t_2|}  .\label{tpacvf}
\end{align}
It is useful to note that in the passive limit, the above expression converges to,
\begin{equation}
   \big \langle   v_{\alpha}(t_1)v_{\alpha}(t_2) \big \rangle= -\frac{D_{\text{eff}}}{N} \sum_{s=0}^{N-1} a_s \left[e^{-a_s|t_1-t_2|}+e^{-a_s(t_1+t_2)} \right] + 2 D_{\text{eff}} \delta(t_1-t_2).
   \label{vcpl}
\end{equation}
We remark that, for any process, the velocity autocorrelation and the mean squared displacement are related by,
\begin{equation}
  \frac{1}{2}  \frac{d }{dt} \, \langle [x(t)-x(0)]^2 \rangle= \int_{0}^{t} dt'  \langle v(t)v(t') \rangle
  \label{vc_msd_relation}
\end{equation}
It can be easily checked that the above equation is satisfied for our case, where $x_{\alpha}(0)=0$, by differentiating \eref{ric_msd} and integrating \eref{tpacvf}. 

It can be immediately seen from \eref{tpacvf} that the velocity autocorrelation with the initial velocity,
\begin{equation}
    \big \langle   v_{\alpha}(0)v_{\alpha}(t) \big \rangle=  \frac{v_0^2}{2N}  \sum_{s=0}^{N-1} \frac{(\dr+2\gamma)\, e^{-(\dr+2\gamma)t}-a_s e^{-a_s t}}{\dr+2\gamma-a_s},
\end{equation}
converges to zero as $t \to \infty$. On the other hand, the limit $\{t_1, t_2 \} \to \infty$, the velocity autocorrelation becomes a function of the difference $\tau=|t_1-t_2|$, given by,
\begin{align}
     F(\tau) & \equiv   \lim_{t \to \infty} \big \langle v_{\alpha}(t)v_{\alpha}(t+\tau) \big \rangle \cr
     & = \frac{v_0^2}{2} \, e^{-(\dr+2\gamma)\tau}  - \frac{v_0^2}{2N}  \sum_{s=0}^{N-1}  \frac{a_s \left[(\dr +2 \gamma) e^{-a_s \tau}- a_s e^{-(\dr+2\gamma) \tau}\right ]}{(\dr+2\gamma)^2-a_s^2}. 
    \label{ltvalpha-a}
\end{align}
This indicates the velocity variables eventually reach a stationary state. In the thermodynamic limit $N \to \infty$, the sum in \eref{ltvalpha-a} can be replaced by an integral, yielding,
\begin{align}
    F(\tau) &= \frac{v_0^2 \, e^{-(\dr+2\gamma)\tau}}{2} + v_0^2 \, e^{-(\dr+2\gamma) \tau} \, \int_{-\pi}^{\pi} \, \frac{dq}{4 \pi} \, \frac{b_q^2 \, }{(\dr+2\gamma)^2-b_q^2} \cr 
     &-   v_0^2(\dr +2 \gamma) \int_{-\pi}^{\pi} \, \frac{dq}{4 \pi} \,  \frac{b_q \,  e^{-b_q \tau}}{(\dr+2\gamma)^2-b_q^2},
    \label{ltvalpha}
\end{align}
where $b_q= 4k\sin^2(q/2)$ [see \eref{eq:bq}].  The short-time behavior can be obtained by expanding \eref{ltvalpha} in Taylor series of $\tau$ and carrying out the $q$-integrals,
\begin{equation}
F(\tau)= 2v_0^2 \Bigg[ \frac{1}{\sqrt{1+\mu}} - \frac{3}{4} \Bigg] - \frac{v_0^2}{2} (\dr+2\gamma) \tau + O(\tau^2) ,
\label{stb_vc}
\end{equation}
where,
\begin{equation}
    \mu=\frac{k}{(\dr+2\gamma)},
    \label{mueqn}
\end{equation}
denotes the ratio of the active time-scale and the trap time-scale. In particular, $F(0)=2v_0^2 (1/\sqrt{1+\mu} - 3/4)$ gives the stationary variance of the velocity $\la v_{\alpha}^2 \ra$.

In fact, the second integral in \eref{ltvalpha} can be performed explicitly for $\dr +2 \gamma > 4 k$. Moreover, for $\dr +2\gamma \gg k$, by ignoring the $-b_q^2$ term in the denominator, the third integral can also be performed, yielding,
\begin{align}
    F(\tau) \simeq \frac{v_0^2}{4} \left[ \frac{1}{\sqrt{1-4 \mu }} + \frac{1}{\sqrt{1+4 \mu }} \right] \, e^{-(\dr+2\gamma)\tau} - v_0^2 \, \mu \, e^{-2 k \tau} \left[I_0\left(2k \tau \right)-I_1\left( 2 k \tau \right)\right], 
    \label{eq:k<d}
\end{align}
where $I_n(z)$ denotes the modified Bessel function of the first kind. Note that, in the passive limit, the first term becomes $2 D_{\text{eff}} \, \delta(\tau)$, and $v_0^2 \, \mu \to  2 D_{\text{eff}} \,  k $ in the second term. Moreover, expanding \eref{eq:k<d} up to $O((\dr+2\gamma)\tau)$ we see that it matches with \eref{stb_vc} to $O(\mu)$. On the other hand, the asymptotic expansion of \eref{eq:k<d} for large $\tau$ results in,
\begin{align}
   F(\tau) & \simeq  -\frac{D_{\text{eff}}}{4 \sqrt{\pi k}} \, \tau^{-3/2}. 
    \label{ltvalpha_approximation}
\end{align}

We illustrate the behavior of $F(\tau)$ in \fref{fig:vc_t}. The correlation function starts from a positive value $F(0)$, and decays to a negative minimum value at some $\tau^*$, and then approaches zero as a power-law $- \tau^{-3/2}$.

\section{Statistics of the separations}\label{sec:separations}

For a chain of passive particles, while the center of mass diffuses freely in the absence of any global confining potential, the separations between the adjacent particles reach an equilibrium state. In fact, for a chain of $N$ passive particles with free boundary conditions (i.e., when $x_0$ and $x_{N-1}$ are only coupled to $x_1$ and $x_{N-2}$ respectively), at a temperature $T$, the $N-1$ separations, $y_{\alpha}=x_{\alpha+1}-x_{\alpha}$, with $\alpha=0,1, \dotsc, N-2$ eventually reach an equilibrium state given by the product measure,
\begin{align}   
P(\{ y_\alpha\}) \propto \exp{\left[-\frac k{2k_{B}T}\sum_{\alpha=0}^{N-2} y_\alpha^2\right ]}. \label{eq:ydist_passive2}
\end{align}
On the other hand, with a periodic boundary condition, the $N$ separations must satisfy $\sum_{\alpha=0}^{N-1} y_\alpha=0$, and hence are not independent. While this global constraint destroys the product measure of the equilibrium state for any finite $N$, one expects that in the limit $N \to \infty$, the product measure is restored for any finite subset of $ \{ y_{\alpha}\}$. In particular, for any $\alpha$,
\begin{equation}
P(y_\alpha) = \sqrt{\frac{k}{2 \pi k_{B}T}} \, \exp{\left[-\frac {k \, y_\alpha^2}{2k_{B}T} \right ]}, \quad\quad   \text{as} ~~ N \to \infty. \label{eq:ydist_passive2}
\end{equation}
It is interesting to ask how activity affects this behavior, which we explore in this section.

Following \eref{eomx}, the equation of motion for the separation for the active chain is given by,
\begin{equation}
\dot{y}_{\alpha}=-k(2y_{\alpha}-y_{\alpha +1}-y_{\alpha -1}) + \eta_{\alpha}(t),
\label{yeqn1}
\end{equation}
where, 
\begin{equation}\eta_{\alpha}(t)=\xi_{\alpha+1}(t)-\xi_{\alpha}(t).
\label{ynoise}
\end{equation}
Since the stationary state is independent of the initial condition, for simplicity, we set  $\{y_{\alpha}(0)=0\}$ for all $\alpha$. Moreover, here, we are using the quenched initial orientation $\{\theta_{\alpha}(0)=\theta_{0}\}$ and $\{\sigma_{\alpha}(0)=1\}$ for all $\alpha$ in the noise $\eta_{\alpha}$ above. We begin by computing two-point spatio-temporal correlation function, $ \big \langle y_{\alpha}(t) y_{\beta}(t+\tau) \big \rangle $. To proceed, we perform DFT [defined in \eref{ft}] on \eref{yeqn1} with respect to $\alpha$, so that we get $N$ decoupled first order differential equations,
\begin{equation}
 \dot {\Tilde{y}}_{s}(t)= -a_{s} \Tilde{y}_{s}(t) + \Tilde{\eta}_{s}(t),~~~ \text{with},\quad  a_{s}=4k\sin^2\Big({\frac{\pi s}{N}}\Big),
 \label{eomFy}
\end{equation}
where $\{ \tilde y_s(t) \}$ and $\{ \Tilde{\eta}_{s}(t) \}$ (for $s= 0,1, \dotsc N-1$) denote the DFT of $\{  \tilde y_\alpha(t)\}$ and $\{\tilde \eta_\alpha(t)\}$ respectively. Since $\{y_\alpha(0)=0\}$, we have $\{ \Tilde{y}_{s}(0)=0\}$. The solution of \eref{eomFy} is given by,
\begin{equation}
  \Tilde{y}_{s}(t)=  e^{-a_{s}t}  \int_{0}^{t} e^{a_{s} t_1} \Tilde{\eta}_{s}(t_1) \, dt_1. \label{sFy}
\end{equation}
Using \eref{ft}, the spatio-temporal correlation can be expressed as,
\begin{equation}
  \big \langle y_{\alpha}(t) y_{\beta}(t+\tau) \big \rangle= \sum_{s=0}^{N-1} \sum_{s'=0}^{N-1}  \exp{\Bigg[\frac{i2\pi (s \alpha - s' \beta)}{N}\Bigg]}\big \langle   \Tilde{y}_{s}(t) \Tilde{y}_{s'}^*(t+\tau) \big \rangle.
  \label{cifs-y}
  \end{equation}
Now, the correlation in the Fourier space, $\big \langle   \Tilde{y}_{s}(t) \Tilde{y}_{s'}^*(t+\tau) \big \rangle$ can be obtained using \eref{sFy},
\begin{equation}
  \big \langle   \Tilde{y}_{s}(t) \Tilde{y}_{s'}^*(t+\tau) \big \rangle =  e^{-(a_{s}t+a_{s'}(t+\tau))}  \int_{0}^{t+\tau} dt_{2} \int_{0}^{t} dt_{1}\, e^{a_{s}t_{1}+a_{s'}t_{2}} \big \langle \Tilde{\eta}_{s}(t_{1}) \Tilde{\eta}^{*}_{s'}(t_{2})\big \rangle, 
\label{cfsy}
\end{equation}
where,
\begin{equation}
 \big \langle \Tilde{\eta}_{s}(\tau_{1}) \Tilde{\eta}^{*}_{s'}(\tau_{2})\big \rangle = \frac{1}{N^2} \sum_{\alpha,\alpha '=0}^{N-1}  e^{\frac{i2\pi }{N}(s' \alpha '-s \alpha)}\big \langle   \eta_{\alpha}(\tau_{1}) \eta_{\alpha'}(\tau_{2}) \big \rangle.
 \label{ynoise2n}
\end{equation}
Here we note that,
\begin{align}
    \big \langle   \eta_{\alpha}(\tau_{1}) \eta_{\alpha'}(\tau_{2}) \big \rangle &= \big \langle \xi_{\alpha+1}(\tau_{1})\xi_{\alpha'+1}(\tau_{2})\big \rangle + \big \langle \xi_{\alpha}(\tau_{1})\xi_{\alpha'}(\tau_{2})\big \rangle \cr &- \big \langle \xi_{\alpha+1}(\tau_{1})\xi_{\alpha'}(\tau_{2})\big \rangle - \big \langle \xi_{\alpha}(\tau_{1})\xi_{\alpha'+1}(\tau_{2})\big \rangle \cr &= [2\delta_{\alpha,\alpha'}-\delta_{\alpha+1,\alpha'}-\delta_{\alpha-1,\alpha'}] G_2(\tau_1,\tau_2),
    \label{ynoise_deltan}
\end{align}
where $G_2(\tau_{1},\tau_{2}) \equiv \big \langle \xi_{\alpha}(\tau_{1})\xi_{\alpha}(\tau_{2})\big \rangle$ is given by,
\begin{align}
G_2(\tau_{1},\tau_{2}) =\frac{v_{0}^2}{2}\Big[ & e^{-(D_{R}+2\gamma) \mid t_{2}-t_{1} \mid} + \cos{ 2\theta_{0}}e^{-\big(D_{R}(  t_{2}+t_{1}+ 2 \, \text{min} [ t_{1},t_{2} ]) +2\gamma \mid t_{2}-t_{1} \mid \big)} \Big].
\label{ficc_sp}
 \end{align}
Substituting \eqref{ynoise_deltan} on \eqref{ynoise2n} and performing the summations over $\alpha,\alpha'$, we get
\begin{equation}
 \big \langle \Tilde{\eta}_{s}(\tau_{1}) \Tilde{\eta}^*_{s'}(\tau_{2})\big \rangle = \frac{4}{N} \sin^2{\Big(\frac{\pi s}{N} \Big)} G_2(\tau_{1},\tau_{2}) \delta_{s,s'}.
 \label{nscy}
\end{equation}

To calculate the spatio-temporal correlation, we substitute \eref{nscy} in \eref{cfsy}, and using \eref{cifs-y} we get,
\begin{align}
  & \big \langle  y_{\alpha}(t) y_{\beta}(t+\tau) \big \rangle = \frac{v^2_0}{2Nk} \sum_{s=0}^{N-1} \frac{\cos{(2 \pi s (\alpha-\beta)/N)} a_s}{a_s-D_{R}-2\gamma} \Bigg[ \frac{e^{-a_s(t+2\tau)}(2\gamma-3\dr+a_s)\cos{2\theta_0}}{(a_s-2\dr)(a_s-3\dr+2\gamma)} 
   \cr 
   & + \frac{a_se^{-(\dr+2\gamma)\tau}-(\dr+2\gamma)e^{-a_s \tau}-e^{-(\dr+2\gamma)(2t+\tau)}+(a_s+\dr+2\gamma)e^{-a_s(2t+\tau)}}{a_s(a_s+\dr+2\gamma)} 
   \cr 
& + \frac{\left[(\dr-2\gamma)e^{-4\dr t-a_s\tau}-(a_s-2\dr) \left(e^{-(\dr+2\gamma)(2t+\tau)}-e^{-(2\gamma \tau+\dr(4t+\tau))} \right) \right] \cos{2 \theta_{0}}}{(a_s-2\dr)(a_s-3\dr+2\gamma)} \Bigg]. \label{stcfyn}
\end{align}

\begin{figure} [t]
            \centering
            \includegraphics[scale=0.7]{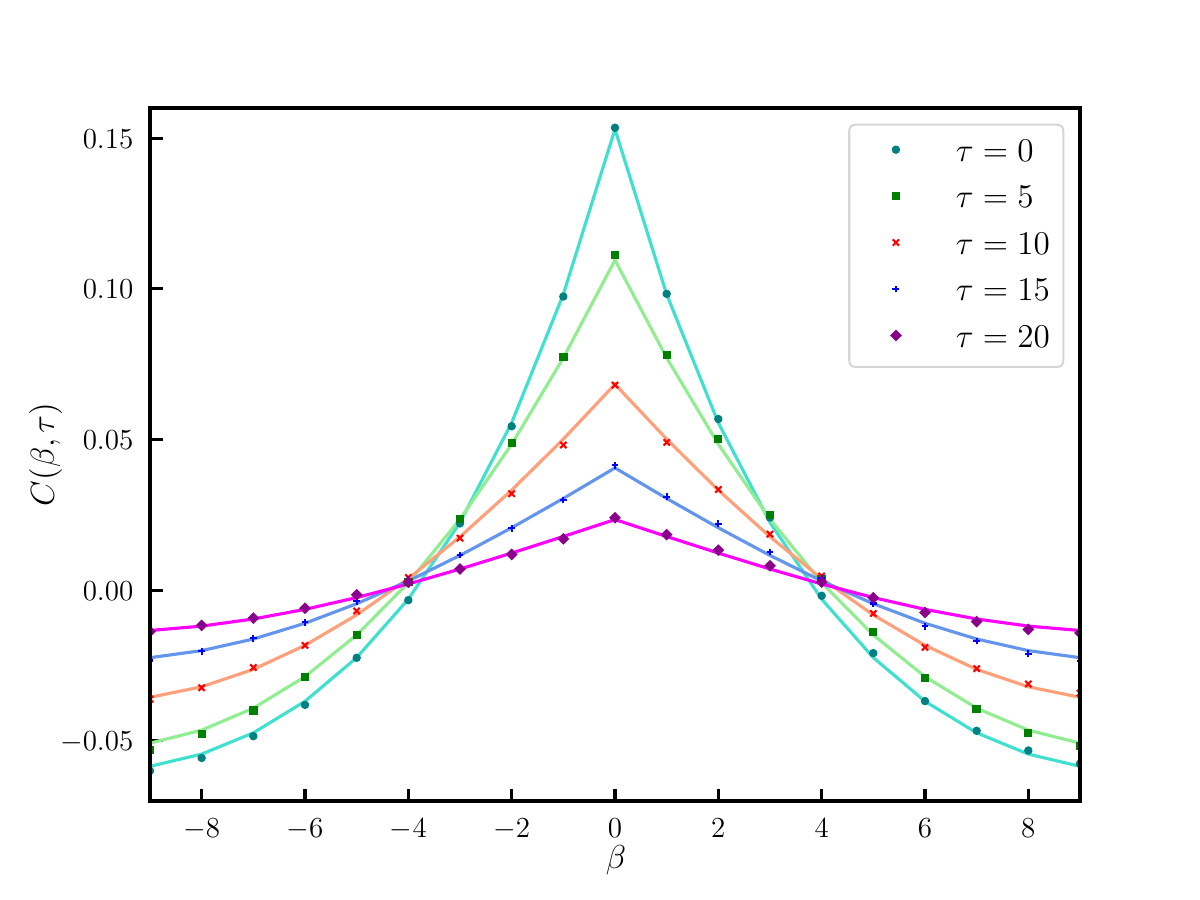}
            \caption{Comparison between the theoretical expression for $C(\beta, \tau)$ (solid lines) \eref{2tcr-an} with the numerical simulation (symbols) for different values of $\tau$, keeping $k=2.0$, $\dr=0.1$, $\gamma=0.01$,  $v_0=1$ and $N=20$ fixed.}
           \label{fig:stc_t}
\end{figure}

\begin{figure} [t]
            \centering
            \includegraphics[scale=0.7]{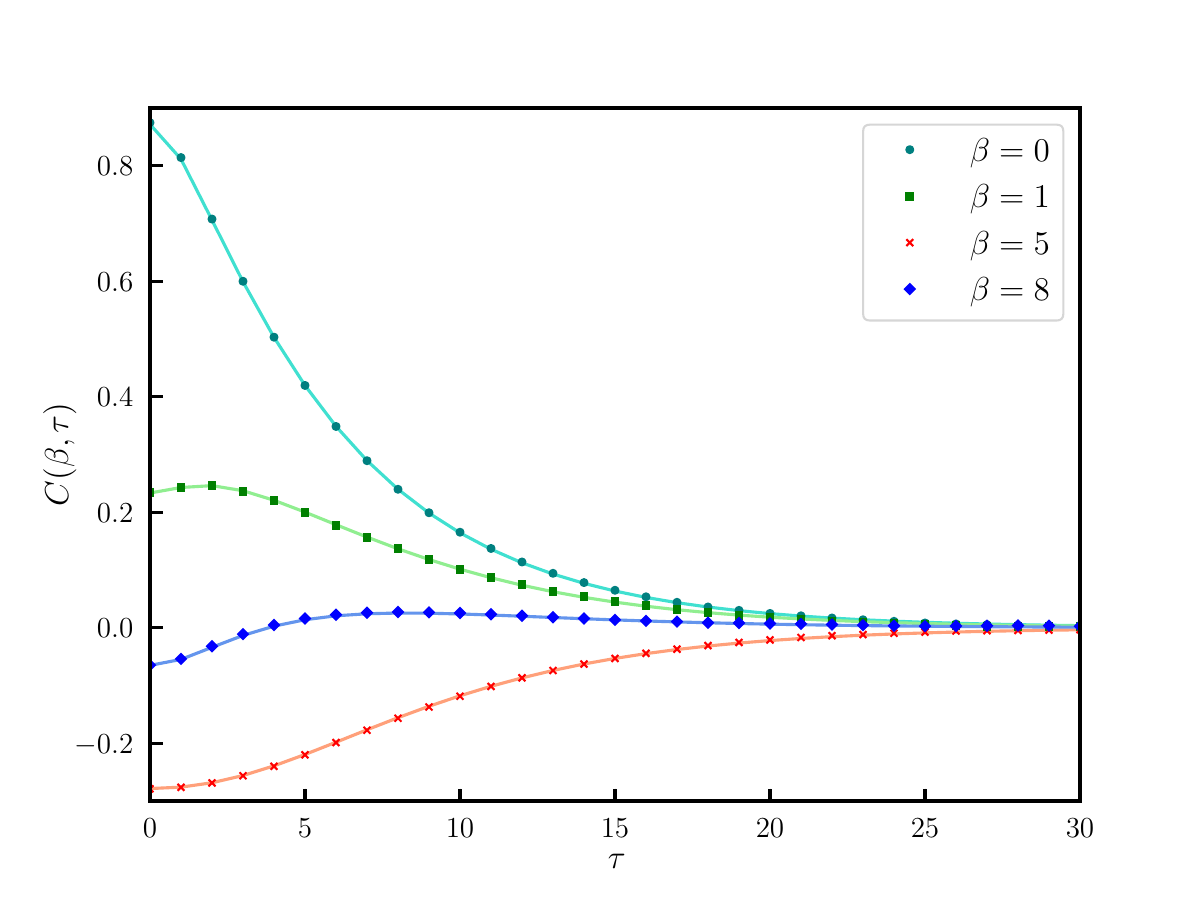}
            \caption{Comparision between the theoretical expression for $C(\beta,\tau)$ (solid lines) \eref{2tcr-an} with the numerical simulation (symbols) for different values of $\beta$, keeping $k=0.5$, $\dr=0.1$, $\gamma=0.1$, $v_0=1$ and $N=10$ fixed}
           \label{fig:stc_b}
\end{figure}

Here we note that the imaginary part of $ \big \langle  y_{\alpha}(t) y_{\beta}(t+\tau) \big \rangle$ turns out to be zero, as expected. To obtain the steady-state behavior of the correlation function, we take the limit $t \to \infty$ in \eref{stcfyn}. Furthermore, since the correlation function depends only on $|\alpha-\beta|$, we set $\alpha=0$ without loss of generality. This yields,
\begin{align}
   C(\beta,\tau) &:=\lim_{t\to\infty} \big \langle  y_{0}(t) y_{\beta}(t+\tau) \big \rangle \cr  &= \frac{v_0^2}{2Nk} \sum_{s=1}^{N-1}  \frac{\cos{(2 \pi s \beta/N)}\left(a_s e^{-(D_{R}+2\gamma)\tau}-e^{-a_s \tau}(\dr+2\gamma) \right)}{(a_s-D_{R}-2\gamma)(a_s+D_{R}+2\gamma)}.
   \label{2tcr-an}
\end{align}
In figures~\ref{fig:stc_t} and \ref{fig:stc_b}, we compare the steady-state behavior of the above-mentioned correlation function \eref{2tcr-an} with numerical simulation for varying $\tau$ and $\beta$ respectively and find very good agreement.

Next, we compute exactly the variance $C(0,0)$, the equal-time spatial correlation $C(\beta,0)$, and the spatio-temporal correction $ C(\beta,\tau)$ in the thermodynamic limit $N \to \infty$. \\

\noindent \textbullet~\textbf{Variance:}
Since the mean separation vanishes, i.e., $\langle y_{\beta}(t \to \infty) \rangle=0$, the variance is given by $C(0,0)$. Substituting $\beta=0$ and $\tau=0$ in \eref{2tcr-an} we get,
\begin{align}
 C(0,0) &= \frac{D_\text{eff}}{Nk}\sum_{s=1}^{N-1} \frac{1}{1+ 4 \mu \sin^2{(\pi s/N)}},
 \label{sumfm}
\end{align}
where $D_{\text{eff}}$ and $\mu$ is given by \eref{eq:deff_def} and \eref{mueqn} respectively. For any finite $N$, the stationary state variance of $y_{\alpha}$ could be calculated by performing the summation in \eqref{sumfm}.

Note that in the passive limit, $v_{0} \to \infty$ and $(D_{R}+2 \gamma) \to \infty$ while keeping $D_{\text{eff}}$ finite and $\mu \to 0$. Consequently, in the passive limit, $C(0,0)=(D_{\text{eff}}/k) (1-1/N) \to D_{\text{eff}}/k$ as $N \to \infty$. Although the potential $U(\{ y_{\alpha} \})=\frac k 2 \sum_{\alpha=0}^{N-1} y_\alpha^2$ suggests a product measure equilibrium state,
\begin{equation}
    P_{\text{eq}}(\{ y_\alpha \}) \propto \exp \left[-U( \{ y_{\alpha} \})/D_{\text{eff}} \right],
    \label{peqy}
\end{equation}
the global constraint $\sum_{\alpha=0}^{N-1} y_\alpha =0$ weakly breaks the product measure, giving rise to the $1/N$ correction in the variance. As expected, the correction disappears in the thermodynamic limit $N \to \infty$.

For any finite activity $\mu$, the summation over $s$ in \eref{sumfm} can be converted to an integration over $q=2\pi s/N$ in the large $N$ limit. Then, we have,
\begin{align}
 C(0,0)= \lim_{t \to \infty\atop N \to \infty }\big \langle  y^2_{\alpha}(t) \big \rangle &= \frac{D_\text{eff}}{2\pi k}  \int_{0}^{2\pi}  \,\  \frac{dq}{1+ 4 \mu \sin^2{(q/2)}}=\frac{D_\text{eff}}{k \sqrt{1+ 4 \mu}},\label{eq:C00}
\end{align} 
which implies that the typical fluctuation of the separations decreases with increasing activity. It is interesting to compare the above result with that of a 
single DRABP in a harmonic trap of strength $k$, where the variance is given by $D_{\text{eff}}/(k(1+\mu))$, which has the same passive limit. It appears that, for large activity, the presence of the interaction allows for larger fluctuations.  \\

\begin{figure} [t]
            \centering
            \includegraphics[scale=0.7]{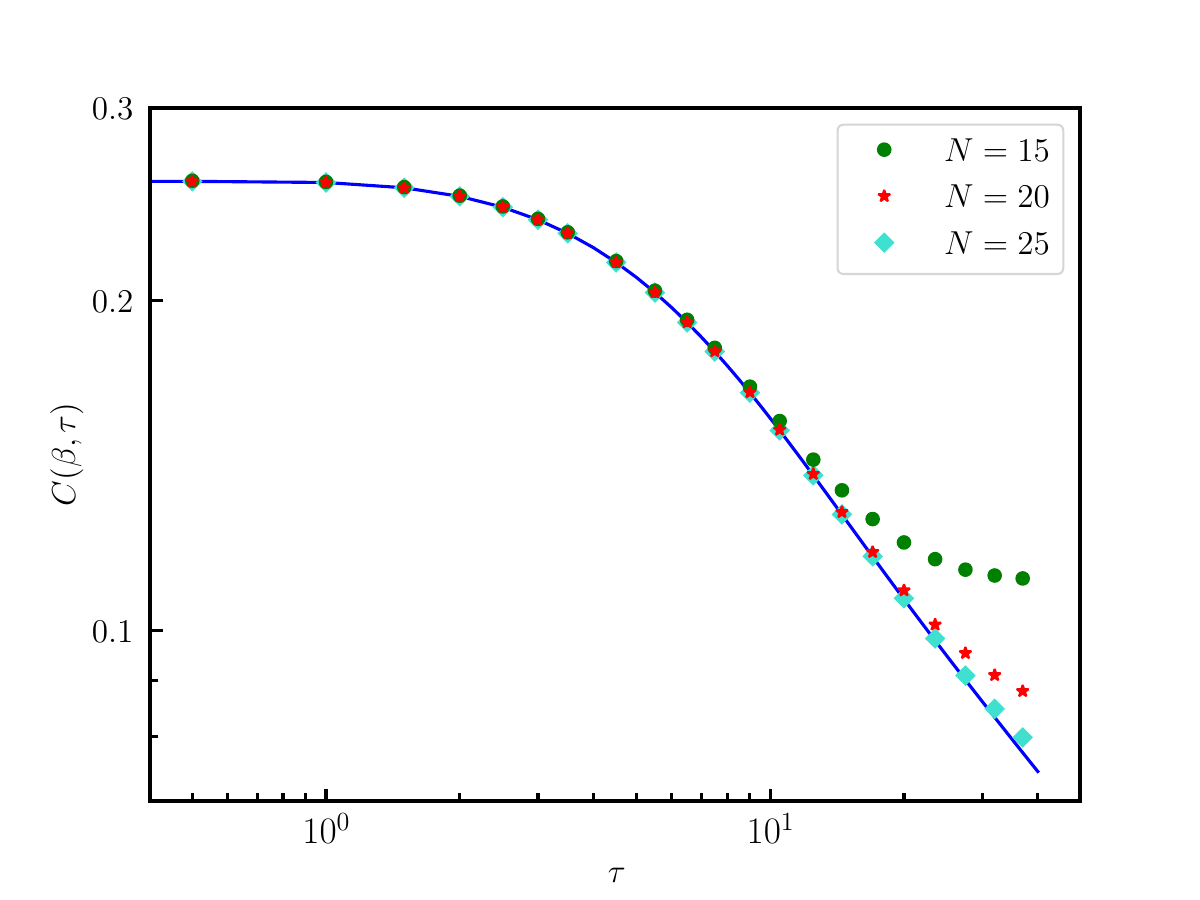}
            \caption{Comparison between the integral expression for $C(\beta, \tau)$ (blue solid line) \eref{2tcr2n} with the summation expression (symbols) \eref{2tcr-an} for different values of $N$, keeping $k=1$, $\dr=0.1$, $\gamma=0.1$, $\beta=1$ and $v_0=1$. For any finite $N$, the integral \eref{2tcr2n} has a correction $-D_{\text{eff}}/(k \, N)$, we have added $D_{\text{eff}}/(k \, N)$ to the summation \eref{2tcr-an} in this plot. }
           \label{fig:intvssum}
\end{figure}

\noindent \textbullet~\textbf{Spatio-temporal correlation:}
Let us now consider the general spatio-temporal correlation function $C(\beta,\tau)$ in the thermodynamic limit. In the limit $N \to \infty$ the summation in the expression \eref{2tcr-an} can be converted to an integral by taking $q=2 \pi s/N$,
\begin{align}
  C(\beta,\tau)&= \frac{v^2_{0}}{4\pi k} \int_{-\pi}^{\pi}  \,\ dq  \cos{(\beta q)} \bigg[ \frac{(\dr+2\gamma)e^{-b_q\tau}-b_{q}e^{-(\dr +2 \gamma)\tau}}{(\dr+2\gamma)^2-b^2_q}  \bigg],
   \label{2tcr2n}
\end{align}
where $b_q$ is given by \eref{eq:bq}. \Fref{fig:intvssum} shows how the summation expression \eref{2tcr-an} approaches the integral expression as $N$ is increased.

The integral is hard to compute explicitly for an arbitrary value of $\tau$. Therefore, we expand the integrand in Taylor's series as a power of $\tau$ and carry out the integral at each order. It turns out that it can be recast in the following form, 
\begin{equation}
    C(\beta,\tau)= \sum_{n=0}^{\infty} C_n(\beta)\, \frac{\big[(\dr+2\gamma)\tau\big]^n}{n!},
    \label{cbt1}
\end{equation}
 where, the coefficients $C_n(\beta)$ are given below. First, for $n=0$, we have,
\begin{align}
  C_0(\beta) \equiv C(\beta,0) &= \frac{D_\text{eff}}{2 \pi k} \int_{-\pi}^{\pi}  \,\ dq \,  \frac{\cos{\beta q} }{4 \mu \sin^2{q/2}+1} \label{2tcqm} \\
   & = \frac{D_\text{eff}}{k (1 + 4 \mu)}  \, {}_3\tilde{F}_2 \left(\frac 12, 1, 1; 1- \beta, 1+ \beta; \frac {4 \mu}{1 + 4 \mu} \right),
   \label{2tcqmFn}    
\end{align}
where $\mu$ is defined by \eref{mueqn}. For a given $\beta$, the above expression of $C_0(\beta)$ has a series expansion in powers of $\mu$, starting from $\mu^{\beta}$.
Next, it is straightforward to see that the coefficient of $\tau$ is zero in the integrand in \eref{2tcr2n}, resulting in $C_1(\beta)=0$. For $n \geq 2$, we have, 
\begin{align}
   C_{n}(\beta)= -(-1)^n \,\frac{v^2_{0}}{4\pi k} \int_{-\pi}^{\pi}  \,\ dq  \cos{(\beta q)} \, S_n(q),
   \label{cnbeta1}
\end{align}
where,
\begin{equation}
   (\dr+2\gamma)^n \, S_n(q) = (D_R+2\gamma) \, b_{q} \, \frac{(b_q)^{n-1}-(\dr+2\gamma)^{n-1}}{b_q^2-(D_R+2\gamma)^2}.
   \label{sf}
\end{equation}
After further algebraic manipulation, the integral can be performed [see \ref{app:Snq}], which yields
\begin{equation}
     C_2(\beta)= C_0(\beta)-\frac{D_{\text{eff}}}{k} \, \delta_{\beta,0},
     \label{c2:eqn}
\end{equation}
and for $n \geq 3$,
\begin{equation}
C_n(\beta)  =\begin{cases} \displaystyle \frac{1+(-1)^n}{2} \,   C_0(\beta) - (-1)^{n+\beta}  \, \frac{D_{\text{eff}}}{k} \, \sum_{l=\beta}^{n-2}   \frac{1+ (-1)^{l+n}}{2}\, \binom{2l}{l-\beta} \,\mu^l  \quad &\text{for} \, \beta \leq n-2 \cr
\displaystyle           \frac{1+(-1)^n}{2} \,   C_0(\beta) \quad &\text{for} \, \beta >n-2.
    \end{cases}
    \label{cnbetaF}
\end{equation}
From the first line of \eref{cnbetaF}, it is apparent that the summation over $l$ contributes only when $l+n$ is even. For example, for $n=3$,  and $4$ we have,
\begin{equation}
\begin{split}
    C_3(\beta) &=\frac{ D_{\text{eff}}}{k} \left( 2 \delta_{\beta,0} - \delta_{\beta,1} \right) \mu, \cr 
    C_4(\beta) &= C_0(\beta) - \frac{D_\text{eff}}{k} \left[ \delta_{\beta,0} +(6 \delta_{\beta,0}- 4 \delta_{\beta,1}+ \delta_{\beta,2})\, \mu^2 \right].
    \end{split}
\end{equation}
In fact, the summation over $l$ in \eref{cnbetaF} can also be carried out, giving a closed-form expression for $C_n(\beta)$ for $n \ge 3$ and $\beta \leq n-2$ as,
\begin{align}
    C_n(\beta)= \frac{1+(-1)^n}{2} \,  C_0(\beta) &\cr  - (-1)^{n+\beta}  \, \frac{D_{\text{eff}}}{2k} \, \Bigg\{&\mu ^{\beta } \bigg[ \, _2F_1\left(\beta +\frac{1}{2},\beta +1;2 \beta +1;4 \mu \right) \cr 
&+ (-1)^{n+\beta} \, _2F_1\left(\beta +\frac{1}{2},\beta +1;2 \beta +1;-4 \mu \right) \bigg] \cr 
&+ \mu ^{n-1} \binom{2 (n-1)}{n-\beta -1} \bigg[ \, _3F_2\left(1,n-\frac{1}{2},n;n-\beta ,n+\beta ;-4 \mu \right) \cr 
&-\, _3F_2\left(1,n-\frac{1}{2},n;n-\beta ,n+\beta ;4 \mu \right) \bigg]      \Bigg\}.
\end{align}

\begin{figure}[t]
\includegraphics[width=0.55\textwidth]{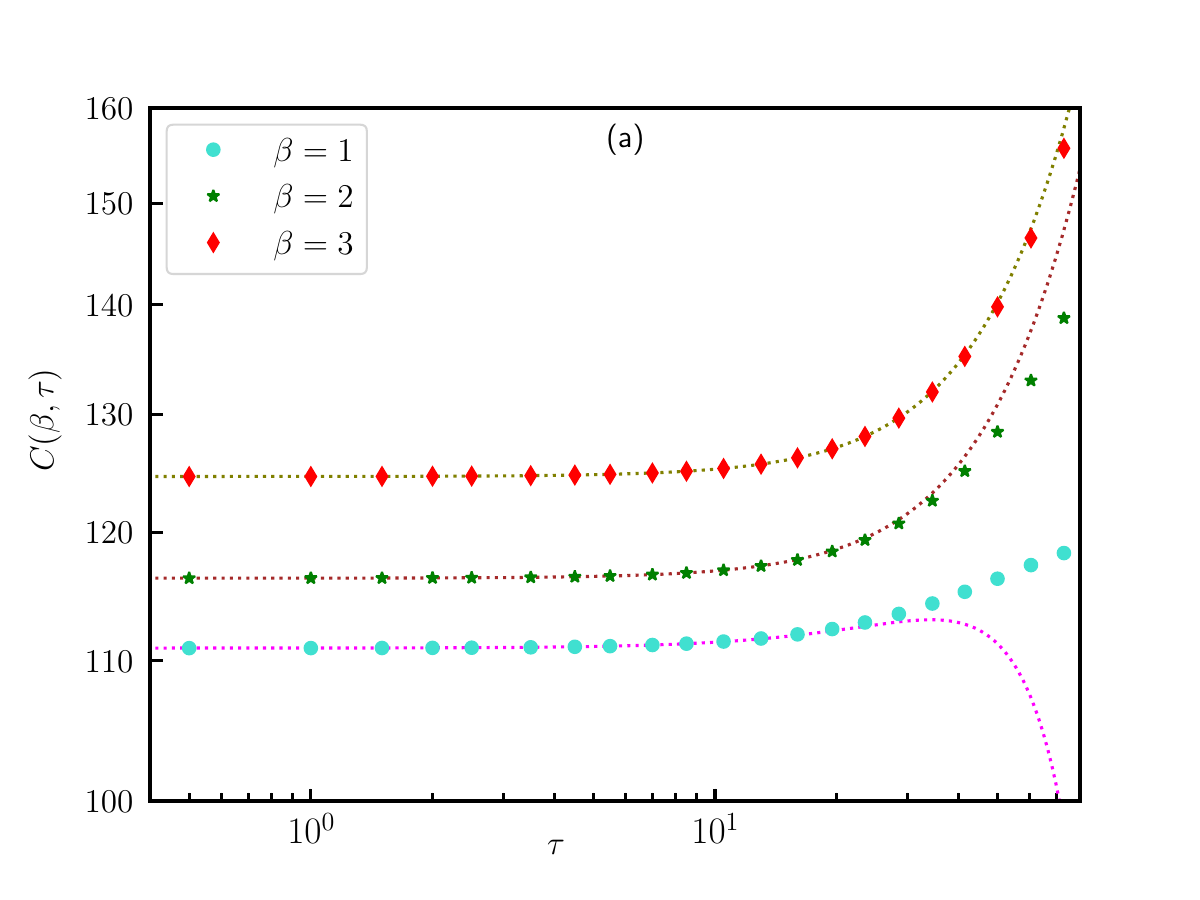}
\hspace*{-0.8cm}
\includegraphics[width=0.55\textwidth]{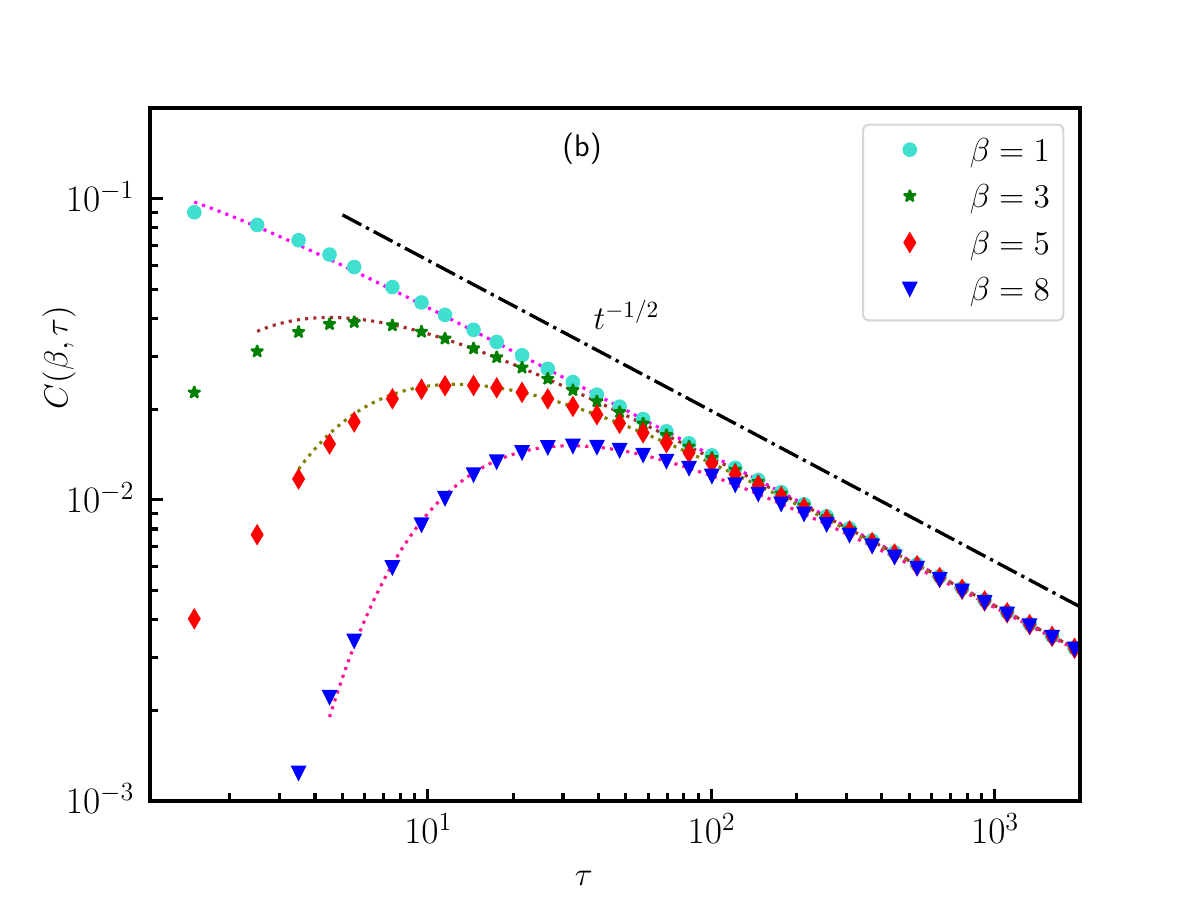}
 \caption{ Behaviour of $C(\beta,\tau)$ for different values of $\beta$ at (a) short-times with $k=0.01$, $\dr=0.005$, $\gamma=0.0025$, and at (b)  late-times with $k=1$, $\dr=0.5$, $\gamma=0.25$. The symbols are obtained by numerically integrating \eref{2tcr2n}. The dotted lines in (a) plot \eref{cbt1}, keeping the first three terms. On the right panel (b), the dotted lines plot the function given by \eref{2tcrAn}. The dash-dotted line indicates the $\tau^{-1/2}$ power law tail. We have taken $v_0=1$ for both the plots.} \label{fig:stc_st_lt}
\end{figure}

For any finite $\tau$, the correlation function $C(\beta,\tau)$ can be evaluated up to arbitrary accuracy by taking sufficient number of coefficients $C_n(\beta)$ in \eref{cbt1}. However, this expansion is not suitable for extracting the asymptotic large $\tau$ behavior of $C(\beta, \tau)$. This can instead be done by taking $\tau \gg \{ (\dr+2\gamma)^{-1}, k^{-1} \}$, where $e^{-(\dr+2\gamma)t} \to 0$ and $e^{-b_{q} \tau} \rightarrow e^{-kq^2 \tau}$ in \eref{2tcr2n}. Consequently, the denominator in \eref{2tcr2n} can be further approximated by $(\dr+2\gamma)^2 -b^2_{q} \rightarrow (\dr+2\gamma)^2$. This yields,
\begin{align}
   C(\beta,\tau) &=  \frac{D_{\text{eff}}}{2 \pi k } \int_{-\pi}^{\pi}  \,\ dq  \cos{(\beta q)} \, e^{-k q^2 \tau} = \frac{D_{\text{eff}} }{2k \sqrt{\pi k }} \frac{e^{-\beta^2/(4k\tau)}}{\sqrt{\tau}}.   
   \label{2tcrAn}
\end{align}

\begin{figure} [t]
            \centering
            \includegraphics[scale=0.7]{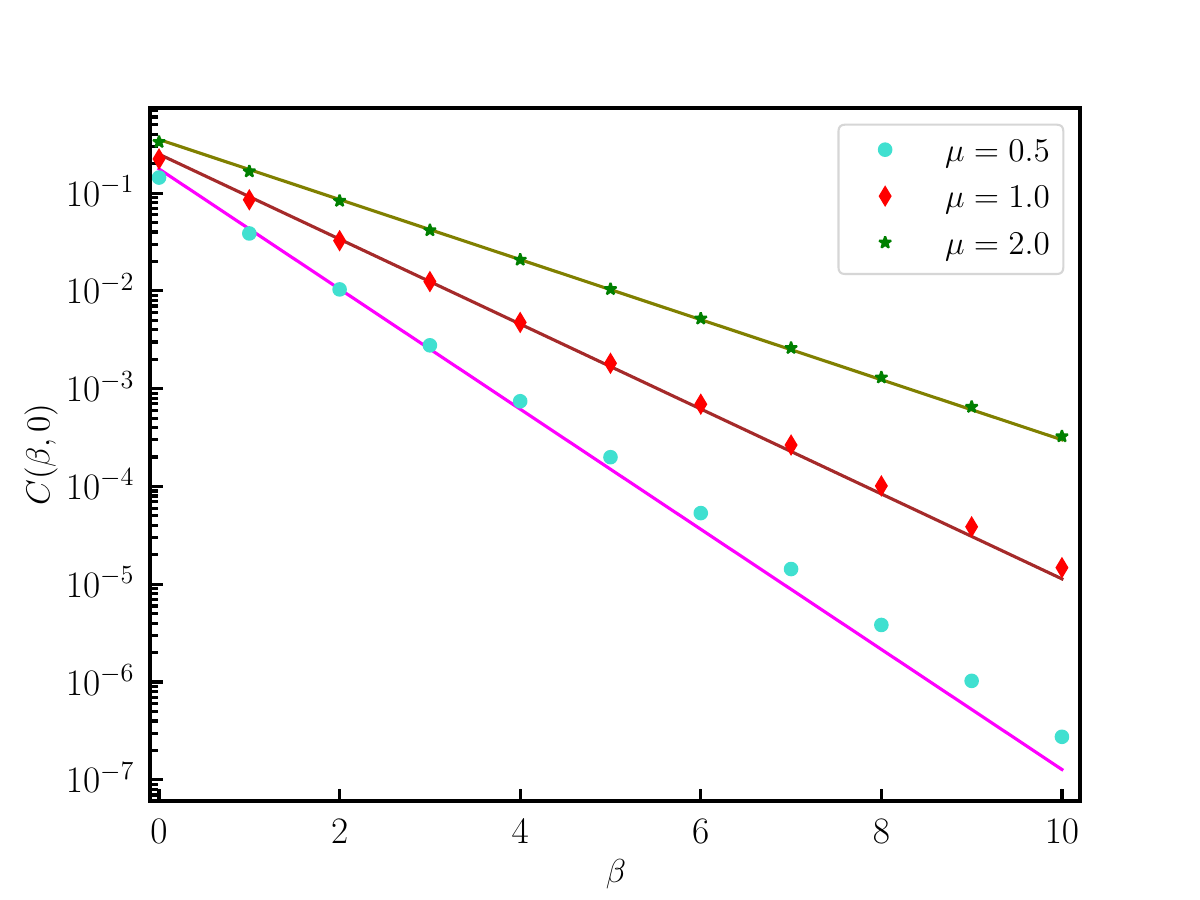}
            \caption{Comparison between the exact expression \eref{2tcqmFn} with the approximate expression \eref{eq:C_b0_active} for $C(\beta, 0)$ for different values of $\mu$, keeping $k=1$, $D_{\text{eff}}=1$, and $v_0=1$ fixed. The solid line shows the approximate expression, while the symbols shows the exact expression evaluated at integer values ($\beta=0, 1, \dots, 10$).}
           \label{fig:ecp_decay}
\end{figure}

In \fref{fig:stc_st_lt}, we compare the short-time and long-time behavior obtained from \eref{cbt1}, by taking first three terms, and \eref{2tcrAn} respectively, with the exact integral expression \eref{2tcr2n} for spatio-temporal correlation. \\


\noindent \textbullet~\textbf{Spatial correlation:}
The equal-time correlation $C(\beta,0)$ is given by \eref{2tcqmFn}.
It is interesting to investigate the behavior of $C(\beta,0)$ in the strongly active limit $\mu \gg 1$. However, it is not straightforward to obtain the behavior of \eref{2tcqmFn} in this limit. Hence, we extract the limiting behavior from the integral in ~\eref{2tcqm}, which is dominated by the contribution from the region $|q| \leq  1/\sqrt{\mu}$. Therefore, for large $\mu$, the integral can be approximated by
\begin{align}
     C(\beta,0) &\simeq  \frac{D_\text{eff}}{2 \pi k} \int_{-\infty}^{\infty}  \,\ dq \,  \frac{\cos{\beta q} }{ \mu q^2+1}  = \frac {D_\text{eff}}{k} \frac{\exp{(- |\beta|/\sqrt{\mu}})}{2 \sqrt{\mu}}.   \label{eq:C_b0_active}
\end{align}

On the other hand, in the passive limit $\mu \to 0$, we get from \eref{2tcqm}, 
\begin{align}
    C(\beta,0) = \frac{D_\text{eff}}{k} \delta_{\beta,0},  \label{eq:C_b0_passive}
\end{align}
consistent with the product measure equilibrium state \eref{peqy} obtained in the thermodynamic limit $N \to \infty$. Interestingly, although \eqref{eq:C_b0_active} is derived using the limit $\mu \gg 1$, taking $\mu \to 0$ in \eqref{eq:C_b0_active} we get \eqref{eq:C_b0_passive}, which is also consistent with \eqref{eq:C00} in large $N$ limit. We see in the thermodynamic limit, the spatial correlation decays exponentially with $\beta$ in the active limit. In \fref{fig:ecp_decay} we compare the approximate expression \eref{eq:C_b0_active} with the exact expression \eref{2tcqmFn} evaluated numerically and find very good agreement for large $\mu$.

\section{Conclusions}
\label{sec-conclusion}

We investigate the dynamical behavior of a harmonic chain of direction reversing active Brownian particles, by characterizing the variance of the position of the tagged particle and the statistics of separation between two consecutive particles. There are three intrinsic time scales in the system  $(k^{-1},\dr^{-1},\gamma^{-1})$, given by the coupling strength $k$, rotational diffusion constant $\dr$, and the direction reversal rate  $\gamma$. The interplay of these time-scales gives rise to multiple dynamical regimes [see Tables \ref{fig:table_FIC} and \ref{fig:table_UIC}], characterized by qualitatively different behavior of the fluctuations, which we analyze in this paper. In the long times $t$ much larger than the intrinsic time-scales, the system behaves similar to a Rouse polymer--- a harmonic chain of Brownian particles---exhibiting a $\sqrt{t}$ subdiffusive behavior of the tagged particle variance. However, in all other dynamical regimes, the signature of activity becomes apparent, giving rise to anomalous behavior $t^\nu$ with a range of values for $\nu$ in the different regimes. In particular, we consider two different scenarios--- one in which all the particles start with a quenched initial orientation $\theta_0$, and another where initial orientations of the particles are chosen randomly from a uniform distribution in $[0, 2\pi]$, independent of each other. 

 For the case of quenched initial orientation and well-separated time-scales, three limiting scenarios emerge, depending on the relative strengths of the coupling time scale $k^{-1}$, and the active time scales $\dr^{-1}$ and $\gamma^{-1}$. We refer to them as strong coupling  [$k \gg (\dr, \gamma)$], weak coupling [$k \ll (\dr, \gamma)$], and moderate coupling [$\gamma \ll k \ll \dr$ or  $\dr \ll k \ll \gamma$] limits. For a given order of coupling, there are four dynamical regimes, which we refer to as short-time, early-intermediate, late-intermediate, and long-time regimes [see Table \ref{fig:table_FIC}].  The short-time ($t \ll \{ k^{-1},\dr^{-1}, \gamma^{-1} \}$) and long-time ($t \gg \{ k^{-1},\dr^{-1}, \gamma^{-1} \}$) behaviors are found to be independent of the coupling limits. In the short-time regime, the position variance has a superdiffusive $t^3$ growth, similar to that of a single DRABP, as expected. On the other hand, the tagged particle exhibits the universal $\sqrt{t}$ subdiffusive behavior observed in single-file systems at long times. Contrarily, in the intermediate dynamical regimes, we observe different growth exponents in the three coupling limits. In the strong coupling limit, the variance grows as $t^{5/2}$ during the early-intermediate regime, followed by a $\sqrt{t}$ behavior in the late-intermediate regime. The interplay of rotational diffusion and direction reversal gives rise to $t^{5/2}$ behavior during the early intermediate regime, which is also observed in active Ornstein–Uhlenbeck particle (AOUP) in the intermediate regime of the strong coupling limit \cite{prashant}. In the moderate coupling limit, the variance grows as $t$ in the early intermediate regime, which crosses over to a subdiffusive $\sqrt{t}$ growth in the late intermediate regime. In the weak coupling limit, irrespective of the relative order of activity time scale, the variance grows linearly with time [see Table~\ref{fig:table_FIC}].  

For annealed initial orientation, interestingly, $\dr$ and $\gamma$ appear together giving rise to one activity time-scale $(\dr+2\gamma)^{-1}$ in addition to the coupling time scale $k^{-1}$. In this case, there are only two limiting scenarios possible, namely, the strong coupling limit, given by $k \gg (\dr+2\gamma)$, and the weak coupling limit, given by $k \ll (\dr+2\gamma)$. Evidently, there are three dynamical regimes in each coupling limit [see Table~\ref{fig:table_UIC}]. The short-time ($t \ll \{k^{-1},(\dr+2\gamma)^{-1}\}$) and long-time ($t \gg \{k^{-1},(\dr+2\gamma)^{-1}\}$) behaviors are again independent of the coupling limits. In the short-time regime, the position variance of the tagged particle grows as $t^2$, similar to that of an independent DRABP with annealed initial orientation. On the other hand, as expected, at long times the variance grows subdiffusively as $\sqrt{t}$. The behavior in the intermediate regime depends on the relative ordering of the time-scales--- in the strong coupling limit, the variance grows superdiffusively as $t^{3/2}$, while in the weak coupling limit, a linear growth is observed.

We also explore the finite-size effect on the tagged particle position variance. For a finite but large system size $N$, there is an additional time scale $t_N=N^2/k$  and the finite size effects appear for $t \gg t_N$. In this regime, the tagged particle exhibits center of mass motion with variance growing linearly with time, as $2(D_{\text{eff}}/N)t $. The crossover from the thermodynamic limit $t \ll t_N$, to the finite size dominated regime $t \gg t_N$ is captured by a scaling behavior of the variance $D_{\text{eff}}\sqrt{t/k}\,f(t/t_N)$, where the crossover function $f(z)$ goes to a constant as $z \to 0$ and it diverges sublinearly as $\sqrt{z}$ for large $z$.

 In addition, we investigate the velocity autocorrelation function $\langle v_{\alpha}(t_1) v_{\alpha}(t_2) \rangle$ . We observe that the velocity autocorrelation function reaches a stationary state $F(\tau)$, in the limit $ \{t_1,t_2 \} \to \infty $, keeping $\tau=|t_1 -t_2|$ finite. At late-times, i.e., $\tau \gg \{ k^{-1}, (\dr+2\gamma)^{-1} \}$, we calculate the closed form expression for $F(\tau) \simeq -\frac{D_{\text{eff}}}{4 \sqrt{\pi k}} \tau^{-3/2}$, and show that velocity autocorrelation decays as power law in $\tau$.  

Finally, we analyze the statistics of the separation $y_{\alpha}$ between two consecutive particles $\alpha$ and $\alpha+1$. For a thermodynamically large harmonic chain of passive Brownian particles, the separation variables reach an equilibrium state given by the Boltzmann distribution, which has a Gaussian product measure. 
We find that, for a harmonic chain of DRABPs, the activity breaks this product measure in the nonequilibrium stationary state, giving rise to nontrivial spatio-temporal correlations among the separation variables. First, we derive a series expansion of the spatio-temporal correlation function $C(\beta,\tau) = \lim_{t \to \infty\atop}\langle y_0(t)y_{\beta}(t+\tau) \rangle$ in powers of $\tau$. We also show that for large $\tau$, $C(\beta,\tau) \sim e^{-\beta^2/(4 k \tau)}/\sqrt{\tau}$. Futhermore, we show that for large activity, $\mu=k/(\dr+2\gamma) \gg 1$, the spatial correlation $C(\beta,0)$ decays exponentially.

The simple model studied here provides a way to explore the behavior of interacting active particles with multiple internal time-scales. It would be interesting to compare our results of different dynamical regimes with those in the presence of short-range interactions, such as the Lennard-Jones potential. Another obvious question is how the system behaves when inertial effects are included. Finally, it would be intriguing to explore the behavior of the tagged particle in a mixture of active and passive particles.

\appendix

\section{Computation of effective noise correlation}\label{t-t-corr}

In this Appendix, we provide the detailed calculation for the auto-correlation of the effective noise $\xi_{\alpha}(t)$ defined in \eref{eq:xi}.  
Since the $\sigma_{\alpha}$ and $\theta_{\alpha}$ processes are independent, it suffices to compute the correlations of $\sigma_{\alpha}(t)$ and $\cos{(\theta_\alpha(t))}$ separately and take the product to get correlation of $\xi_{\alpha}(t)$. The propagator for the $\theta_{\alpha}(t)$ and $\sigma(t)$ processes are given by,
\begin{align}
P_{\theta}(\theta_{\alpha} ,t|\theta_{0},0) &=\frac{1}{\sqrt{4\pi \, \dr \, t}} \exp \left[-{\frac{(\theta_{\alpha}-\theta_{0})^2}{4 \, \dr \, t }} \right], 
\label{tp}
\intertext{and}
P_{\sigma}(\sigma_{\alpha},t|\sigma_0,0) &=  \frac{1}{2}\bigg[1+\sigma_{\alpha} \sigma_0 \, e^{-2\gamma t }\bigg].
\label{sp}
\end{align}
Using the above propagators, we get the mean and the auto-correlation of the noise for both quenched and annealed initial orientations below. \\

\noindent (i) \emph{Quenched initial orientation}: From \eref{tp} and \eref{sp}, we respectively get, 
\begin{align}
 \big \langle  \cos{ \theta_{\alpha}(t)} \big \rangle &=  \int_{-\infty}^{\infty} \frac{\cos{\theta_{\alpha}} }{\sqrt{4 \, \pi \, \dr \, t}} \exp \left[-{\frac{(\theta_{\alpha}-\theta_{0})^2}{4 \, \dr \, t }} \right] \, d\theta_{\alpha} = \cos\theta_{0}\,e^{-\dr \, t}, \label{ctavg_f} 
\intertext{and}
 \big \langle  \sigma_{\alpha}(t) \big \rangle &= \sum_{\sigma_{\alpha}= \pm  1}^{} \frac{\sigma_{\alpha}}{2}\big[1+\sigma_{\alpha} \, e^{-2\gamma t }\big] =  e^{-2\gamma t}.
 \label{stavg_f}
\end{align}
Therefore, from \eref{eq:xi}, we have,
\begin{equation}
\big \langle  \xi_{\alpha}(t) \big \rangle=  v_0 \big \langle  \cos{\theta_{\alpha}(t)} \big \rangle \,\big \langle  \sigma_{\alpha}(t) \big \rangle = v_{0} \, \cos{\theta_{0}} \,e^{-(2\gamma +\dr)t}.
\label{xiavg_f}
\end{equation}
The two-time auto-correlations can also be calculated similarly, yielding,
\begin{align}
 \big \langle   \cos{ \theta_{\alpha}(t_1)} \cos{ \theta_{\alpha}(t_2)} \big \rangle &=   e^{-D_{R} \mid t_{1}-t_{2} \mid} + \cos{2\theta_{0}} \, e^{-D_{R} \left( t_{1}+ t_{2}+ 2 \min [ t_{1},t_{2} ]\right)},
\label{c_tp} \\
   \big \langle  \sigma_{\alpha}(t_{1})  \sigma_{\alpha}(t_{2}) \big \rangle &= e^{-2\gamma \mid t_{2}-t_{1} \mid },
   \label{s_tpc} 
\end{align}
Consequently,
\begin{align}   
 \big \langle   \xi_{\alpha}(t_1) \xi_{\alpha}(t_2)  \big \rangle &=   \frac{v_{0}^2}{2} \left[e^{-(D_{R}+2\gamma) \mid t_{1}-t_{2} \mid} + \cos{ 2\theta_{0}}\,e^{-D_{R}(t_{1}+ t_{2}+ 2 \min [ t_{1},t_{2} ]) -2\gamma \mid t_{1}-t_{2} \mid } \right].
\label{ficc_tp}
\end{align}

\noindent (ii) \emph{Annealed initial orientation}: 
To find the correlations for annealed initial orientations, we average over $\theta_0$ with respect to a uniform distribution in $[0,2 \pi]$ in \eref{xiavg_f} and \eref{ficc_tp}. Noting $\langle \cos \theta_0 \rangle =\langle \cos 2 \theta_0 \rangle =0$,   
\begin{align}
\big \langle  \xi_{\alpha}(t) \big \rangle = 0 \quad\text{and}\quad \big \langle   \xi_{\alpha}(t_1) \xi_{\alpha}(t_2)  \big \rangle =   \frac{v_{0}^2}{2} e^{-(D_{R}+2\gamma) \mid t_{1}-t_{2} \mid}.
\label{ficc_tpr}
\end{align} 

From \eref{xiavg_f}, \eref{ficc_tp}, and  \eref{ficc_tpr},  it is clear that $G(t_1,t_2)\equiv\big \langle   \xi_{\alpha}(t_1) \xi_{\alpha}(t_2)  \big \rangle - \big \langle   \xi_{\alpha}(t_1) \big \rangle  \big \langle   \xi_{\alpha}(t_2)  \big \rangle$, is given by \eref{ficc} and \eref{g_ric} for the quenched and annealed initial orientations respectively.

\section{Computation of $C_n(\beta)$}\label{app:Snq}
In \eref{cbt1}, we have expressed the spatio-temporal correlation function of the separation variables $C(\beta,\tau)$ in a power series of $\tau$ with coefficients $C_n(\beta)$. The goal of this appendix is to evaluate $C_n(\beta)$, starting from \eref{cnbeta1}. These coefficients involve $S_n(q)$, defined in \eref{sf}, can be expressed as
\begin{align}
a^n \, S_n(q) &= \sum_{r=0}^{n-2} \frac{ b_q^{r+1} \, a^{n-r-1}}{b_q+a}, \quad\text{with} \quad a=\dr+2\gamma.
\end{align}
 Writing $b_q^{r+1}=(b_q+a-a)^{r+1}$ and performing binomial expansion in powers of $b_q+a$ and $a$ we get,
\begin{align}
 a^n \, S_n(q) &= \sum_{r=0}^{n-2} \, \sum_{m=0}^{r+1} \, (-1)^{r+1-m}\, \binom{r+1}{m} \, (b_q+a)^{m-1} \, a^{n-m}. 
\end{align}
Separating the $m=0,1$ terms and performing another binomial expansion of $(b_q+a)^{m-1}$ for $m>1$ we get,
\begin{align}
a^n \, S_n(q) &= -\frac{(-a)^n}{b_q+a} \cdot \frac{1+(-1)^n}{2} + \frac{(-1)^n \,a^{n-1}}{4} \bigl[1+(2n-1)(-1)^n \bigr] \cr 
 & - \sum_{r=1}^{n-2} \, \sum_{m=2}^{r+1} \, \sum_{l=0}^{m-1} \, (-1)^{r-m}  \binom{r+1}{m} \, \binom{m-1}{l} \, b_q^l \, a^{n-l-1}.
\label{snmrs}
\end{align}
Now, inserting the above expression of $S_n(q)$ in \eref{cnbeta1} and carrying out the integration over $q$ we get,
\begin{align}
&C_n(\beta) = \frac{1+(-1)^n}{2} \,  C_0(\beta)  - \left[ n-\frac{1-(-1)^n}{2} \right] \frac{D_{\text{eff}}}{2k} \, \delta_{\beta,0} + \Tilde{S}_n,
\label{cnbetaf6}
\end{align}
where we used,
\begin{equation}
 \int_{-\pi}^{\pi} \, dq \, \cos{(\beta q)} \, b_{q}^l=  \begin{cases}
\displaystyle \frac{(-1)^{\beta}  \,  (2l)! \, 2 \pi k^l }{(l-\beta)!(l+\beta)!}  &\text{for}~ l \ge \beta, \\
0 & \text{for}~ l < \beta,
 \end{cases}
  \label{cnbia}
\end{equation} 
and
\begin{equation}
\Tilde{S}_n = \frac{D_{\text{eff}}}{k} \, \sum_{r=1}^{n-2} \, \sum_{m=2}^{r+1} \, \sum_{l=0}^{m-1} \, (-1)^{n+\beta+r-m}  \binom{r+1}{m} \, \binom{m-1}{l} \, \binom{2l}{l-\beta} \, \mu^l.
\label{cnbetafapp6}
\end{equation}
To evaluate this sum, it is useful to change the order of the summations, which yields,
\begin{align}
\Tilde{S}_n  = (-1)^{n+\beta} \, \frac{D_{\text{eff}}}{k} \, \sum_{l=0}^{n-2} \binom{2l}{l-\beta} \, \mu^l \, \sum_{r=1}^{n-2} \, \sum_{m=2}^{r+1} \, (-1)^{r+m} \,  \binom{r+1}{m} \, \binom{m-1}{l}, 
\label{cnbetafapp7}
\end{align}
where $\binom{m-1}{l}=0$ for $m<l+1$. Furthermore, it is convenient to separate the $l=0$ term as,
\begin{align}
\Tilde{S}_n  =  (-1)^{n+\beta} \, \frac{D_{\text{eff}}}{k} \Bigg[ \sum_{l=1}^{n-2} \binom{2l}{l-\beta} &\, \mu^l \, \sum_{r=1}^{n-2} \, \sum_{m=l+1}^{r+1} \, (-1)^{r+m} \,  \binom{r+1}{m} \, \binom{m-1}{l} \cr 
&+ \delta_{\beta,0} \sum_{r=1}^{n-2} \sum_{m=2}^{r+1} (-1)^{r+m} \binom{r+1}{m} \Bigg].
\label{cnbetafapp8}
\end{align}
The summations over $r$ and $m$ can be performed explicitly, yielding,
\begin{align}
\Tilde{S}_n  = (-1)^{n+\beta+1}  \, \frac{D_{\text{eff}}}{k} \, \Bigg[\sum_{l=1}^{n-2}   \frac{1+ (-1)^{l+n}}{2}\, \binom{2l}{l-\beta} \,\mu^l  +   \frac 14 \delta_{\beta,0} \big(1- (2n-3)(-1)^n\big) \Bigg].
\label{cnbetaf10}
\end{align}
Combining \eref{cnbetaf6} and \eref{cnbetaf10} we get,
\begin{align}
C_n(\beta) = \frac{1+(-1)^n}{2} \,  \left[ C_0(\beta)  -  \frac{D_{\text{eff}}}{k} \, \delta_{\beta,0} \right] - (-1)^{n+\beta}  \, \frac{D_{\text{eff}}}{k} \, \sum_{l=1}^{n-2}   \frac{1+ (-1)^{l+n}}{2}\, \binom{2l}{l-\beta} \,\mu^l,
\label{cnbetaF1}
\end{align}
where, $\binom{2l}{l-\beta}=0$ for $l<\beta$. Note that the term containing the summation over $l$ contributes only for $n \geq 3$. For $n=2$,  we have \eref{c2:eqn}  whereas for  $n\ge 3$ we get \eref{cnbetaF} in the main text. \\

\section{Details about numerical simulation}
To numerically simulate the interacting system, we discretize the Langevin equation \eref{eomx} in the time step of $\Delta t$, to the first order in $\Delta t$. During the time interval $[t,t+\Delta t]$, we update the position of each particle parallelly as,
\begin{equation}
    x_{\alpha}(t+\Delta t)= x_{\alpha}(t) + \Delta x_{\alpha}(t), \label{eq:discr_x}
\end{equation}
where, 
\begin{equation}
     \Delta x_{\alpha}(t)= k \left [  x_{\alpha+1}(t)+ x_{\alpha-1}(t)-2 x_{\alpha}(t) \right ] \Delta t  + v_0 \, \sigma_{\alpha}(t) \, \cos{\theta_{\alpha}(t)} \Delta t \label{eq:delta_x}
\end{equation}
We update $\theta(t)$ dynamics using
the Euler-Maruyama method, 
 \begin{equation}
     \theta_{\alpha}(t+\Delta t)=\theta_{\alpha}(t)+\sqrt{2 \dr \Delta t} \,\, W_{\alpha}(t), \label{eq:discr_theta}
 \end{equation}
where $\{W_{\alpha}(t)\}$ are independent and identically distributed random numbers, drawn from a Gaussian distribution with zero mean and unit variance. Moreover, we have,   
\begin{equation}
    \sigma_{\alpha}(t+\Delta t)= \begin{cases}
        - \sigma_{\alpha}(t)  \quad & \text{with probability} ~~\gamma \Delta t \\
        \sigma_{\alpha}(t) &  \text{with probability} ~~1-\gamma \Delta t,    
    \end{cases}
    \label{eq:discr_sigma}
\end{equation}
independently for each $\alpha$. We use $\Delta t= 10^{-3}$ in all our simulations.  

The expectation value  $\la \mathcal{O} \ra$ of an observable $\mathcal{O}$ is numerically evaluated using 
\begin{equation}
    \la \mathcal{O} \ra = \frac 1{\mathcal N}\sum_{\mathcal R=1}^{\mathcal N} \mathcal O(\mathcal R)
    \label{avg_eqn}
\end{equation}
where $\mathcal{O}(\mathcal R)$ represents the value of the observable for a given realization $\mathcal R$ and $\mathcal N$ represents the number of independent realizations.  Using the discretized dynamics \eref{eq:discr_x}-\eref{eq:discr_sigma}, we numerically evaluate the expectation values $\big \langle  x_{\alpha}(t) \big \rangle$, $\big \langle  x_{\alpha}^2(t) \big \rangle$, $\big \langle  x_{\alpha}(t) x_{\beta}(t') \big \rangle$,  etc., by averaging over $10^{5}$-$10^{6}$ realizations, \`a la \eref{avg_eqn}.

\section{Some useful integrals and sums}\label{aap:int} 

For the convenience of the readers, we present a collection of integrals and summations, that appear in the main text, below. \\

\reqnomode

\begin{align}
 \hspace{5.5em} & \int_{-\infty}^{\infty} dz \frac{e^{-(z^2+a^2)}}{a^2+z^2} = \frac{\pi}{a} \left(1-\text{erf}{(a)}\right),
 \label{AI1} \\
 & \int_{-\infty}^{\infty} dz \frac{(e^{-a^2}e^{-z^2}-e^{-2 z^2})}{a^2-z^2} = \frac{\pi}{a} e^{-2a^2} \left[\text{erfi}(a)- \text{erfi}(\sqrt{2}a) \right], 
 \label{AI2} \\
 & \int_{-\infty}^{\infty} dz \frac{e^{-a^2}-e^{-2 z^2}}{a^2-2z^2} = -\frac {\pi}{\sqrt{2}a}  e^{-a^2} \text{erfi}(a), \label{AI3} \\
 & \int_{-\infty}^{\infty} dz \frac{e^{-(z^2+b^2)}-e^{-a^2}}{a^2-b^2-z^2} = \frac {\pi}{\sqrt{b^2-a^2}}  e^{-a^2} \text{erf} \left(\sqrt{b^2-a^2} \right),  \label{AI4} \\
 & \begin{aligned}
  \int_{-\infty}^{\infty}  dz\left[\frac{e^{-a}-e^{-z^2}}{a-z^2} \right]^2 &= -\frac{\pi e^{-2a}}{a^{3/2}} \Bigg[ (2a+1)\,\text{erfi}\left(\sqrt{a}\right)
    - \left(2a+\frac{1}{2} \right) \,\text{erfi}\left(\sqrt{2a}\right) \Bigg] \\
   & -\frac{\sqrt{2\pi} }{a}\left( 1 -\sqrt{2} e^{-a} \right),
 \end{aligned} 
 \label{AI5} \\
 & \sum_{m=0}^{\infty}\frac{\Gamma \left(m+\frac{1}{2}\right)}{ \sqrt{\pi } \, m!} \, a^m = \frac{1}{\sqrt{1-a}}, \label{exp_sum1}   \\
 & \sum_{m=0}^{\infty} \frac{\Gamma \left(m+n+\frac{1}{2}\right)}{ \sqrt{\pi } \, (m+n)!} \, a^m  =\frac{\Gamma(n+1/2)} {\sqrt{\pi} \, n!}\, {}_2F_1 \left(1,n+ \frac 12 ;1+n;a \right).
    \label{expansion_sum}
\end{align}

Integrals \eref{AI1}-\eref{AI5} are used in Sec.~\ref{scl-sc} to derive \eref{scEq} in the strong coupling limit, by taking appropriate limit in \eref{zeqn}. Summations \eref{exp_sum1}-\eref{expansion_sum}  is used in Sec.~\ref{wcl-sc}, i.e., weak coupling limit, to evaluate the sum in \eref{bqeqnee}.

\vspace*{0.5 cm}

\section*{References}


\bibliographystyle{iopart-num}
\bibliography{ref}

\end{document}